\documentclass[a4paper, amsfonts, amssymb, amsmath, preprint, showkeys, nofootinbib, twoside]{revtex4-2}
\usepackage{array}[=2016-10-06]
\usepackage{multirow, makecell, booktabs, siunitx, tabularx}
\usepackage[version=3]{mhchem}
\usepackage[table,dvipsnames]{xcolor} 
\usepackage{colortbl}
\usepackage{graphicx, float}
\usepackage{textcomp}
\usepackage{chemformula, gensymb}
\usepackage[T1]{fontenc}
\usepackage{subcaption}
\usepackage{bm}   
\usepackage{bbm}  
\usepackage{xpatch}

\newcolumntype{L}{>{\raggedright\arraybackslash}X}
\newcommand{\diag}{\operatorname{diag}}

\newcommand{\sinc}{\operatorname{sinc}}
\newcommand{\rect}{\operatorname{rect}}
\newcommand{\E}{\mathbb{E}}
\DeclareMathOperator{\Var}{Var}

\makeatletter
\patchcmd{\@ssect@ltx}
  {\addcontentsline{toc}{#1}{\protect\numberline{}#8}}
  {}
  {}
  {}
\makeatother

\begin{document}


\author{Josh Leeman}
    \affiliation{Department of Chemistry, Princeton University, Princeton, NJ 08540, USA}

\author{Sudipta Chatterjee}
    \affiliation{Department of Chemistry, Princeton University, Princeton, NJ 08540, USA}


\author{Max Hirschberger}
    \affiliation{Department of Applied Physics and Quantum-Phase Electronics Center (QPEC), The University of Tokyo, Bunkyo-ku, Tokyo, 113-8656 Japan}

\author{Leslie M. Schoop}
    \email{lschoop@princeton.edu}
    \affiliation{Department of Chemistry, Princeton University, Princeton, NJ 08540, USA}
    \affiliation{Princeton Materials Institute, Princeton University, Princeton, NJ 08540, USA}

\title
  {From raw magnetoresistance to reliable cyclotron masses: A practical guide to processing quantum oscillation data}
\begin{abstract} 
Quantum oscillations are an essential probe of the Fermi surface in conductors, and cyclotron masses extracted using the Lifshitz–Kosevich (LK) formula are used to evaluate correlation strength and the assignment of topological states.
These masses depend on choices in signal processing, including the field window, method of background subtraction, and spectral window applied before the Fourier transform. These choices are often guided by visual inspection, yet they can alter spectral features and bias the extracted masses. We show that (1) representing a finite field interval by a single effective field can underestimate the cyclotron mass; (2) polynomial background subtraction can attenuate and shift low-frequency peaks; and (3) spectral tapering can reduce mass bias from leakage between neighboring peaks, at the cost of frequency resolution. We demonstrate these effects using numerical calculations and Shubnikov–de Haas experiments on the rare-earth tritellurides $R$Te$_3$ ($R$ = Pr, Nd, Gd, Tb) with a dense low-frequency spectrum in the charge ordered state. The examples reveal masses shifting by tens of percent with the field window, apparent peak shifts produced only by the background fit, and a reported departure from LK behavior that resolves into two neighboring frequencies.

\end{abstract}

\maketitle

\section{Introduction} 
Quantum oscillations describe the periodic oscillation of observables as a function of inverse magnetic field. They are among the most powerful experimental probes of the electronic structure of metals and semimetals, providing information about the geometry of the Fermi surface, cyclotron masses, and quantum scattering lifetimes of electrons. These quantities are important for understanding materials relevant to electronics~\cite{Huber2023, QO_HighSens_Zhang, 10.1021/acsaelm.4c01089, 10.1021/acsnano.5b02020}, spintronics~\cite{PhysRevB.109.134404, 10.1002/advs.202303978, v73x-5ptd}, and other emerging quantum technologies~\cite{PhysRevB.80.085303, QO_Heterostruct, PhysRevB.99.165133, Zhao2019}. However, reliably extracting these quantities depends on a sequence of data processing choices. Some of these choices include selecting the range of magnetic field over which an oscillation amplitude is determined, estimating a non-oscillatory background to be subtracted before processing, and using a spectral window before Fourier transform of oscillation patterns. These choices can change oscillation amplitudes and redistribute spectral weight. Here, we examine how they affect the extracted quantities and how the resulting biases can be diagnosed.

Quantum oscillations can be measured in several material properties, but most commonly in electrical resistivity and magnetization. Since the initial observation of these oscillations in bismuth~\cite{dehaas1930dependence, 10.1038/srep18674, Shubnikov1930Nature}, they have been detected in many other materials, predominantly in single crystals of metals and semimetals~\cite{Potter2014_10.1038, PhysRevLett.117.077201, PhysRevB.109.155117, PhysRevB.96.045127, Matusiak2017_ncomms15219, PhysRevLett.114.176601, PhysRevB.107.104421}, but also in thin films~\cite{PhysRevB.58.R13318, PhysRevB.93.081103, PhysRevLett.91.226801, Liu2015_am.2015.110, Li2015_10.1038}, low-dimensional systems~\cite{QO_MagneticBreakdown,QO_PrTe3, QO_NdTe3, PhysRevB.110.L161108}, and bulk topological insulators~\cite{PhysRevB.80.085303, PhysRevLett.116.046404, PhysRevB.100.085124}. Advancements in instrumentation and experimental techniques have also broadened the range of accessible observables; for example, lock-in detection improves sensitivity~\cite{PhysRevB.71.081303, nv42-w4t7, Chen2024_s42005-024-01687-6}, modulation techniques permit measurements of derivative signals~\cite{Hollricher2026Cyclotron, Oder1965Temperature, PhysRevX.14.021012}, and high-field facilities can reveal oscillations that are strongly damped at lower fields~\cite{PhysRevLett.100.047003, VIGNOLLE2011446, Chang2026_s41928-026-01590-8}.

Figure 1 shows the general steps for extracting quasiparticle parameters from quantum oscillations. While the principles of quantum oscillation analysis are well established, processing choices are often guided by visual inspection of the extracted oscillations or Fourier spectrum. The extracted frequencies and amplitudes can be particularly sensitive to these choices when the oscillations have low frequencies or the background contribution is not smooth. Small pockets produce low-frequency oscillations, so only a few periods may be observed over the accessible field range. A low-degree polynomial may fail to capture sharp features or changes in curvature in the background, whereas a higher degree fit may absorb the low-frequency oscillations themselves. 

Different but equally reasonable processing choices can yield substantially different effective masses from the same experimental data. A result may appear reasonable even when the extracted mass changes strongly with the processing parameters. Therefore, systematically varying these parameters is important for assessing its reliability. Reporting these parameters is equally important for distinguishing physical differences in masses across studies from those introduced by the analysis.


The rare-earth tritellurides are a useful testbed for these processing steps. They are highly conducting, with large residual resistivity ratios and small cyclotron masses, so oscillations are visible even at modest fields~\cite{PhysRevMaterials.4.114005, QO_GdTe3_Shiming, QO_NdTe3, QO_PrTe3, PhysRevB.73.033101, Yumigeta_AS_2004762, 10.1063/5.0306708, PhysRevB.77.035114, PhysRevB.67.144417, PhysRevB.104.155147, 10.1063/5.0007517, Zaitsev-Zotov2025}. Their closely related crystal structures and differing charge density wave and magnetic states allow us to test how the processing choices affect the analysis across several related compounds.


\begin{figure}[!htbp]
    \centering
    \includegraphics[width=3in]{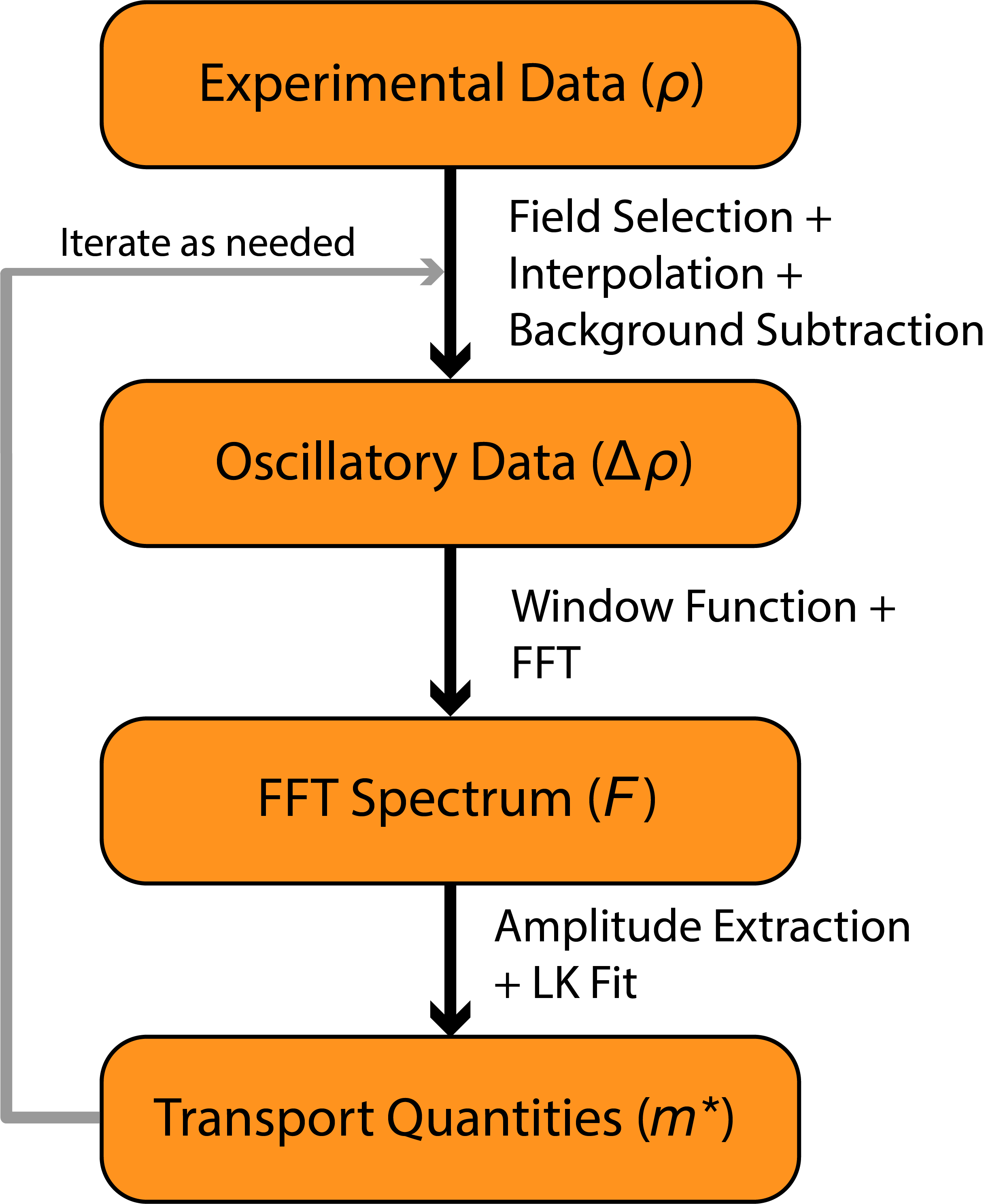}
    \caption{Extracting quasiparticle parameters from quantum oscillations. This work focuses on extracting the cyclotron effective mass $m^*$ from resistivity $\rho$.}
    \label{Fig1}
\end{figure}

This paper is organized as follows. Section II introduces the theory behind quantum oscillations, with emphasis on magnetotransport. Section III examines the effects of field range selection, background subtraction, and windowing on the extracted quantities. Lastly, Section IV uses measurements of the rare-earth tritellurides to show how processing choices can explain apparent anomalies and affect the interpretation of quantum oscillations.


\section{Introduction to Quantum Oscillations}
\label{sec:qo_background}
Quantum oscillations come from the quantization of electronic states in a magnetic field. In the semiclassical picture, charge carriers move along closed orbits on the Fermi surface in planes perpendicular to an applied field. An applied magnetic field quantizes the cyclotron motion into Landau levels, and sweeping a sufficiently strong field drives successive levels through the Fermi energy, producing quantum oscillations observable in thermodynamic and transport quantities. Oscillations observed in magnetization are referred to as de Haas--van Alphen (dHvA) oscillations, and those in electrical resistivity as Shubnikov--de Haas (SdH) oscillations. The following section generally follows the standard treatment in Shoenberg~\cite{shoenberg1984magnetic}.

To observe quantum oscillations, thermal broadening and disorder-induced scattering must not completely obscure the Landau level structure. Consequently, measurements are generally performed at low temperatures and sufficiently strong magnetic fields using samples with long quasiparticle-scattering times. Although our examples focus on SdH oscillations, the same processing considerations apply to oscillations in magnetization, heat capacity, and torque.

\subsection{Oscillation frequencies and the Fermi surface}
\label{sec:onsager_relation}

\begin{figure}[!htbp]
    \centering
    \includegraphics[width=5.64in]{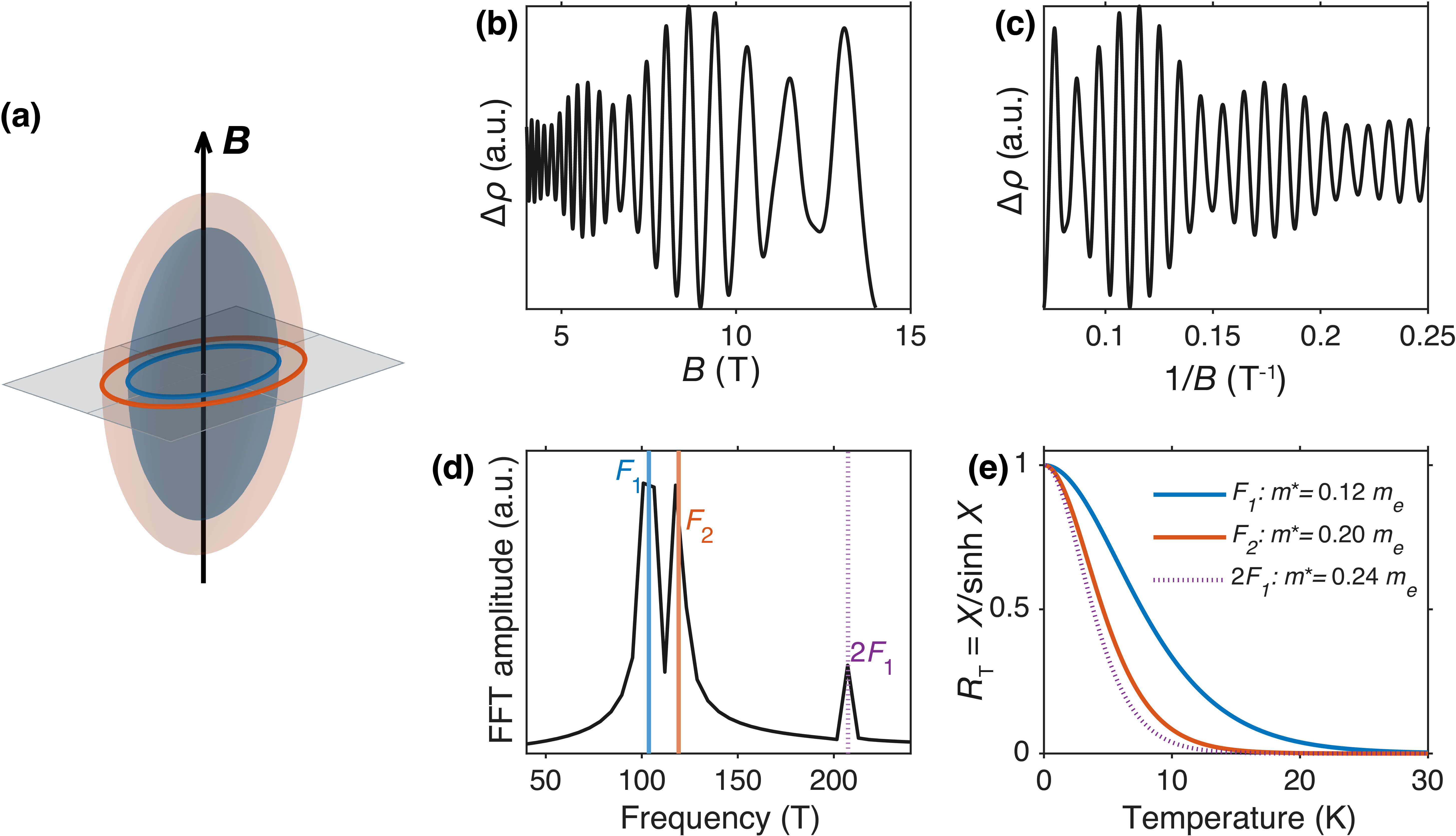}
    \caption{
    \textbf{(a)}~Schematic three-dimensional Fermi surface comprising two sheets (blue and orange ellipses) intersected by a plane perpendicular to the applied magnetic field, defining two extremal orbits (blue and orange lines). Simulated oscillations associated with the two orbits, including the field-dependent damping from thermal broadening and quasiparticle scattering, plotted against \textbf{(b)}~field and \textbf{(c)}~inverse field. \textbf{(d)}~Fast Fourier transform (FFT) of the finite oscillatory signal, showing the two fundamental frequencies $F_1$ and $F_2$ together with a higher-harmonic contribution at $2F_1$. \textbf{(e)}~LK thermal damping factors for the two orbits at fixed magnetic field illustrating the more rapid suppression of the oscillation amplitude associated with the larger $m^*$. The second harmonic fulfills the relation $m^*_{2F} = 2 m^*_{F}$.}
    \label{Fig2}
\end{figure}

The frequency of a quantum oscillation is related to an extremal cross-sectional area of the Fermi surface through the Onsager relation~\cite{onsager1952} given by
\begin{equation}
    F
    =
    \frac{\hbar}{2\pi e}A_{\mathrm{ext}},
    \label{eq:onsager}
\end{equation}
where \(F\) is the oscillation frequency, \(A_{\mathrm{ext}}\) is an extremal cross-sectional area of the Fermi surface perpendicular to the applied magnetic field, \(\hbar\) is the reduced Planck constant, and \(e\) is the elementary charge. An extremal orbit is a maximum or minimum cross section of the Fermi surface along the magnetic field direction (Fig.~\ref{Fig2}a), where the cross-sectional area is stationary. A single Fermi surface pocket can therefore produce more than one oscillation frequency when it contains both minimum and maximum extremal cross sections. The extremal area depends on the field direction, and rotating the sample traces out different cross sections, mapping out the three-dimensional shape of the Fermi surface in \textit{k}-space.

Note that while the data are generally recorded linearly in field, the oscillatory signal is periodic in \textit{inverse} magnetic field (see Fig.~\ref{Fig2}b--c). In its simplest form, a single oscillatory component $i$ can be written as
\begin{equation}
    \Delta \rho_i(B) =
    A_i(B)\cos\left[2\pi\left(\frac{F_i}{B}-\gamma_i\right)\right],
    \label{eq:simple_oscillation}
\end{equation}
where $A_i(B)$ is the field-dependent amplitude, $F_i$ is the oscillation frequency, and $\gamma_i$ represents the phase. The phase can contain contributions from Landau quantization, the dimensionality and curvature of the Fermi surface, and the Berry phase~\cite{mikitik1999}. It is often extracted from a Landau fan plot, in which the index assigned to each oscillation extremum is plotted against its inverse field. The slope gives the frequency, while the intercept gives the phase according to the chosen indexing convention. Processing can shift the apparent positions of these extrema and bias the extracted phase~\cite{Fominykh_2026, 6tq6-sr6w}. A seemingly linear Landau fan does not by itself establish a reliable Berry phase, particularly when only a few oscillations are resolved or multiple components overlap. A detailed discussion of phase extraction is beyond the scope of this work.

For a material containing several extremal orbits, the oscillatory response is therefore a sum of components $i$,
\begin{equation}
    \Delta \rho_{\mathrm{osc}}(B)
    =
    \sum_i
    A_i(B)
    \cos\left[
        2\pi
        \left(
            \frac{F_i}{B}
            -
            \gamma_i
        \right)
    \right].
    \label{eq:multiband_oscillation}
\end{equation}
A Fourier transform with respect to $1/B$ reveals the frequencies $F_i$, and through the Onsager relation, the corresponding extremal areas in the Fermi surface. In an ideal measurement over an infinitely large inverse-field interval, distinct oscillatory components would appear at sharply defined frequencies. However, experimental measurements span a finite field interval and thus contain a limited number of oscillation periods. The resulting Fourier peaks consequently have finite widths that may overlap with neighboring peaks (see Sec.~\ref{Field range selection}). Furthermore, the $p$th harmonic at integer multiples $pF_i$ of the fundamental frequency may be visible in the FFT spectrum, though generally at higher fields and lower temperatures. 

Physically, the harmonic peaks reflect how sharply the Landau levels are resolved as they pass through the Fermi energy. The sharpness of these level crossings, and thus how many harmonics survive, depends on thermal broadening, and scattering-induced broadening; these will be discussed in the next section. As scattering-induced broadening is affected by field and not by temperature, lowering the temperature alone does not recover higher harmonics---high magnetic fields are required as well.


\subsection{Lifshitz--Kosevich formalism}
\label{sec:lk_background}

The field- and temperature-dependent amplitudes of quantum oscillations are commonly described using the Lifshitz--Kosevich (LK) formalism~\cite{LIFSHITS19581, shoenberg1984magnetic}. The oscillation amplitude of an orbit $i$ may be expressed as
\begin{equation}
    A_i(B,T)
    =
    A_{i,0}(B) \, R_{T,i} \, R_{D,i} \, R_{S,i},
    \label{eq:lk_amplitude}
\end{equation}
where $R_T$ and $R_D$ describe thermal and scattering-induced damping, respectively, and $R_S$ is the spin-splitting factor. Each factor depends on the orbit through its cyclotron mass and scattering rate; for readability we suppress the orbit index $i$, and all amplitudes and damping factors refer to a single extremal orbit unless stated otherwise. The $A_0(B)$ term collects the remaining temperature-independent prefactors, which may depend on the measured quantity, as well as Fermi surface dimensionality and experimental geometry.

The thermal damping factor is
\begin{equation}
    R_{T}
    =
    \frac{X}{\sinh X}, \qquad
            X
    =
    \frac{2\pi^2 k_{\mathrm{B}}T m^*}
         {\hbar eB},
    \label{eq:thermal_damping}
\end{equation}
where $m^*$ is the cyclotron effective mass~\cite{LIFSHITS19581}. The cyclotron effective mass is related to the energy dependence of the corresponding extremal orbit area by
\begin{equation}
    m^*
    =
    \frac{\hbar^2}{2\pi}
    \left.
    \frac{\partial A(E)}{\partial E}
    \right|_{E=E_{\mathrm{F}}}.
    \label{eq:cyclotron_mass}
\end{equation}
Therefore, $m^*$ reflects the quasiparticle band dispersion near the Fermi energy.

Experimentally, $m^*$ can be obtained by measuring the oscillation amplitude as a function of temperature and fitting it to Eq.~\ref{eq:thermal_damping}. Since the Fourier transformation is not calculated at a fixed magnetic field value but over a finite  window, the Fourier amplitude is often evaluated using the mean value $1/B_{\mathrm{eff}}$ of the chosen field window~\cite{PhysRevLett.115.226401, PhysRevB.96.045127, PhysRevB.95.205134, PhysRevB.81.235103, 10.1038/srep18674, PhysRevLett.110.257002, PhysRevLett.110.257002, 10.1038/srep18674, Pezzini2018}. This effective field $B_{\mathrm{eff}}$ is given by 
\begin{equation}
    B_{\mathrm{eff}}
    =
    \left[
        \frac{1}{2}
        \left(
            \frac{1}{B_{\min}}
            +
            \frac{1}{B_{\max}}
        \right)
    \right]^{-1}
    =
    \frac{2}
    {\displaystyle \frac{1}{B_{\min}}
     +\frac{1}{B_{\max}}},
    \label{eq:nominal_effective_field}
\end{equation}
where $B_{\text{min}}$ and $B_{\text{max}}$ are the minimum and maximum field of the chosen window, respectively, which is often constrained by instrument limitations. The temperature-dependent Fourier amplitude is then fitted using
\begin{equation}
    A_{\mathrm{FFT}}(T)
    =
    A_{\mathrm{FFT}}(0)
    \frac{X_{\mathrm{eff}}}{\sinh X_{\mathrm{eff}}},
    \qquad
    X_{\mathrm{eff}}
    =
    \frac{2\pi^2 k_{\mathrm{B}}T m^*}
    {\hbar e B_{\mathrm{eff}}}.
    \label{eq:fft_lk_fit}
\end{equation}
The thermal damping of the $p$th harmonic of an orbit is governed by the product $p m^*$. Thus, if a Fourier peak at $pF$ is a harmonic of a fundamental orbit at $F$, the corresponding cyclotron mass is approximately equal to~\cite{leeb2025fieldguide}
\begin{equation}
    m^{*}_{pF}
    \approx
    p\,m^{*}_{F}.
    \label{eq:harmonic_mass}
\end{equation}
Figure~\ref{Fig2}d--e shows a harmonic pair, $F_1$ and $2F_1$, and their expected amplitude decay according to the LK formula. This mass relationship can help distinguish harmonics from unrelated extremal orbits, although overlapping peaks and finite-window effects may complicate the comparison. 

In addition to harmonics, combination peaks such as $F_1+F_2$ and $|F_1-F_2|$ can arise from magnetic breakdown, chemical potential oscillations, and scattering rate oscillations. A comprehensive guide to non-Onsager quantum oscillations is given by Leeb \textit{et al.}~\cite{leeb2025fieldguide}. Frequency relationships alone cannot distinguish these contributions; their masses and field and temperature dependencies must also be considered. Beyond these physical mechanisms, finite field windows and background subtraction can distort the Fourier spectrum and introduce spurious peaks, as discussed in Sec.~III.

The Dingle factor describes broadening of the landau levels due to scattering by
\begin{equation}
    R_D
    =
    \exp\left(
        -\frac{2\pi^2 k_{\mathrm{B}}T_{D} m^*}
              {\hbar eB}
    \right),
    \label{eq:dingle_factor}
\end{equation}
where $T_{D} =\hbar / (2\pi k_{\mathrm{B}}\tau_{q})$ is the Dingle temperature and \(\tau_{q}\) is the quantum scattering lifetime~\cite{10.1098/rspa.1952.0055, 10.1098/rspa.1952.0056}.

As both $R_T$ and $R_D$ decrease with decreasing magnetic field, quantum oscillations generally have lower amplitudes at low fields. The extent of this attenuation can differ between Fermi surface pockets because each may possess a different $m^*$ and quantum scattering lifetime. In particular, larger $m^*$ and shorter quantum lifetimes produce stronger damping. Therefore, higher harmonics are more strongly attenuated at low fields~\cite{leeb2025fieldguide}.

\begin{figure}[!htbp]
    \centering
    \includegraphics[width=6.25in]{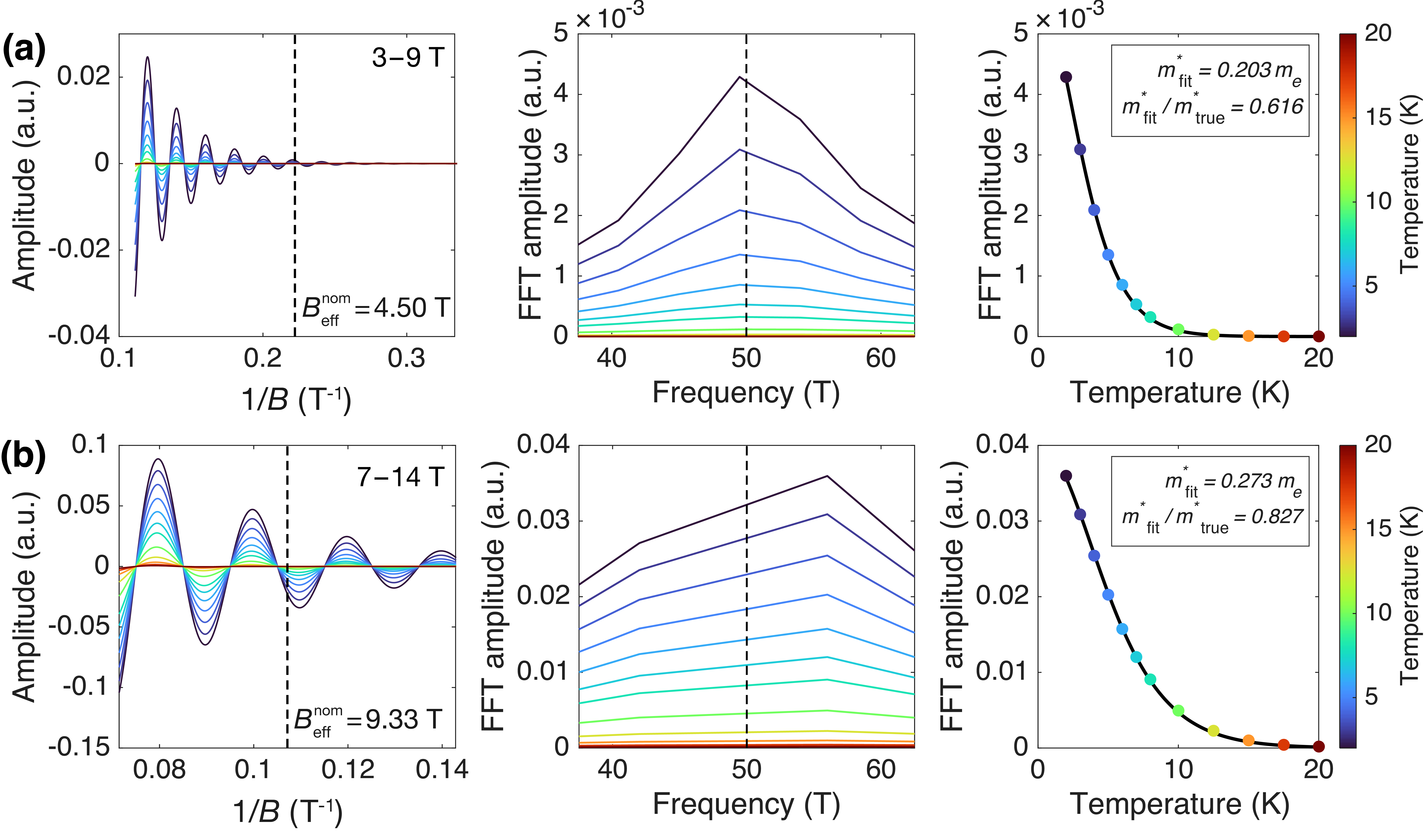}
    \caption{Effect of a finite field window on an LK fit. Simulated signal against inverse field (left), corresponding Fourier spectrum (middle) and LK fit (right) for $F = 50$~T, $m^{*}_{\mathrm{true}} = 0.33~m_{e}$ and $T_{D} = 8$~K, with a field window of \textbf{(a)}~3--9~T and \textbf{(b)}~7--14~T. No additional window function or zero padding was applied to emphasize the underlying frequency resolution. Vertical dashed lines in the left and middle panels indicate the effective field $B_{\mathrm{eff}}$ and true oscillation frequency, respectively.}
    \label{Fig_LK}
\end{figure}

Figure~\ref{Fig_LK} demonstrates the effect of a finite field window on the cyclotron masses extracted using an LK fit. The quantum oscillations were simulated for a single orbit with frequency $F = 50$~T, cyclotron mass $m^{*}_{\mathrm{true}} = 0.33\,m_{e}$ and Dingle temperature $T_{D} = 8$~K. The signal was analyzed using a field window of 3--9~T (see Fig.~\ref{Fig_LK}a) and 7--14~T (see Fig.~\ref{Fig_LK}b). Fits over 3--9~T (Fig.~\ref{Fig_LK}a) and 7--14~T (Fig.~\ref{Fig_LK}b) yield $m^{*}_{\mathrm{fit}} = 0.203\,m_e$ and $0.273\,m_e$, underestimating the true mass by 38\% and 17\%, respectively. This is a direct result of the chosen window parameters: the oscillations are strongly damped at lower fields but remain prominent at higher fields.

In both intervals, the oscillation amplitude increases monotonically with field. The resulting amplitude envelope weights the Fourier signal toward the high-field end of the interval. This choice of $B_{\mathrm{eff}}$ does not account for this unequal weighting, and using this value in the LK fit leads to an underestimated cyclotron mass.

The spin-splitting factor
\begin{equation} 
R_S^{(p)} = \cos(\pi p S), \qquad S = g m^*/2m_e , 
\end{equation} 
where $g$ is the effective $g$-factor of the orbit, reflecting the phase interference of the spin-up and spin-down quasiparticles. As $R_S$ is temperature independent, it does not bias a fitted cyclotron mass. However, it can suppress a peak entirely when $pS$ approaches a half-odd integer. As a result, an orbit expected from ARPES measurements or band-structure calculations may be absent from the quantum oscillation spectrum for reasons unrelated to noise or background subtraction~\cite{shoenberg1984magnetic}.

\section{Processing Oscillation Data}

\begin{figure}[!htbp]
    \centering
    \includegraphics[width=6.5in]{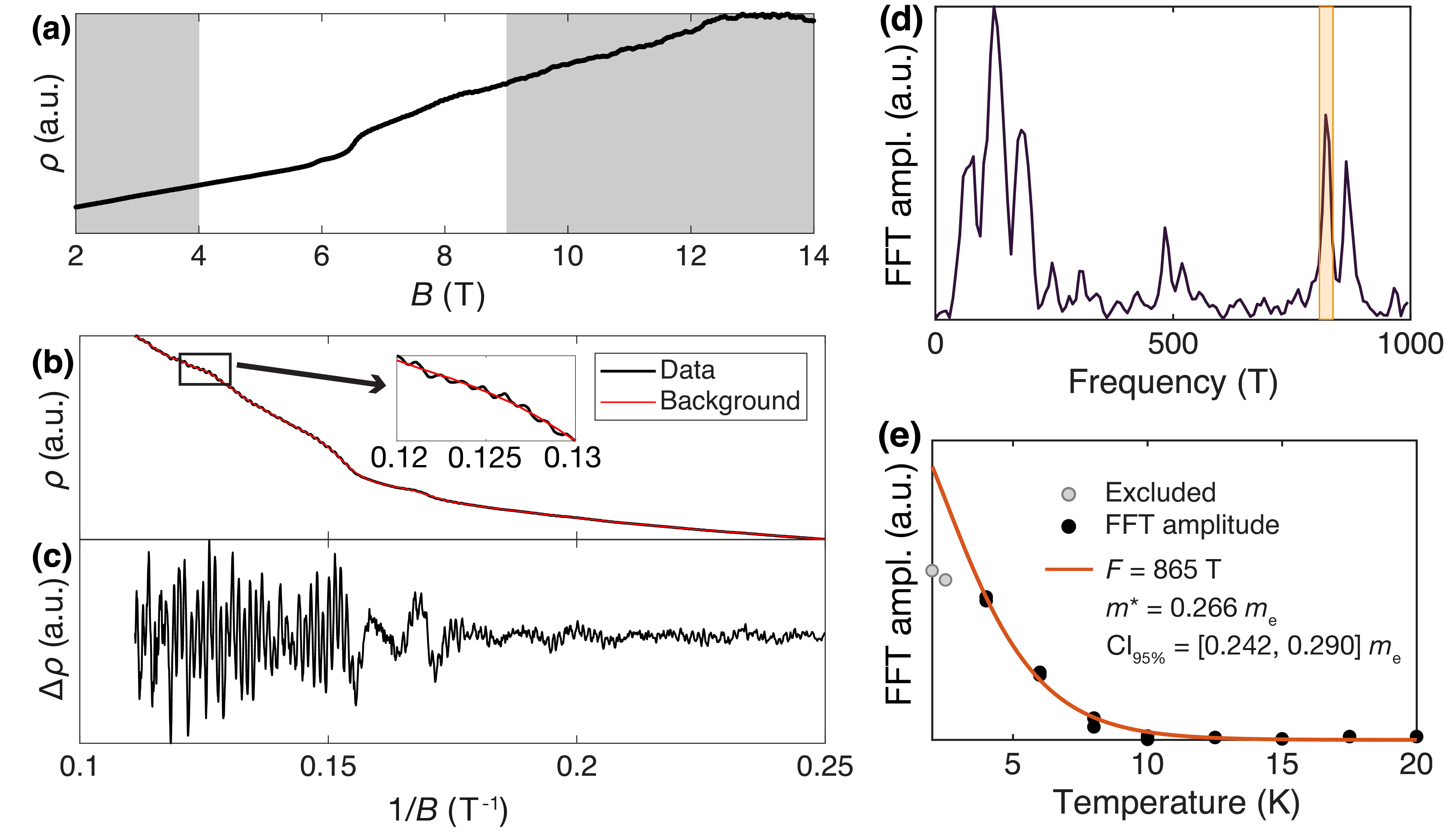}
    \caption{Procedure for extracting the cyclotron effective mass in TbTe$_3$. \textbf{(a)}~Raw resistivity as a function of field at 2~K. The white area from 4 to 9~T shows the rectangular field window chosen for the analysis, which is broad enough to resolve the target frequency without introducing magnetic breakdown peaks. \textbf{(b)}~Data corresponding to (a) plotted against $1/B$, with a background fitted using the Savitzky--Golay (SG) method. \textbf{(c)}~Oscillatory component $\Delta\rho$ after background subtraction and \textbf{(d)}~its Fourier spectrum without zero padding. The frequency peak at 865~T is marked in orange. \textbf{(e)}~LK fit of the orbit corresponding to 865~T, giving $m^{*} = 0.266(24)\,m_e$. The gray data points were excluded because TbTe$_3$ orders magnetically below 5~K~\cite{ma15248772}.}
    \label{Fig_Steps}
\end{figure}

Extracting electronic properties and Fermi surface characteristics from quantum oscillation measurements requires several processing steps, each of which can influence the recovered frequency spectrum. Figure~\ref{Fig_Steps} shows the full cyclotron mass analysis using the 865~T peak in TbTe$_3$, which includes field range selection, interpolation of the data onto a uniform inverse-field grid, background subtraction, field windowing, and the Fourier transformation. Repeating this procedure at each temperature yields the amplitude decay used to extract the cyclotron mass of the respective orbit. In this section, we discuss how these analysis choices affect the extracted frequencies and amplitudes.
\subsection{Field range selection}
\label{Field range selection}
The magnetic field range is often adjusted to enhance spectral features; however, it also affects the frequency resolution, consequently influencing the entire analysis. A finite inverse-field interval,
\begin{equation}
    \Delta x
    =
    \frac{1}{B_{\min}}
    -
    \frac{1}{B_{\max}},
    \label{eq:inverse_field_interval}
\end{equation}
determines the Fourier frequency resolution $\delta F \sim 1 / \Delta x$, which defines a characteristic resolution scale for a chosen window of [$B_{min}, B_{max}$]. This effect is visible in the FFT spectra in Fig.~\ref{Fig_LK}. The frequency resolution worsens rapidly as $B_{\min}$ is increased but improves only gradually with increasing $B_{\max}$ since at higher fields, equal changes in field correspond to progressively smaller changes in $1/B$. The 3--9~T window has a characteristic frequency resolution of $\delta F \approx 4.5$~T, while the narrower inverse-field interval corresponding to 7--14~T gives $\delta F \approx 14$~T. For the simulated frequency $F=50$~T, the FFT grid of the 3--9~T window contains a bin at $49.5$~T, close to the true peak center. On the other hand, the two nearest FFT bins for the 7--14~T window occur at $42$ and $56$~T. Therefore, identifying the peak from the maximum FFT amplitude would misassign the peak to a frequency of $56$~T. As the true frequency lies between the available FFT bins, the sampled amplitude at $56$~T is smaller than the actual peak maximum. This amplitude underestimation is known as scalloping loss~\cite{proakis2007digital}. If the spectral profile changes with temperature, measuring the amplitude at this offset frequency can also distort the temperature dependence used in the LK fit.


Zero padding adds zeros to the data after windowing, giving more closely spaced frequency points in the FFT spectrum. This does not add information or improve the underlying frequency resolution~\cite{proakis2007digital}. In practice, resolving two nearby peaks may require their separation to exceed $\delta F$, depending on their relative amplitudes, phases, signal-to-noise ratio, and the applied window function. Nevertheless, the more closely spaced frequency bins allow the peak position and amplitude to be estimated more accurately, reducing the amplitude underestimation associated with scalloping loss.

\begin{figure}[!htbp]
    \centering
    \includegraphics[width=3.41in]{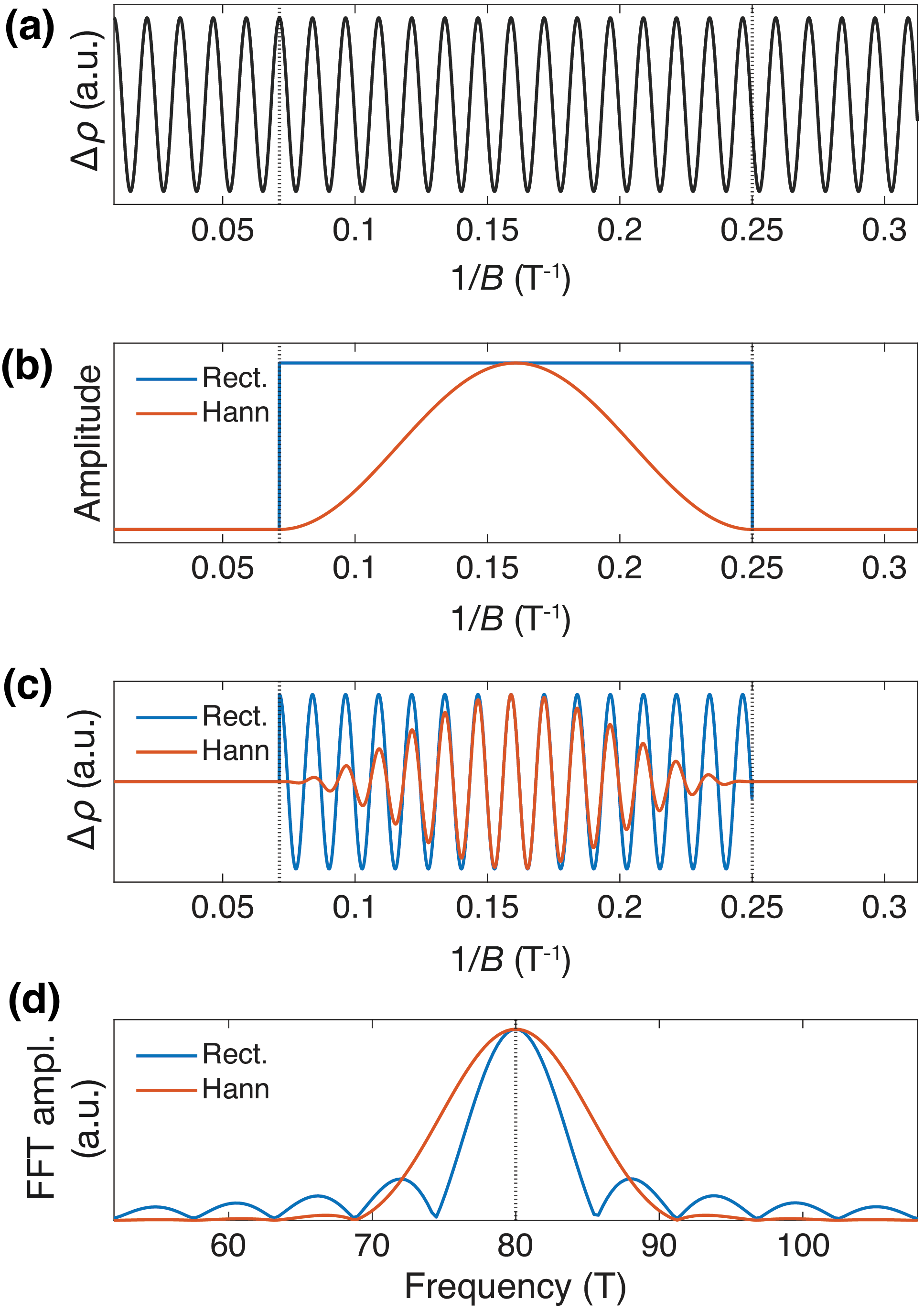}
    \caption{Effect of a window function on an FFT peak. \textbf{(a)}~Calculated sine wave with frequency \(F_0=80\ \mathrm{T}\). The dotted lines mark \(1/B_{\max}\) and \(1/B_{\min}\). \textbf{(b)}~Rectangular and Hann window functions, with \textbf{(c)}~showing the resulting windowed oscillatory signals. \textbf{(d)}~Independently normalized Fourier-amplitude spectra. The rectangular window produces a narrow central peak with pronounced sidelobes, while the Hann window suppresses the side windows but has a broader central peak.}
    \label{Fig3}
\end{figure}

Restricting the measurement to a finite inverse-field interval is equivalent to multiplying the signal by a rectangular window $w(x)$,
%
\begin{equation}
    s_{\mathrm{obs}}(x)
    =
    w(x)s(x).
    \label{eq:windowed_signal}
\end{equation}
By the convolution theorem, multiplication in inverse-field space corresponds to convolution in Fourier space,
\begin{equation}
    \widetilde{s}_{\mathrm{obs}}(F)
    =
    \widetilde{w}(F)
    *
    \widetilde{s}(F),
    \label{eq:window_convolution}
\end{equation}
where the tilde denotes a Fourier transform and $*$ is the convolution operator~\cite{proakis2007digital}. 

Figure~\ref{Fig3}a shows an ideal sinusoidal signal, while Fig.~\ref{Fig3}b shows the rectangular and Hann window functions. Multiplying the sine wave by the window functions produces the windowed signals shown in Fig.~\ref{Fig3}c. For the rectangular window, the convolution produces the sinc-shaped spectral response in Fig.~\ref{Fig3}d. The sinc function, defined here as sinc$(x)$ = sin$(\pi x)$/$(\pi x)$, has a central lobe surrounded by sidelobes whose amplitude decays away from the true oscillation frequency. These sidelobes indicate that spectral weight can leak into neighboring frequencies, complicating peak identification and biasing quantities that involve the extraction of temperature-dependent amplitudes~\cite{1455106, 6768513}. How window functions address spectral leakage is discussed later in Sec. \ref{window-func}.

Several experimental limitations complicate this analysis. The field at which each component becomes detectable depends on variables such as the cyclotron mass and scattering rate. At low fields, the quantum oscillation amplitudes become too weak to resolve against the electrical noise. Extending to a higher field range includes regions of larger oscillation amplitude, making distinct oscillatory components easier to identify. However, increasing $B_{\max}$ can also introduce magnetic breakdown orbits and higher harmonics, crowding the Fourier spectrum and complicating the assignment of individual peaks. Therefore, the selected field range should balance the visibility of the target frequencies against the spectral complexity that emerges at higher fields, while retaining enough inverse-field span for adequate frequency resolution.

\begin{figure}[!htbp]
    \centering
    \includegraphics[width = 5.09in]{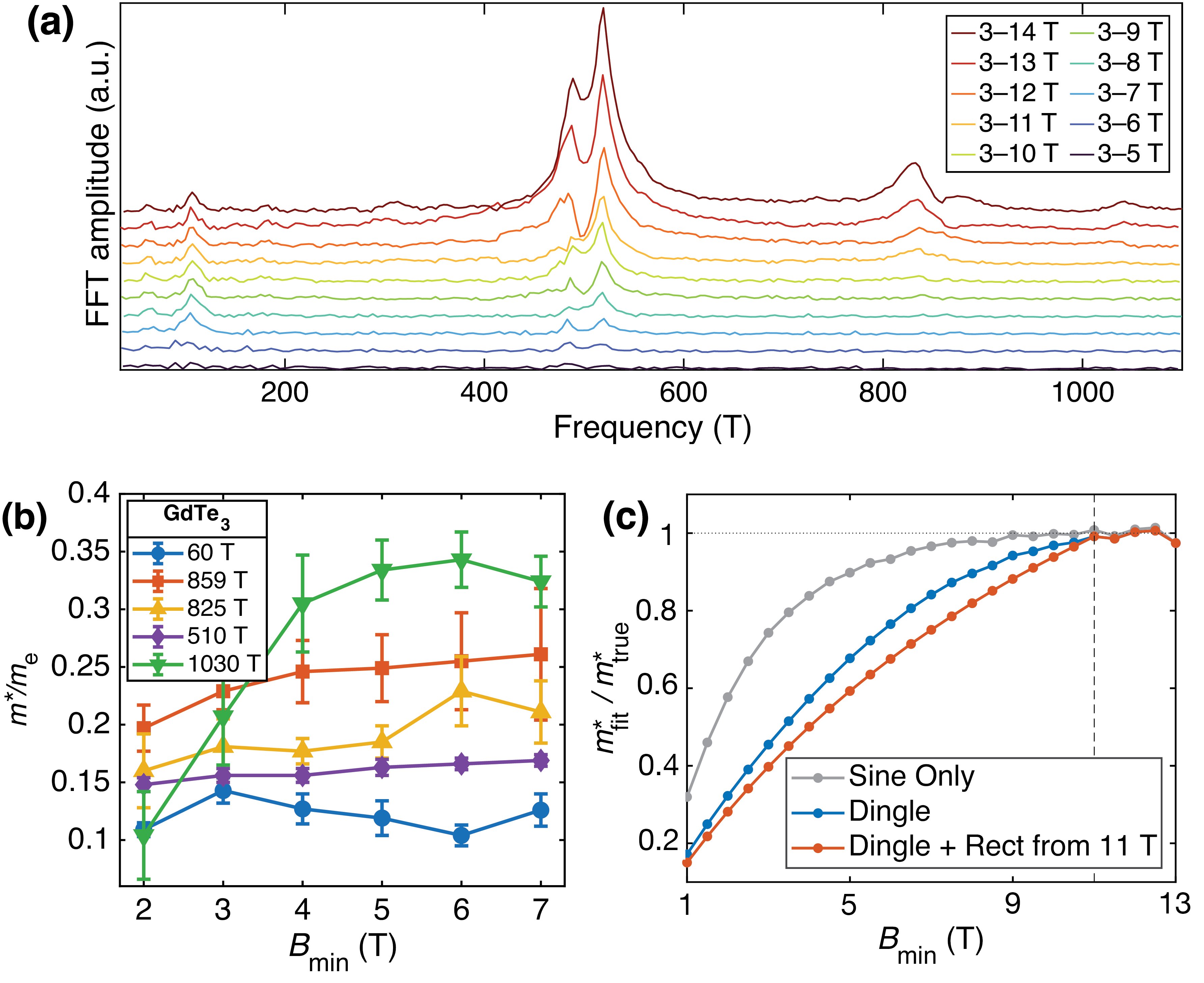}
    \caption{\textbf{(a)} Fourier spectra of quantum oscillations in GdTe$_3$ measured at $T=2.0~\mathrm{K}$ with $B_{\min}$ fixed at 3~T while $B_{\max}$ is incremented from 5 to 14~T. No additional window function was applied, and the spectra are vertically offset for clarity. $B_{\min}$ was fixed but $B_{\max}$ varied to successively add a higher-field segment, effectively testing the presence of a peak over increasing field ranges. \textbf{(b)} The calculated cyclotron mass of several pockets based on a fixed $B_{\max}=14.0$ T but a variable $B_{\min}$. Every window contains the high-field region where these oscillations are known to be present, so the comparison isolates the effect of using a larger window and extending the fit toward a lower field. \textbf{(c)} The calculated ratio of fitted to known effective mass ($m^* = 0.33 m_e$) for varying $B_{\min}$ to 14~T under three conditions: a sine wave, a sine wave with Dingle damping, and a sine wave with Dingle damping, present only above 11~T with $T_D$ = 6~K.}
    \label{Fig4}
\end{figure}

To investigate how the choice of field window affects the extracted effective mass, consider the quantum oscillation data obtained from GdTe$_3$ in Fig.~\ref{Fig4}. Inspecting Fig.~\ref{Fig4}a, the peak amplitude of a given frequency tends to increase as $B_{\max}$ is raised. This indicates that the oscillatory amplitude tends to increase with increasing field, as expected. Furthermore, several peaks become visible only when $B_{\max}$ is sufficiently large, such as the $830~\mathrm{T}$ and $1000~\mathrm{T}$ peaks. The extracted effective masses for select peaks under various field windows are given in Fig.~\ref{Fig4}b. The calculated effective mass trends downward as $B_{\min}$ is lowered, most evidently for the $1000~\mathrm{T}$ peak, which drops by almost a factor of three. Together, these observations suggest that the extracted effective mass is likely sensitive to both the oscillatory amplitude at different fields and the choice of the effective field.


The bias introduced by using $B_{\mathrm{eff}}$ to represent a finite-window Fourier amplitude has been reported. Mercure showed, using simulated dHvA oscillations containing thermal and Dingle damping, that fitting from a lower $B_\mathrm{min}$ systematically underestimates $m^*$~\cite{mercure2010haasvanalpheneffect}. Additionally, Audouard and Fortin derived the effect analytically, showing that the bias is set by the product of the inverse-field window width and a decay rate containing the Dingle temperature, and reaches tens of percent for a rectangular window~\cite{Audouard2018Fourier}. Both treat a single isolated component, for which the prescription is to narrow the inverse-field window as far as the available oscillation periods allow. Here we separate the contributions to that bias and examine how the prescription changes when the spectrum contains closely spaced orbits.

Figure ~\ref{Fig4}c explores three different scenarios in which $B_{\min}$ is varied but $B_{\max}$ is fixed. Firstly, the gray curve considers the effect of performing the LK fit on a sine wave under thermal damping only. Because the thermal damping factor depends on field, it is itself influenced by the chosen window. For this analysis, we use $B_{\mathrm{eff}}$ in Eq. \ref{eq:nominal_effective_field}. Next, the blue curve includes the Dingle damping factor to help investigate how it may further dampen the low-field oscillations and consequently affect the fitted $m^*$. Lastly, motivated by the high-field onset of the high frequency peaks, the orange curve considers the case where the oscillatory signal is absent below $B^*_\mathrm{min}$. While the numerical results may vary with the choice of $m^*$ and $T_D$, they are independent of the oscillation frequency. 

The thermal damping factor in gray can significantly underestimate $m^*$ at low $B_{min}$; in the example given, the underestimate can exceed 40\% when a pocket with an effective mass of 0.33~$m_{e}$ is fitted from 2--14~T. Because the thermal damping factor depends on $Tm^*/B$, representing the field-weighted signal by a single effective field gives the approximate relation
\begin{equation}
    \frac{m_{\mathrm{fit}}^{*}}
         {B_{\mathrm{eff}}}
    \approx
    \frac{m_{\mathrm{true}}^{*}}
         {B_{\mathrm{eff}}^{\mathrm{wt}}}.
\end{equation}
Here, $B_{\mathrm{eff}}^{\mathrm{wt}}$ represents the effective field after accounting for the field-dependent weighting of the oscillation amplitude. Therefore, if the oscillatory signal is weighted toward the high field end of the interval, $B_{\mathrm{eff}}^{\mathrm{wt}}>B_{\mathrm{eff}}$, then it is approximately true that $m_{\mathrm{fit}}^{*} < m_{\mathrm{true}}^{*}$.

A claim of field-dependent mass enhancement is one case where accounting for field range effects matters. Because a mass extracted over a high-field interval is systematically less biased than one extracted over a wider, lower-field interval, a comparison between orbits observed over different intervals can mimic a genuine field dependence~\cite{Pezzini2018}. Reports of such enhancement should state the interval over which each mass was obtained so that the two effects can be separated.

Dingle damping suppresses the oscillation amplitude more strongly at low fields, shifting the weighting toward higher fields and, as shown in Fig.~\ref{Fig4}c, further increasing the mass underestimate. Let $B_{\min}^{*}$ denote the lowest field at which the target oscillation becomes experimentally resolvable. This threshold depends on measurement sensitivity and does not necessarily reflect an intrinsic onset of the oscillation. In the simulated example, extending the fit below $B_{\min}^{*}$ further increases the mass underestimate, though by less than the Dingle contribution. These findings demonstrate a tradeoff between mass bias and frequency resolution: a narrower inverse-field interval may reduce the bias in $m^*$ but leave neighboring peaks unresolved, so their combined amplitude can depart from LK behavior. A wider interval improves frequency resolution but at the cost of exacerbating the mass underestimate.

Restricting the fit to fields above $B_{\min}^{*}$ can reduce this additional bias. A rough estimate of $B_{\min}^{*}$ can be obtained from the peak width if the resolvable signal is approximated as an isolated, constant-amplitude sinusoid confined to $[B_{\min}^{*},B_{\max}]$. In this approximation, the Fourier-amplitude full width at half maximum (FWHM) is
\begin{equation}
    \mathrm{FWHM}
    \approx
    \frac{1.207}
    {\displaystyle \frac{1}{B_{\min}^{*}}
     - \frac{1}{B_{\max}}},
    \label{eq:fwhm_effective_range}
\end{equation}
which can be inverted to estimate $B_{\min}^{*}$ for a given $B_{\max}$. The derivation is provided in Sec.~\ref{sec:si-fwhm} of the Supplemental Material. This estimate can guide field-window selection, although raising $B_{\min}$ too far reduces the number of oscillation periods and the frequency resolution, making the extracted mass less reliable.

\subsection{Background subtraction}
\label{background-subtract}
In a quantum oscillation measurement, the oscillatory contribution is measured together with a generally much larger, non-oscillatory background. For SdH oscillations, the measured longitudinal resistivity may be represented as
\begin{equation}
    \rho(B,T)
    =
    \rho_{\mathrm{bg}}(B,T)
    +
    \Delta\rho_{\mathrm{osc}}(B,T)
    +
    \epsilon(B,T),
    \label{eq:measured_resistivity}
\end{equation}
where \(\rho_{\mathrm{bg}}\) is the non-oscillatory background and \(\epsilon\) represents an additive noise term. The background may vary among samples because of differences in sample quality and geometry, contact configuration and quality, or field-dependent deformations to the sample. The oscillatory component $\Delta\rho_{\mathrm{osc}}$ needs to be extracted by subtracting a model of the background,
\begin{equation}
    \Delta\rho_{\mathrm{est}}(B,T)
    =
    \rho(B,T)
    -
    \rho_{\mathrm{bg,est}}(B,T).
    \label{eq:estimated_oscillation}
\end{equation}
Any mismatch between the physical background and the estimated background remains in the extracted oscillatory component \(\Delta\rho_{\mathrm{est}}\) and can influence the resulting Fourier spectrum. This issue is especially important for low-frequency oscillations, whose slow variation with \(1/B\) can be difficult to distinguish from a smoothly varying magnetoresistive background. In this section, we examine how a polynomial background subtraction can alter the Fourier spectrum and motivate the Savitzky--Golay (SG) method as an alternative that better preserves low-frequency oscillations.

Because the FFT requires uniformly sampled data but an experimental sweep is generally uniform in $B$, the measured signal is first interpolated uniformly in $x = 1/B$. Note that interpolation is itself a linear operator with its own frequency response. In addition to linear interpolation, cubic spline interpolation is also commonly used as it passes through every measured point and preserves most of the oscillatory amplitude~\cite{PhysRevLett.128.176603, ythr-l17b, PhysRevB.60.13371, PhysRevB.96.241401, PhysRevB.33.5585, 10.1063/1.365244}. The non-oscillatory background is then estimated and subtracted after interpolation. 

\begin{figure}[!htbp]
    \centering
    \includegraphics[width=3in]{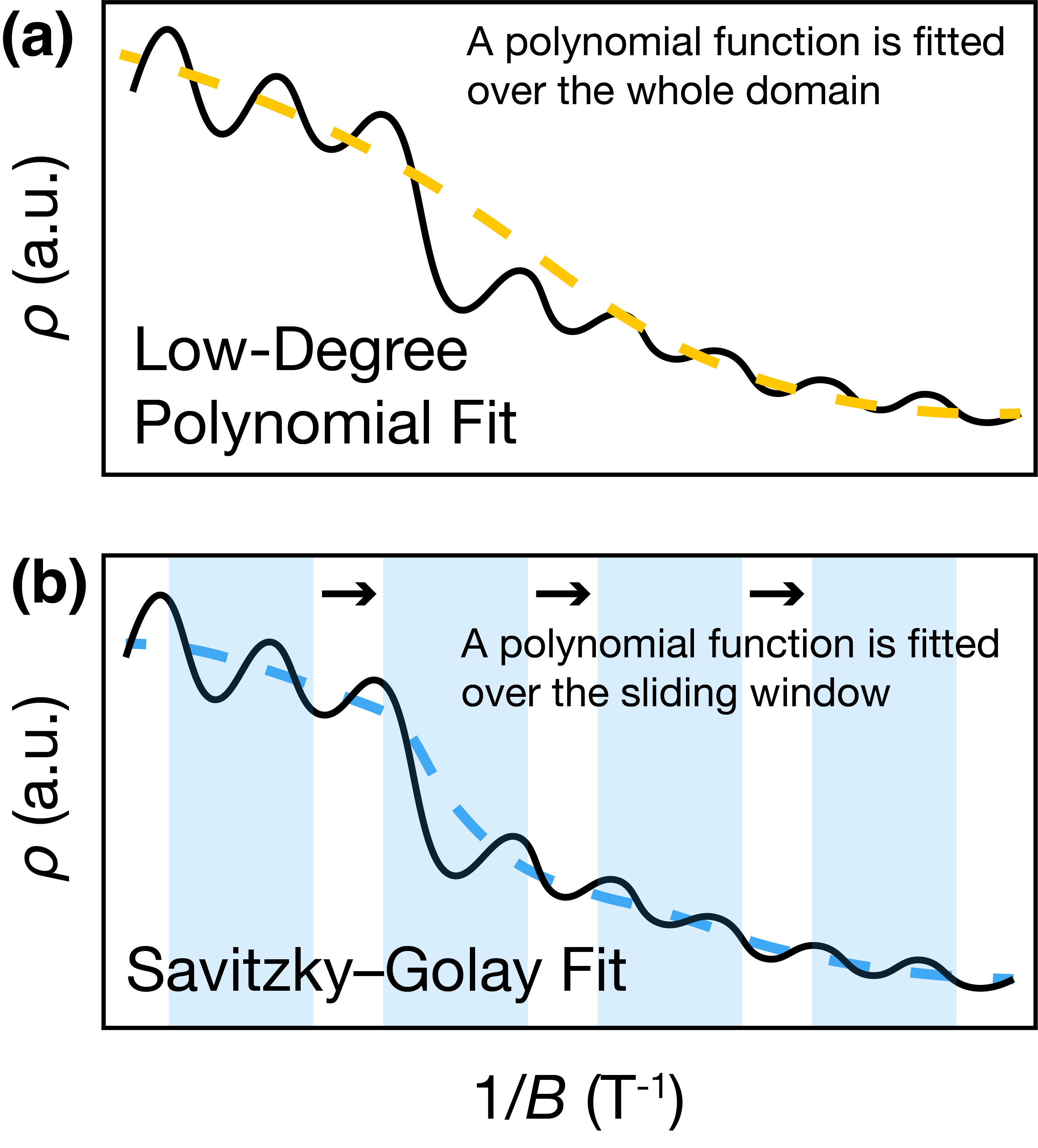}
    \caption{Schematic of a background fit to an oscillatory signal containing a sharp feature, using \textbf{(a)} a low-degree polynomial and \textbf{(b)} the SG algorithm. The blue shaded regions mark three representative positions of the window, which advances one point at a time.}
    \label{Fig_Background}
\end{figure}

Low-order polynomials, commonly of degree 2--5, are frequently used to fit the background signal because they adapt to variations in the background structure while limiting the risk of overfitting~\cite{hein1985dehaas, pc57-schm, PhysRevB.109.205115, doi:10.1126/science.1242247, QO_PrTe3}. However, a low-degree polynomial may not adequately capture more complicated or sharp features in the background structure; this is illustrated in Fig.~\ref{Fig_Background}a. Increasing the polynomial degree gives the fit greater flexibility but can also absorb a greater range of low-frequency oscillations. For example, the magnetoresistance background in Fig.~\ref{Fig_Steps}a shows a kink-like feature at 6.5 T that is associated with a magnetic phase transition and is difficult to fit with a low-degree polynomial~\cite{ma15248772}. We therefore examine how the recovered signal changes with polynomial degree. 

\begin{figure}[!htbp]
    \centering
    \includegraphics[width = 3.17 in]{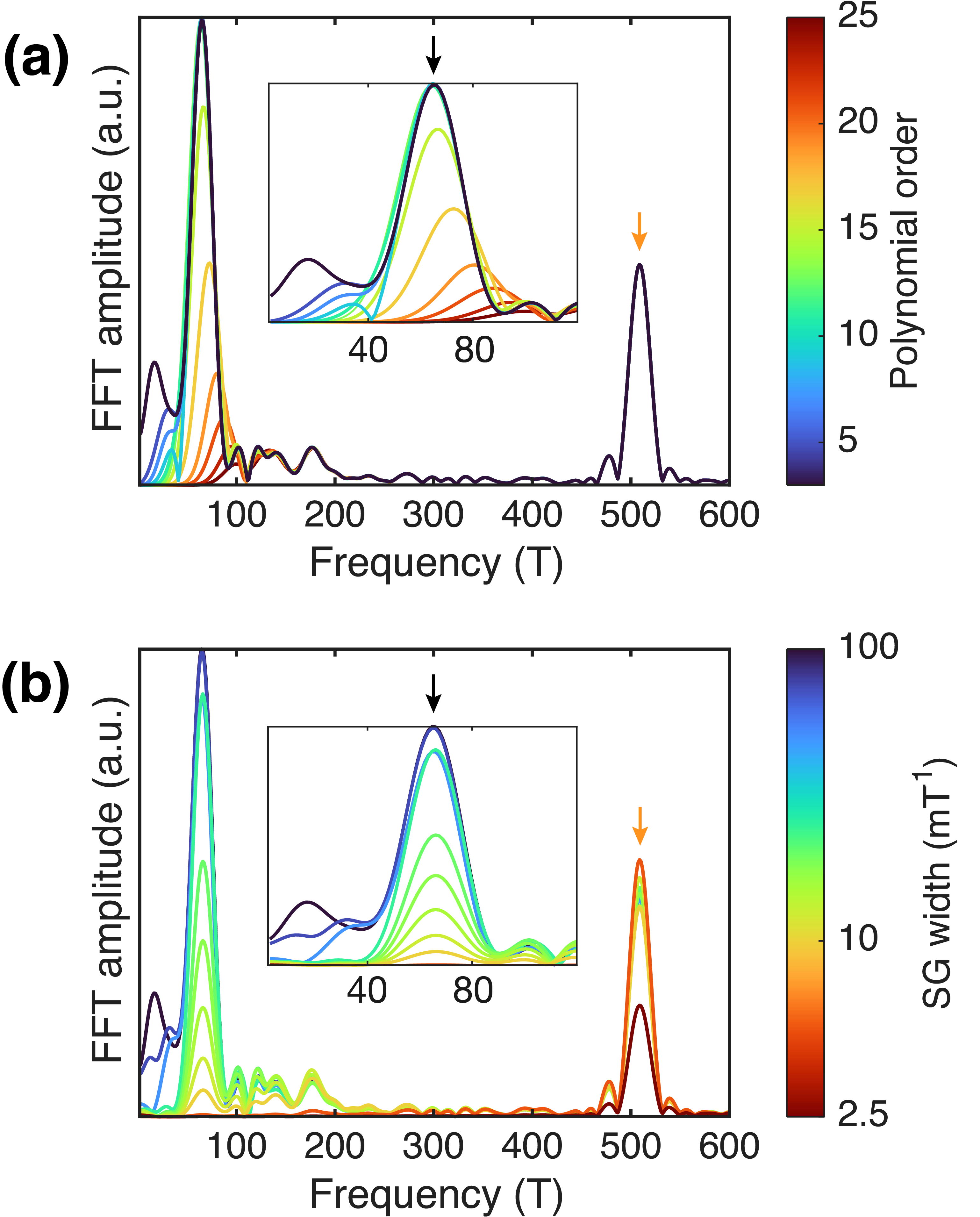}
    \caption{The FFT spectra of PrTe$_3$ over 4--14 T subject to \textbf{(a)} polynomial fits of degree 3--25 and \textbf{(b)} SG fits of polynomial degree 3 and windows of 2.5--100 mT$^{-1}$. A Hann window and a zero padding factor of 8 were applied. The color bar represents the strength of the background subtraction, where red (blue) represents a more (less) aggressive fit. The black arrow denotes the 65 T peak, which appears to slide with increasing polynomial degree, but is only rescaled under a stronger SG fit. The orange arrow denotes the 500 T peak, which is insensitive to the polynomial degrees tested but whose amplitude is rescaled by the SG window.}
    \label{Fig8}
\end{figure}

Increasing the polynomial degree attenuates low-frequency components and can redistribute spectral weight between frequencies. A more detailed analysis is available in the Supplemental Material (Sec.~\ref{sec:common_response}). A commonly overlooked effect of polynomial background subtraction is that it redistributes spectral weight between frequencies rather than solely rescaling each component. This effect can be seen in the low-frequency spectrum of PrTe$_3$ in Fig.~\ref{Fig8}a. At first glance, it seems that increasing the polynomial degree shifts the 65 T peak (denoted by the black arrow) to the right. However, this apparent shift can arise from the combined attenuation of the original component and redistribution of its spectral weight to surrounding frequencies. As the original peak is suppressed, the redistributed contributions can become relatively more prominent, moving the spectral maximum without changing the underlying oscillation frequency. Therefore, polynomial subtraction can produce an apparent peak shift through attenuation and frequency mixing. The magnitude of these effects depends on the polynomial degree, fitting interval, sampling grid, and frequency under consideration.


A background subtraction is a linear operation on the sampled data and can be written as a matrix acting on the data vector. In the Fourier basis, the diagonal elements rescale individual frequency components, while the off-diagonal elements mix contributions between frequencies. Avoiding frequency mixing therefore requires a diagonal matrix in this basis. This motivates considering convolution-based methods such as SG method, which applies the same local polynomial fit as a window moves across the data~\cite{10.1021/ac60214a047}.

As illustrated in Fig.~\ref{Fig_Background}b, the SG method estimates the background at each point by fitting a polynomial within a window of $m$ data points. Away from the record boundaries, repeating this local fit is equivalent to the convolution introduced above. However, this sliding window fails around the window boundaries where the fitting window cannot span $(m-1)/2$ data points in both directions. The end points contribute non-diagonal component that introduces frequency mixing. This contribution is not always small: a $15~\mathrm{mT^{-1}}$ window spans roughly a fifth of a $7$--$14~\mathrm{T}$ record. We evaluate how this boundary-induced mixing depends on the SG window size in the Supplemental Material (Sec.~\ref{sec:common_response}). We show that in general, the mixing increases with window size but remains smaller than that produced by polynomial subtraction of comparable strength. The choice of an optimal window size is discussed by Sadeghi \textit{et al.}~\cite{8957550}.

The FFT spectrum of PrTe$_3$ after SG background subtraction is shown in Fig.~\ref{Fig8}b, where a smaller window corresponds to a more aggressive background subtraction. The Supplemental Material quantifies the attenuation at each frequency bin using the amplitude response $A_j$. As the SG window is reduced, the peak near $65$~T decreases in amplitude but remains at approximately the same frequency, as shown in the inset. In contrast, increasing the polynomial degree produces an apparent shift of this peak toward higher frequencies in Fig.~\ref{Fig8}a. The SG spectrum also reveals a robust peak near $100$~T that would likely be dismissed as noise after polynomial subtraction. The attenuation without a measurable shift in peak position makes the SG method useful for analyzing low frequency and closely spaced peaks.

The two methods therefore excel in different regimes. For high frequency pockets, the polynomial spectrum in Fig.~\ref{Fig8}a is robust in both peak position and amplitude across the degrees tested, whereas changing the SG window rescales amplitudes across the whole spectrum (Fig.~\ref{Fig8}b). At low frequencies, changing the SG window alters peak amplitudes but preserves their positions, whereas polynomial subtraction increasingly redistributes spectral weight between frequencies as the fitting degree increases. These results favor polynomial subtraction for high frequency oscillations and SG filtering for low-frequency oscillations, provided that the SG window is long enough to preserve the oscillations.


\subsection{Window functions and the Fourier transform}
\label{window-func}
In digital signal processing, window functions are a standard tool to suppress spectral leakage, including the sidelobe structure that surrounds a peak~\cite{1455106}. This technique has been used in quantum oscillation analysis~\cite{Chang2026_s41928-026-01590-8, Gomez-Alvarado:2026aa}. Here, we extend the windowing formalism of Sec. \ref{Field range selection} from the implicit rectangular window to an applied taper.

Tapering the interval suppresses sidelobes at the cost of a broader main lobe (Fig.~\ref{Fig3}d). As an LK analysis tracks peak amplitude against temperature, any redistribution of spectral weight biases the extracted cyclotron mass.

\begin{figure}[!htbp]
    \centering
    \includegraphics[width=4.75in]{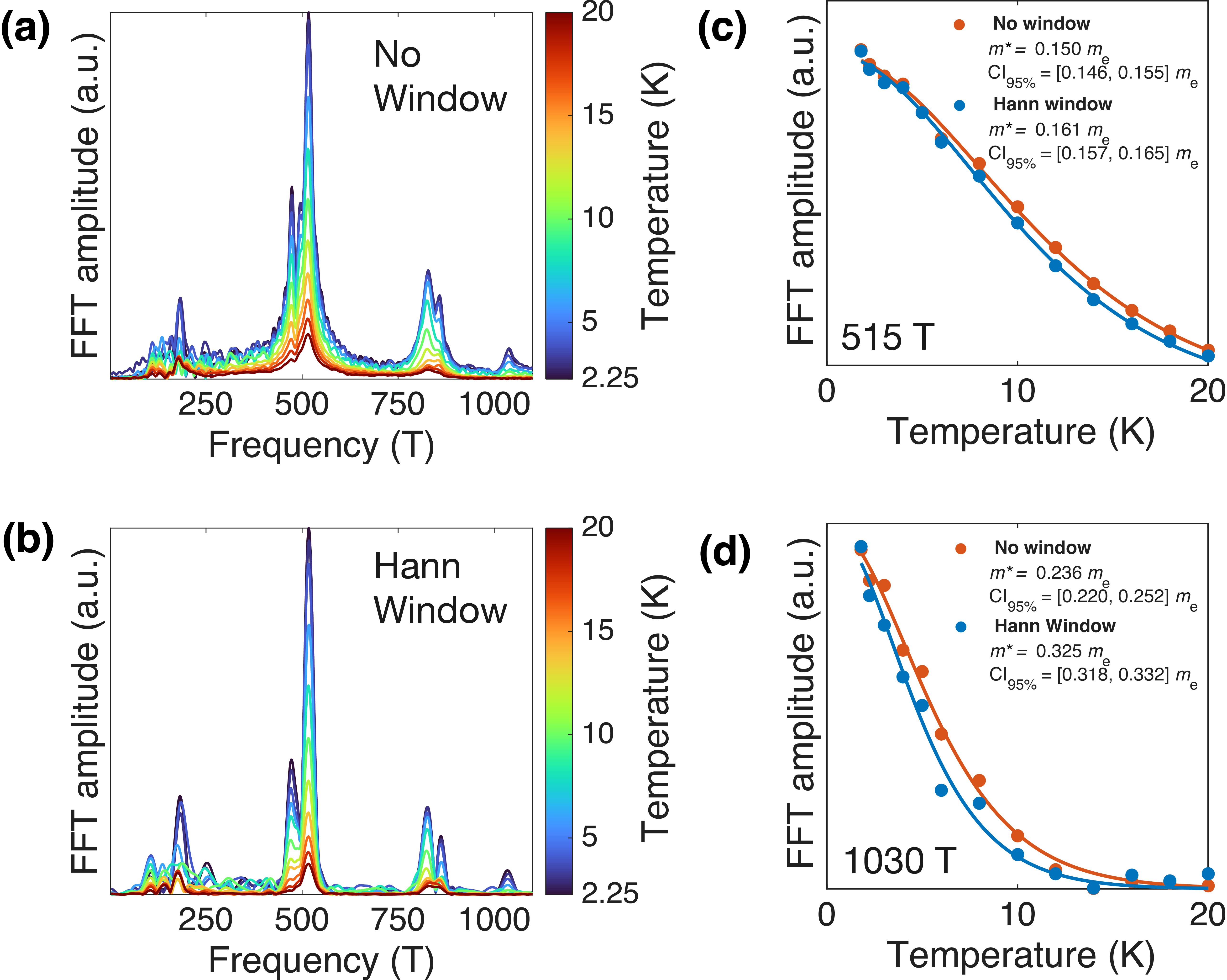}
    \caption{The FFT spectrum of PrTe$_3$ over 8--14 T at various temperatures after an SG background subtraction with a second-degree polynomial and window size 15 mT$^{-1}$ \textbf{(a)} without windowing and \textbf{(b)} with Hann windowing. LK fits to the peak amplitudes in (a)-(b) for the \textbf{(c)} 515 T and \textbf{(d)} 1030 T peaks.}
    \label{Fig6}
\end{figure}

Consider the example with PrTe$_3$ presented in Fig.~\ref{Fig6}a--b, which compares the FFT spectrum with and without an additional window function and shows its impact on the calculated $m^*$ of the 515 T $\beta$ pocket and its second harmonic at 1030 T. Visually, the peaks in Fig.~\ref{Fig6}a have much broader shoulder features that extend well past the peak center; these are largely suppressed in Fig.~\ref{Fig6}b. The difference between the two spectra is the presence or absence of a commonly used window function, the Hann function. A comparison of the two figures shows that there is substantial spectral leakage when no windowing is applied. Critically, the effective mass fitting in Fig.~\ref{Fig6}c--d shows that the effective mass of the 1030 T peak is 2.02 times greater than that of the 515 T peak with windowing, but only 1.57 times without windowing. As shown in Eq. \ref{eq:harmonic_mass}, the masses of the two peaks should be related by a factor of 2. Without windowing, as temperature rises and the true signal decays, the leaked contribution in Fig.~\ref{Fig6}a makes up a growing share of what is measured, consequently flattening the apparent decay curve and biasing the fit toward a smaller mass.

A window function can also reduce the effective-field bias discussed in Sec.~III.A by attenuating the signal at both ends of the inverse-field interval, consequently bringing $B^{\mathrm{wt}}_{\mathrm{eff}}$ closer to $B_{\mathrm{eff}}$. Audouard and Fortin report that a Hann window reduces the mass underestimate from roughly 30\% to 15\%~\cite{Audouard2018Fourier}. However, windowing also broadens spectral peaks, so components that are just resolved without tapering may become unresolved. Several studies have used the Tukey window, whose adjustable taper interpolates between rectangular and Hann windows~\cite{Gomez-Alvarado:2026aa, Chang2026_s41928-026-01590-8}. Adjusting this taper may help balance leakage suppression and frequency resolution when analyzing closely spaced peaks.


\subsection{Practical workflow and diagnostic questions}
\label{sec:guidelines}
\begin{figure}[!htbp]
    \centering
    \includegraphics[width=6.25in]{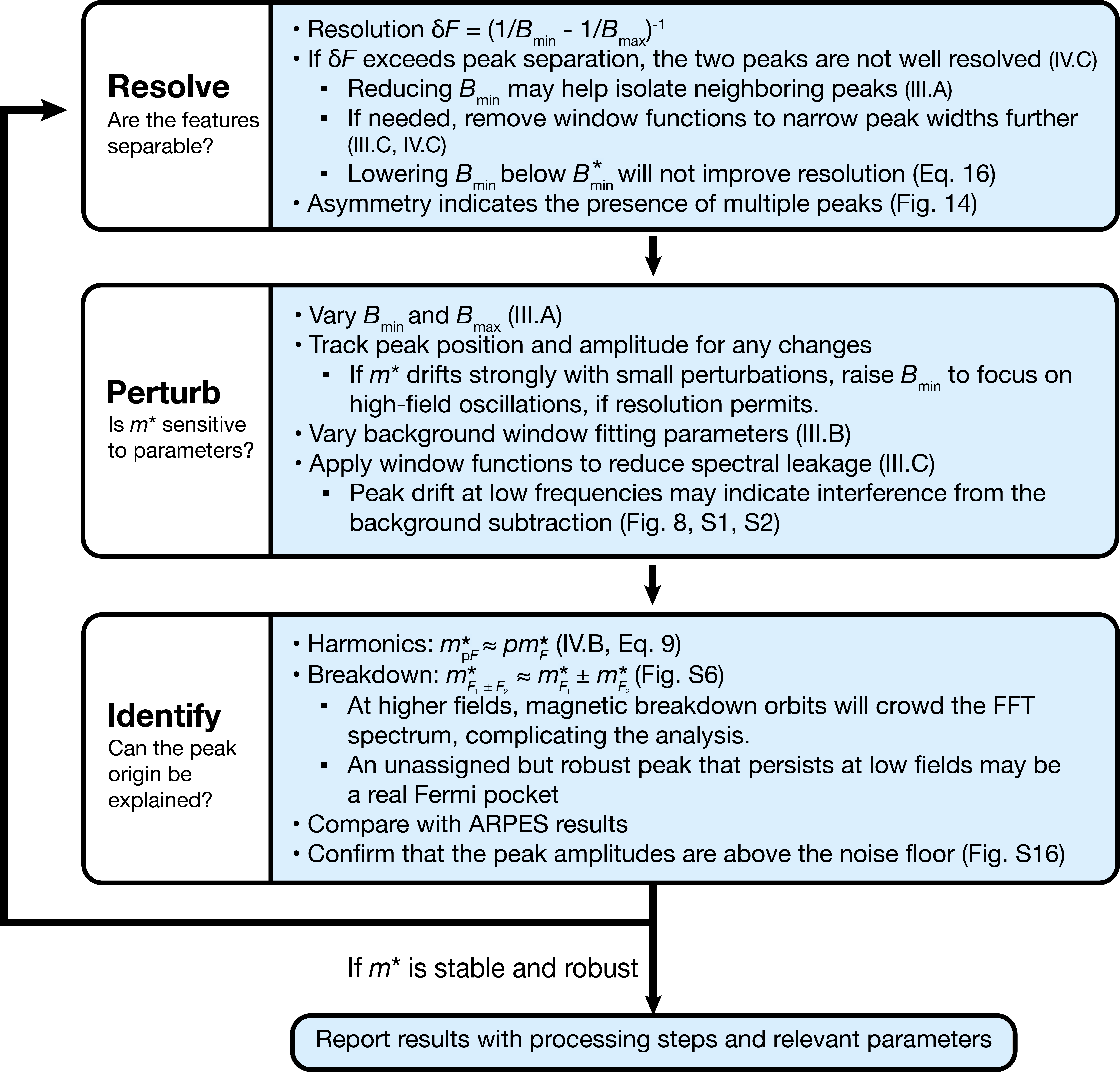}
    \caption{A three-step process for identifying and fitting a spectral feature to extract effective masses from the LK expression. The steps include resolving the spectral features, perturbing the parameters to ensure robustness, and identifying the origin of the peak. Resolving the peak requires being conscious of the intrinsic resolution $\delta F$ from the chosen field range if the peak of interest is surrounded by closely spaced neighboring peaks. Perturbing the fitting parameters, such as the field ranges and the background fit, can help determine whether a spectral feature is real or a product of the data processing steps. Lastly, identifying the peak involves asking whether a feature satisfies an independent relation expected of a harmonic, a breakdown orbit, or another combination mechanism.}
    \label{Fig_Flowchart}
\end{figure}

No single combination of parameters is optimal for every dataset, and several iterations may be needed to find an appropriate one. The flowchart in Fig.~\ref{Fig_Flowchart} organizes that iteration into three steps. The aim of varying the processing parameters is to assess the reliability of the extracted mass, not to select the most convincing spectrum. The resulting spread in $m^*$ indicates sensitivity to the analysis and should not automatically be interpreted as a change in electronic structure. A peak that is robust to perturbations in the background model and the window function is more likely to be a feature of the measured signal than an artifact of the processing, and the mass extracted from it correspondingly more reliable. Field-range dependence requires particular care because it can also reflect physical changes, such as magnetic breakdown, or an oscillation becoming detectable above the noise floor.

Whatever combination of parameters is chosen, the analysis is only reproducible if the processing choices are stated. We suggest that a reported cyclotron mass be accompanied by:

\begin{itemize}
\item the field interval $[B_{\mathrm{min}}, B_{\mathrm{max}}]$ used for the transform, and the resulting $\delta F$;
\item the interpolation scheme and the number of points on the inverse-field grid;
\item the background method and its parameters (polynomial degree, SG parameters, etc.), together with the range of values over which the result was checked;
\item the window function, its taper parameter, and the zero padding factor;
\item how the peak amplitude was extracted;
\item the effective field used in the fit, the temperature range, any excluded points and the criterion for excluding them, and a confidence interval on $m^{*}$.
\end{itemize}

Below, we provide a non-exhaustive set of diagnostic questions to aid a careful case-by-case analysis. These are ordered by processing step rather than by observation and are meant to be worked through before an analysis rather than consulted after one. The goal is to identify where a given dataset is most vulnerable to processing-induced artifacts so that the final conclusions are as robust as possible.

\paragraph{To what extent does the noise floor limit the observed spectrum?} 

A peak whose amplitude approaches the noise floor at high temperatures may seem to deviate from LK behavior because of thermal noise. To check, estimate the local noise floor over a frequency window containing no peaks and compare it with the amplitudes being fitted. Reducing the noise floor before subtraction---for example, by coherent averaging---can extend the usable temperature range (see Fig.~\ref{SI_TbTe3_2_CA})~\cite{rompelman1986coherent}.

\paragraph{Does the field range used for the FFT match the range appropriate for the LK fit?} 

A wider interval improves $\delta F$ but averages the amplitude over a larger field range. If a substantial portion of the interval lies at a lower field range, the extracted $m^{*}$ will be biased downward, thus highlighting the importance of reporting the field interval used.

\paragraph{Which temperature points enter the LK fit, and on what criterion?} 

Points may be excluded because the sample orders magnetically, or because the amplitude has fallen to the noise floor and is no longer measurable. When excluding points, state the criterion rather than the symptom.

\paragraph{Are the values being compared extracted under the same
conditions?} 

The bias introduced by a given field window differs from peak to peak, so a ratio of two masses (Eq.~\ref{eq:harmonic_mass}) is only as reliable as the consistency of the processing behind it. Higher harmonics often become visible only at higher fields, which can make a comparison across different windows unavoidable. Therefore, both field windows should be reported. If the fundamental mass is more strongly underestimated over a lower or wider field interval than the harmonic mass measured at higher fields, the apparent mass ratio will be inflated.

Taken together, these questions are best applied iteratively as a change made in response to one (e.g., restricting the field range) can alter the answer to another (e.g., the achievable frequency resolution). They are presented in processing order, but there is no correct sequence in which to ask them.

\section{Application to Experimental Data}
\begin{figure}[!htbp]
    \centering
    \includegraphics[width=3.25in]{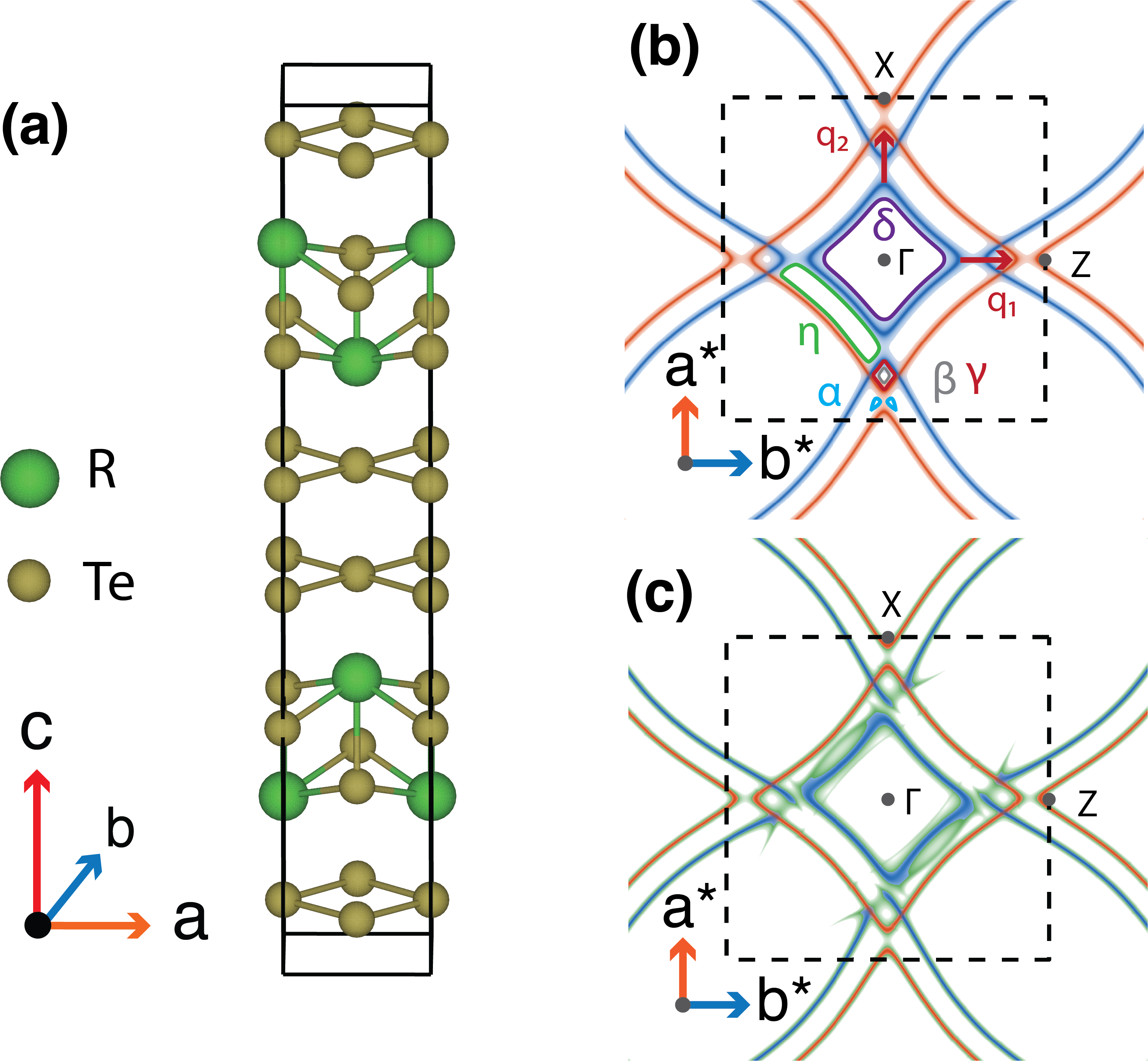}
    \caption{ \textbf{(a)} Crystal structure of $R\mathrm{Te}_3$, consisting of $R$--Te slabs separated by pairs of conducting Te square-net layers. \textbf{(b)} Fermi surface contours obtained from a tight binding model (TBM) of the coupled Te square nets in the absence of charge density wave order. The blue and orange contours denote the bonding and antibonding bilayer-split branches produced by hybridization between the two Te square nets. The two CDW q-vectors that may nest this system are denoted with the arrows in the \textit{a}$^*$ and \textit{b}$^*$ directions. The Fermi surface pocket assignments are shown, where the $\alpha, \beta, \gamma, \eta,$ and $\delta$ pockets correspond to frequencies near 40, 500, 850, 2000, and 3600 T, respectively~\cite{PhysRevB.110.L161108, j.newton.2025.100320}. \textbf{(c)} Representative Fermi surface reconstruction produced by a single CDW $\mathbf{q}_{\mathrm{1}}$, based on the tight binding model in Brouet \textit{et al.}~\cite{PhysRevB.77.235104}. The green spectral weight highlights CDW-folded and hybridized states, which gap portions of the unreconstructed Fermi surface and generate smaller reconstructed pockets.}
    \label{Fig7}
\end{figure}

We now apply the preceding analysis to Shubnikov--de Haas measurements of four $R$Te$_3$ compounds, whose conducting Te square nets (Fig.~\ref{Fig7}a) give rise to strongly two-dimensional metallic behavior. Larger \textit{R} are expected to have one q-vector $\mathbf{q}_{\mathrm{1}}\approx(2/7)\,b^{*}$, while smaller \textit{R} have a second q-vector $\mathbf{q}_{\mathrm{2}}\approx(1/3)\,a^{*}$. Fermi surface nesting may contribute to the formation of the charge density waves (CDW), which partially gap and reconstruct the underlying Fermi surface~\cite{chikina2023ndte3, PhysRevB.77.235104, PhysRevLett.93.126405, PhysRevB.110.014111, PhysRevB.104.195153, PhysRevMaterials.8.080301, PhysRevB.104.155147, PhysRevLett.99.046401}. As shown in Fig.~\ref{Fig7}c, a tight binding model predicts that the CDW will gap the Fermi surface, introducing the smaller pockets highlighted in green. The Fermi surface pocket assignments commonly used to characterize the $R$Te$_3$ system are shown in Fig.~\ref{Fig7}b. Quantum oscillation measurements can help provide detailed information about the reconstructed Fermi surface geometry and associated quasiparticle properties.

Several of these compounds have been previously characterized by quantum oscillations~\cite{QO_NdTe3, QO_GdTe3_Shiming, QO_PrTe3, QO_MagneticBreakdown}. Our aim is not to revise previously reported Fermi surface assignments but to show how the processing choices can affect what those assignments would be. As an extracted frequency or mass depends on the processing choices taken, values obtained under different choices are not directly comparable, and variations from previously reported numbers are expected. More information on the samples measured and their extracted quantities is available in the Supplemental Material (Sec.~\ref{sec:si-data}). 

\subsection{Resolving weak oscillations at low fields}
High frequency quantum oscillations can be present at relatively low magnetic fields while remaining experimentally undetected. Several possible reasons were discussed in Sec.~III.A, including strong thermal and Dingle damping, a small magnetic breakdown probability, or limited voltage sensitivity. Consequently, Fermi surface orbits observed by ARPES or predicted by electronic structure calculations may be absent from the measured Fourier spectrum, even if the corresponding oscillatory contribution exists. This distinction between an oscillation being present and experimentally detectable is particularly important when interpreting magnetic breakdown. The apparent onset depends on measurement sensitivity and therefore does not directly determine the characteristic magnetic breakdown field~\cite{PhysRevLett.7.231, leeb2025fieldguide}.

We illustrate this problem using TbTe$_3$, which is reported to have a low-temperature CDW transition at 41 K~\cite{PhysRevB.87.155131}. The onset of $\mathbf{q}_{\mathrm{2}}$ is expected to reconstruct the Fermi surface and suppress the high-frequency $\beta$ and $\gamma$ orbits~\cite{QO_MagneticBreakdown}. At sufficiently high magnetic fields, the high-frequency orbits may be recovered through magnetic breakdown. The magnetic breakdown probability is commonly approximated by
\begin{equation}
    P(B) = \exp\left(-\frac{B_0}{B}\right),
    \label{eq:MB_probability}
\end{equation}
where $B_0$ is the characteristic magnetic breakdown field. Because $P(B)$ rises continuously rather than switching on abruptly, there is no field below which the breakdown orbit is strictly absent.
 
\begin{figure}[!htbp]
    \centering
    \includegraphics[width=2.75in]{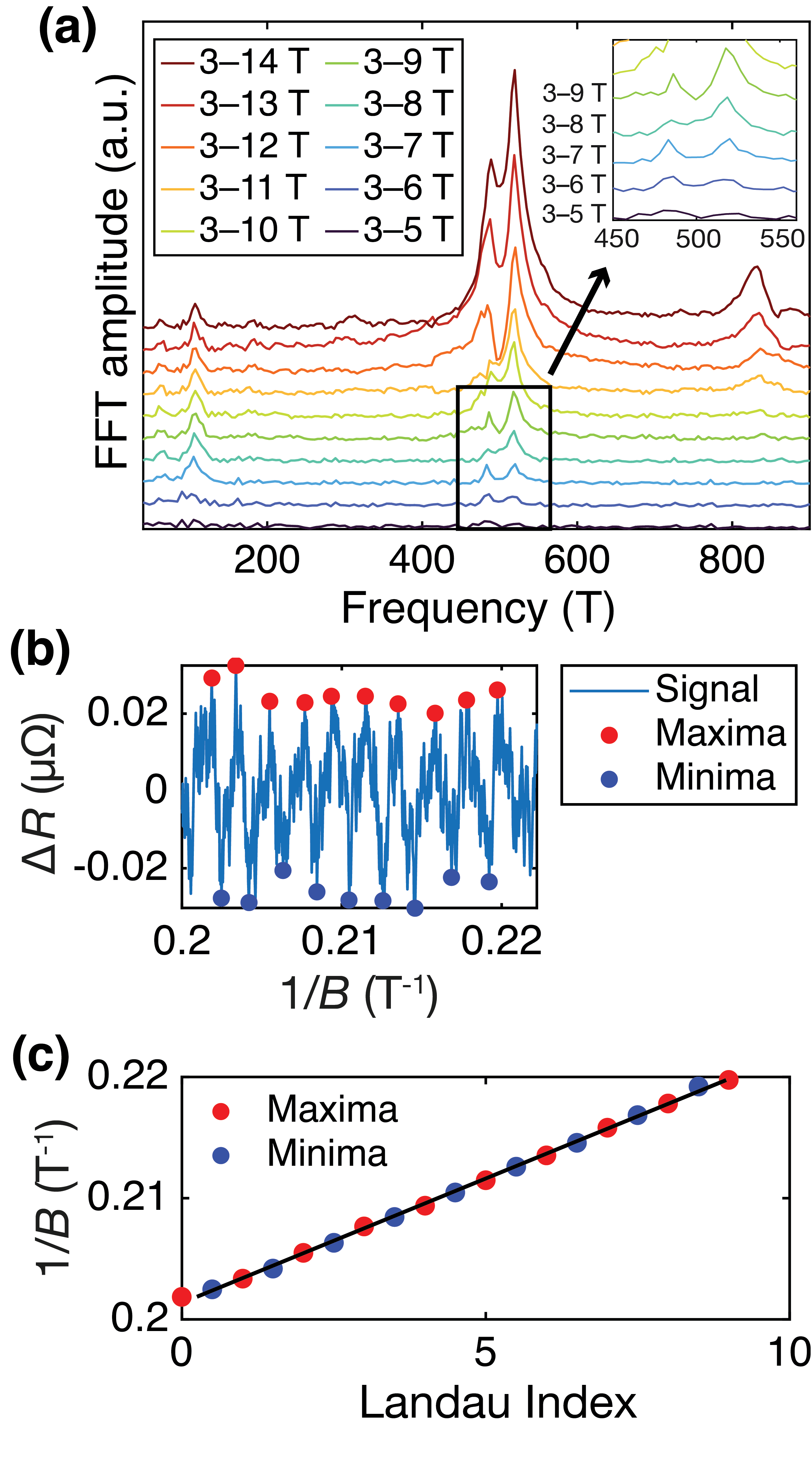}
    \caption{\textbf{(a)} The FFT of TbTe$_3$ from various field ranges showing how different peaks become visible at different field ranges. The inset shows the onset of the $\beta$ peaks from as low as 3--6 T. \textbf{(b)} The weak oscillatory signal from 4.5--5 T and its corresponding \textbf{(c)} Landau fan plot fitting the red maxima and blue minima points, showing linear behavior consistent with the 481 T $\beta_1$ peak.}
    \label{Fig_TbTe3}
\end{figure}

Figure~\ref{Fig_TbTe3}a shows the FFT spectra calculated using a fixed $B_{min}$ of $3~\mathrm{T}$ and $B_{max}$ values ranging from $5$ to $14~\mathrm{T}$. Note that the $\beta$ pocket is visible beginning over $3$--$6~\mathrm{T}$. To test whether this structure represents an oscillatory component rather than an isolated Fourier artifact, we fitted the oscillatory signal directly over the field interval $4.5$--$5~\mathrm{T}$. The extrema fall on an approximately straight line in the Landau fan plot, as shown in Fig.~\ref{Fig_TbTe3}b--c, supporting the presence of a component that is periodic in $1/B$ at fields below $5~\mathrm{T}$.

A second TbTe$_3$ sample (see Fig.~\ref{SI_TbTe3_2_CA}) provides supporting evidence, with oscillatory components from the $\beta$ and $\gamma$ pockets detected in data restricted to fields below $5~\mathrm{T}$. Detection of these components was improved using coherent averaging, a technique which involves averaging over multiple field sweeps; this preserves reproducible quantum oscillations while suppressing uncorrelated noise.

These observations demonstrate that the apparent onset field of a weak high frequency oscillation can be limited by experimental sensitivity. However, they do not uniquely determine the origin of the $\beta$ and $\gamma$ pockets. Alternate interpretations include magnetic breakdown with a lower characteristic field than previously inferred, incomplete removal of the corresponding Fermi surface states by the second CDW, or a CDW reconstruction that primarily affects the crowded low-frequency region.

For a semiclassical magnetic breakdown orbit composed of several orbit segments, the measured frequency and cyclotron mass should be consistent with the corresponding signed orbit areas and mass contributions~\cite{falicov1966,pippard1962}. This provides an important consistency check for a magnetic breakdown assignment, although simple addition or subtraction rules may be complicated by electron--hole trajectories, quantum-interference paths, harmonics, or chemical-potential effects. The Landau fan therefore establishes that the feature is periodic in $1/B$, but not its origin.

\subsection{Fundamental frequencies and harmonics}

As discussed in the previous sections, assigning low-frequency peaks can be difficult, especially in the \textit{R}Te$_3$ system where such small pockets are abundant. A measured peak may represent a fundamental orbit, a harmonic, a combination frequency, or a processing-induced artifact~\cite{leeb2025fieldguide}. Here, we apply the previously introduced harmonic effective mass relations to compare low-frequency features in PrTe$_3$ and NdTe$_3$. 

\begin{figure}[!htbp]
    \centering
    \includegraphics[width=4.48in]{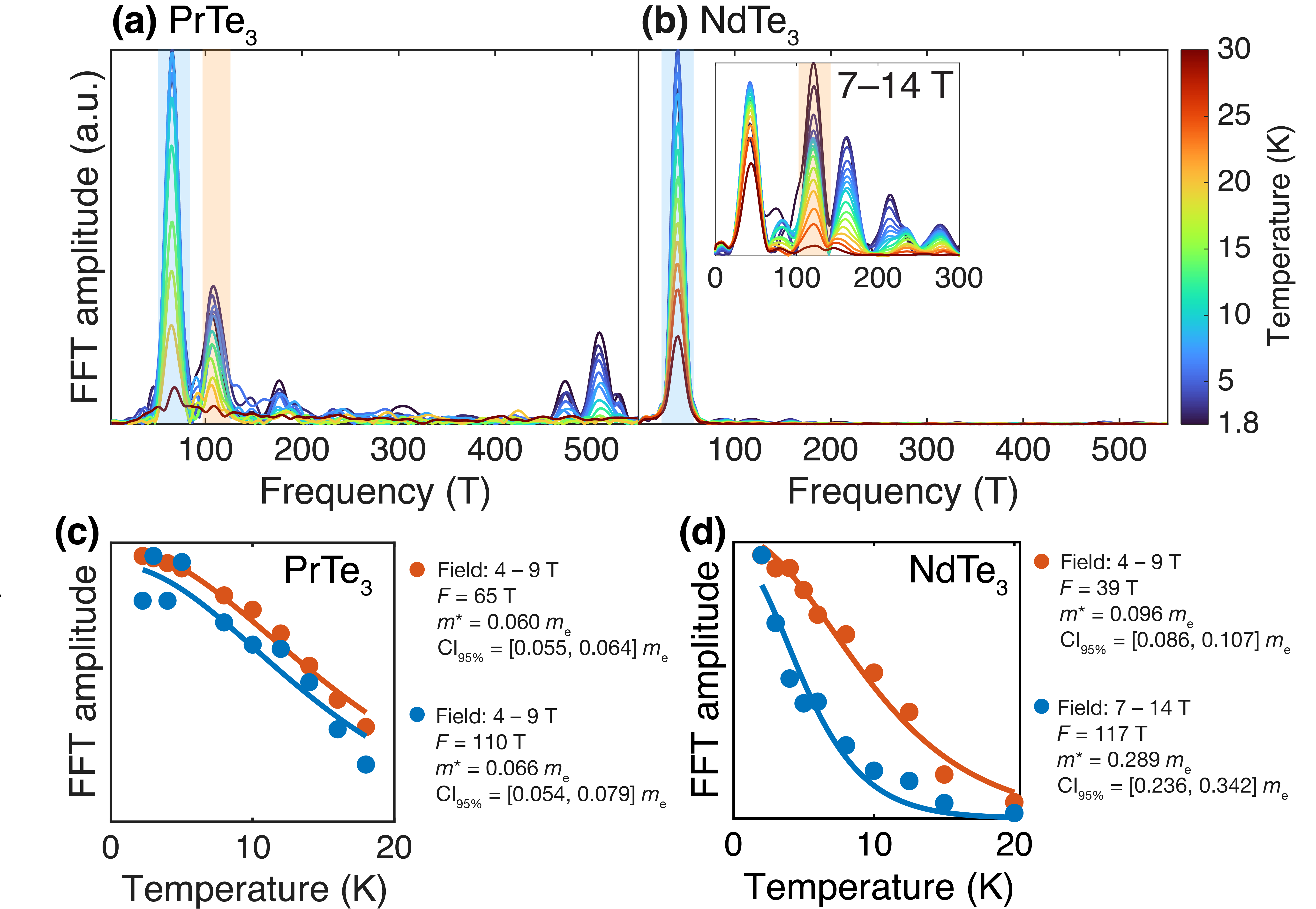}
    \caption{\textbf{(a)} The FFT spectrum of PrTe$_3$ taken over a field range of 4--9T after a fourth-degree polynomial background subtraction. \textbf{(b)} The FFT spectrum of NdTe$_3$ taken over a field range of 4--9T after a fourth-degree polynomial background subtraction, with the inset showing a 7--14T fit after an SG background subtraction of degree 2 and window size of 15 mT$^{-1}$. The LK fits for \textbf{(c)} PrTe$_3$ and \textbf{(d)} NdTe$_3$ corresponding to the shaded regions in the FFT spectrum. Note that the 117 T peak in NdTe$_3$ was fitted over 7--14 T, while the remaining peaks were fitted over 4--9 T.}
    \label{Fig9}
\end{figure}

The $4$--$9~\mathrm{T}$ Fourier spectrum of PrTe$_3$ in Fig.~\ref{Fig9}a contains two prominent low-frequency features near $65$ and $110~\mathrm{T}$. Their temperature dependences, shown in Fig.~\ref{Fig9}c,  give effective masses of 0.060 $m_e$ and 0.066 $m_e$, respectively. This gives a corresponding frequency and mass ratios of $\frac{F_{110}}{F_{65}} \approx 1.69$ and $\frac{m_{110}^*}{m_{65}^*} \approx 1.10.$ The $110~\mathrm{T}$ feature is therefore not the second harmonic of the observed $65~\mathrm{T}$ component. Such an assignment would instead require a frequency near $130~\mathrm{T}$ and an apparent mass near $2m_{65}^* \approx 0.120\,m_e$.

The relatively small mass of the 110~T component also disfavors the simplest sum-orbit interpretation involving the 65~T orbit. A combination at $65 + 45 = 110$~T would require the 45~T peak to have $m^* \approx 0.006\,m_e$, roughly an order of magnitude below the lightest mass observed anywhere in this family. The 110~T feature is also already prominent in the 4--9~T analysis, and its low-field visibility makes a high-field magnetic breakdown origin less likely because the magnetic breakdown probability generally increases with field. As the 110~T component is not related to the observed 65~T component by an integer frequency or mass relationship and is observable at relatively low fields, it is likely a distinct fundamental extremal orbit.

Alternative interpretations of the $110$~T feature remain possible. A combination frequency cannot be excluded solely on the basis of mass relations since they do not always obey a simple sum or difference rule~\cite{leeb2025fieldguide}. Its visibility over 4--9~T also does not exclude a magnetic breakdown orbit with a sufficiently small characteristic breakdown field. However, it is unlikely a second harmonic peak as it requires a fundamental peak near $55$~T.

The extracted mass may also be affected by neighboring components. The 100--200~T region contains many unassigned features, and the frequency resolution over 4--9~T is $\delta F \approx 7.2$~T. Closely spaced components may therefore remain unresolved and contribute to the amplitude tracked at $110$~T, potentially explaining the poorer LK fit obtained for this peak.


The corresponding behavior in NdTe$_3$ is qualitatively different; the $4$--$9~\mathrm{T}$ spectrum in Fig.~\ref{Fig9}b contains a prominent feature near $39~\mathrm{T}$ with an effective mass of 0.096 $m_e$ in Fig.~\ref{Fig9}d. A feature near $117~\mathrm{T}$ becomes visible in the higher-field $7$--$14~\mathrm{T}$ analysis and has an extracted effective mass of 0.289 $m_e$. Therefore, we obtain $\frac{F_{117}}{F_{39}} = 3.00$ and $\frac{m_{117}^*}{m_{39}^*} \approx 3.01$, suggesting that the $117~\mathrm{T}$ feature is possibly third harmonic of the $39~\mathrm{T}$ peak. Its emergence at higher fields provides additional support for this interpretation because higher harmonics are more strongly attenuated by thermal and Dingle damping at lower fields.

This comparison nevertheless contains an important systematic limitation. The $39$ and $117~\mathrm{T}$ masses were extracted over different field intervals using different background-subtraction procedures, complicating the comparison of the extracted masses. Furthermore, the $117~\mathrm{T}$ mass has a relatively large confidence interval. The measured ratios should therefore be interpreted as being consistent with a third-harmonic assignment rather than as a precise verification of the harmonic mass relation. A more stringent test would extract the 39 T and $117~\mathrm{T}$ masses over the same field interval using identical processing parameters. The next subsection shows that lowering B$_{\text{min}}$ resolves this feature into multiple components.

Although PrTe$_3$ and NdTe$_3$ both contain a spectral feature near $110~\mathrm{T}$, their frequencies, effective masses, and field dependence indicate that the two features likely have different origins. The comparison demonstrates how effective mass scaling can differentiate fundamental from harmonic features when it is combined with field dependence and careful consideration of the processing conditions.

\subsection{Apparent deviations from LK behavior}
Experimental factors such as sample imperfections, insufficient signal-to-noise ratios, or phase transitions can introduce complications that prevent the amplitude decay from following the LK form~\cite{Huber2023}. Using NdTe$_3$, we show that another commonly overlooked reason is the resolvability of closely spaced peaks.

\begin{figure}[!htbp]
    \centering
    \includegraphics[width=4.17in]{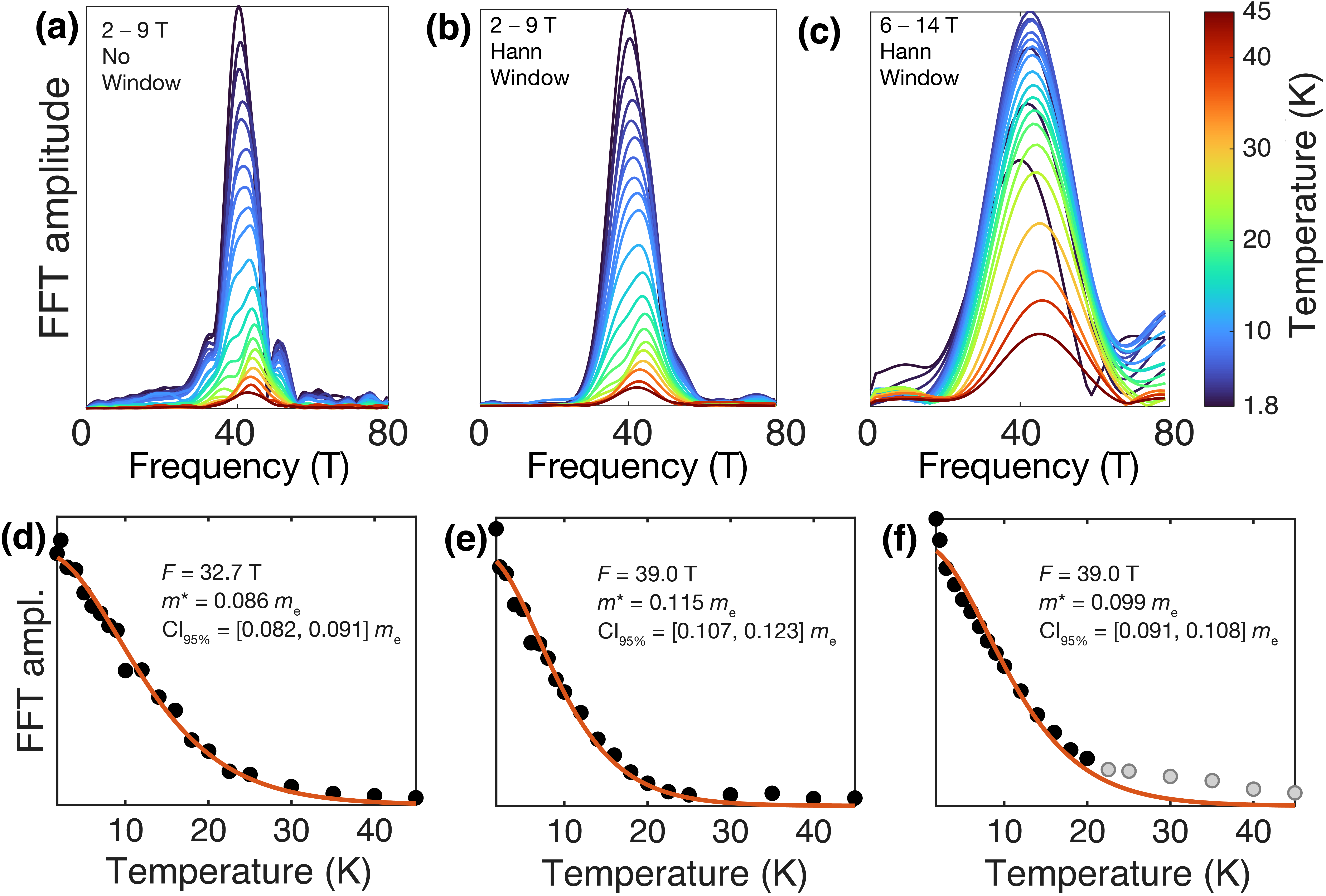}
    \caption{The experimental FFT spectrum of NdTe$_3$ obtained from \textbf{(a)} 3--9 T without Hann windowing, \textbf{(b)} 3--9 T with Hann windowing, and \textbf{(c)} 6--14 T with Hann windowing. The LK fits correspond to the \textbf{(d)} 32.7 T and \textbf{(e)} 39 T peak in (a), and the \textbf{(f)} 39 T peak in (b). The points excluded in (f) lie in the temperature range where anomalous SdH behavior was reported by Dalgaard et al.~\cite{QO_NdTe3}.}
    \label{Fig_NdTe3}
\end{figure}

NdTe$_3$ is reported to have a deviation from LK amplitude decay for the low-frequency $\alpha$ peaks above 20~K~\cite{QO_NdTe3}. To investigate this, we measured the SdH oscillations from 3--14~T across different temperatures. There are three different fitting conditions in Fig.~\ref{Fig_NdTe3}: (a) 3--9~T without a Hann window, (b) 3--9~T with a Hann window, and (c) 6--14~T with a Hann window. As discussed in Sec. III.C, the Hann window in (b) suppresses the sidelobe at the expense of a broader main lobe. The higher $B_\mathrm{min}$ in (c) coarsens the frequency resolution scale from $\delta F \approx 2.6$ T in the 3--9~T case to $\delta F \approx 10.5$~T in the 6--14~T case; consequently, the peaks near 40~T merge into a single broad feature. On closer inspection of (c), the tracked frequency does not sit at a local maximum of the lowest-temperature spectra, unlike in (a) and (b), and the shape of the merged feature changes with temperature. Furthermore, the peaks in (b) and (c) have an asymmetric shape, which indicates that more than one frequency contribution may be present. The changing shape of this feature suggests that its measured amplitude contains multiple oscillatory components. Note also that because the 40~T peak spans fewer than four oscillation periods over the 6--14~T interval, the amplitudes are likely sensitive to background subtraction.

When the $32.7$ and $39.0$~T components are analyzed separately without Hann windowing, their temperature dependencies follow LK damping, giving $m^*=0.086\,m_e$ and $0.115\,m_e$, respectively (Fig.~\ref{Fig_NdTe3}d--e). However, with Hann windowing, the two peaks are less clearly resolved. Consequently, their combined amplitude does not follow the expected LK damping curve because they represent multiple masses with potentially different phases.

Resolving the low-frequency feature also affects the harmonic assignment discussed in Sec.~IV.B. The $117$~T frequency matches the third harmonic of the $39.0$~T component. However, the uncertainty in the fitted $117$~T mass is large enough that three times either the $32.7$~T or $39.0$~T mass falls within its confidence interval. The mass comparison alone therefore cannot distinguish between these two possible parent orbits. A stronger test would compare the resolved fundamental and harmonic using the same field interval and processing procedure.


The hump that appears at 20~K in Fig.~\(\ref{Fig_NdTe3}\)f is thus consistent with an unresolved superposition of neighboring frequencies, and does not necessarily indicate a deviation from LK behavior. This demonstrates that an anomalous LK curve alone is insufficient to establish an intrinsic electronic transition when the corresponding Fourier feature contains unresolved components. Nevertheless, interference should be distinguished from alternative effects such as the positive FFT noise floor and residual background contributions.

\section{Conclusion}
We demonstrated and analyzed the steps for processing quantum oscillations using Shubnikov--de Haas measurements across the rare-earth tritellurides $R$Te$_3$ ($R$ = Pr, Nd, Gd, Tb). In TbTe$_3$, techniques to raise the signal-to-noise ratio of weak oscillatory signals revealed high frequency orbits that were not detected in previous reported measurements. In PrTe$_3$ and NdTe$_3$, the mass relationship between different harmonics provides a way to distinguish a harmonic from independent orbits. Lastly, deviations from LK temperature dependence in NdTe$_3$ were shown to be consistent with multiple unresolved neighboring frequencies.

Although these examples focus on $R$Te$_3$, the effects of field range selection, background subtraction, and spectral windowing apply to all quantum oscillation experiments. Even when the temperature dependence is well described by the LK formula, the extracted mass may depend strongly on the chosen processing parameters. Testing this dependence and reporting the parameters used are therefore essential for assessing the reliability of the mass and comparing results across studies.

\begin{acknowledgments}
This work was supported by the NSF CAREER grant (DMR-2144295) to LMS, by the Air Force Office of Scientific Research under award number FA9550-24-1-0110, by the Gordon and Betty Moore Foundation's EPiQS Initiative through Grant GBMF9064 to LMS, and by the Princeton Center for Complex Materials (PCCM), a National Science Foundation (NSF) Materials Research Science and Engineering Center (MRSEC; DMR-2614599). The authors acknowledge the use of Princeton's Imaging and Analysis Center (IAC), which is partially supported by the PCCM MRSEC (DMR-2614599). This work was supported by the University of Tokyo/Princeton University Strategic Partnership fund. MH acknowledges support from the Japan Society for the Promotion of Science (JSPS) under Grant Nos.~JP25K24552, JP26H00644 and JP26H01290. This work was partially supported by the Japan Science and Technology Agency via JST FOREST (JPMJFR2238). It was also supported by Japan Science and Technology Agency (JST) as part of Adopting Sustainable Partnerships for Innovative Research Ecosystem (ASPIRE), Grant Number JPMJAP2426.
\end{acknowledgments}

\bibliography{main}

\clearpage
\section*{Supplemental Material}
\setcounter{section}{0}
\setcounter{figure}{0}
\setcounter{table}{0}
\setcounter{equation}{0}
\renewcommand{\thesection}{S\Roman{section}}
\renewcommand{\thefigure}{S\arabic{figure}}
\renewcommand{\thetable}{S\Roman{table}}
\renewcommand{\theequation}{S\arabic{equation}}

\section{Deriving the lobe structure and FWHM of a central peak}
\label{sec:si-fwhm}

In this section, we provide the proof on how we get a sinc-shaped peak after a Fourier transform of a finite period sample. Then, we show how we can use the full-width half-maximum (FWHM) of the sinc wave to estimate how many periods of oscillations are in the data. This can then be used to estimate at what minimum field the oscillations become resolvable, assuming that the oscillations persist at the maximum field of the data collection. 

\addtocontents{toc}{\protect\setcounter{tocdepth}{0}}

\subsection*{Deriving the sinc-wave shaped peak}
We first define the following functions and conventions:
\begin{equation}
  \rect(u)=
  \begin{cases}
    1, & |u|\leq \tfrac12,\\
    0, & |u|>\tfrac12.
  \end{cases}
  \qquad
  \sinc(u)=\frac{\sin(\pi u)}{\pi u},
  \qquad \sinc(0)=1,
\end{equation}

The Fourier transforms are given by:
\begin{equation}
  \mathcal F\{f\}(B)=\int_{-\infty}^{\infty}f(x)e^{-i2\pi Bx}\,dx,
  \qquad
  \mathcal F^{-1}\{F\}(x)=\int_{-\infty}^{\infty}F(B)e^{i2\pi Bx}\,dB.
\end{equation}

For a rectangle of width $W>0$ centered at the origin,
\begin{equation}
  f(x)=\rect\!\left(\frac{x}{W}\right),
\end{equation}

The rectangle is nonzero only on the domain $[-W/2,W/2]$.  Therefore

\begin{align}
  \mathcal F\{f\}(B)
  &=\int_{-W/2}^{W/2}e^{-i2\pi Bx}\,dx \\
  &=\left.\frac{e^{-i2\pi Bx}}{-i2\pi B}\right|_{-W/2}^{W/2} \\
  &=\frac{e^{-i\pi BW}-e^{i\pi BW}}{-i2\pi B} \\
  &=\frac{\sin(\pi BW)}{\pi B}
   =W\frac{\sin(\pi BW)}{\pi BW}
   =W\sinc(WB).
\end{align}

Note that by Fourier duality, the inverse transform of a rectangular spectrum is also a sinc function. 

\subsection*{FWHM of a peak produced by the field window}

Quantum oscillations are periodic in inverse field $x=1/B$.
Define the inverse-field window and its center by
\begin{equation}
  \Delta x=\frac{1}{B_{\min}}-\frac{1}{B_{\max}},
  \qquad
  x_c=\frac{1}{2}
  \left(\frac{1}{B_{\min}}+\frac{1}{B_{\max}}\right).
\end{equation}

A pocket with oscillation frequency \(F_0\), measured over this finite window, is represented by
\begin{equation}
  s(x)=S_0e^{i2\pi F_0x}
  \rect\!\left(\frac{x-x_c}{\Delta x}\right).
\end{equation}

Taking the Fourier transform and substituting \(u=x-x_c\) gives
\begin{align}
  S(F)
  &=S_0\int_{x_c-\Delta x/2}^{x_c+\Delta x/2}
    e^{-i2\pi(F-F_0)x}\,dx \\
  &=S_0\Delta x\,
    e^{-i2\pi(F-F_0)x_c}
    \sinc\!\left[\Delta x(F-F_0)\right].
\end{align}

Therefore, the Fourier peak is a sinc function centered at \(F_0\) whose width is controlled only by \(\Delta x\). Since the exponential has unit magnitude, \(F_0\) determines the peak center, whereas \(\Delta x\) determines its width. The normalized amplitude envelope is
\begin{equation}
  A_{\mathrm n}(F)
  =
  \left|\sinc\!\left[\Delta x(F-F_0)\right]\right|.
\end{equation}

Let \(u_{1/2}>0\) satisfy \(\sinc(u_{1/2})=1/2\); then, \(u_{1/2}\approx0.6034\). Therefore, the points at which the amplitude is half are given by $F=F_0\pm\frac{u_{1/2}}{\Delta x}$, and
\begin{equation}
  \boxed{
  \operatorname{FWHM}_{\mathrm{amplitude}}
  \approx
  \frac{1.207}{
  \dfrac{1}{B_{\min}}-\dfrac{1}{B_{\max}}
  }.}
\end{equation}

Therefore, a larger inverse-field interval produces a narrower Fourier peak.
\clearpage
\newpage

\section{Background subtraction methods}
\label{sec:si-background}
An experimental dataset contains finite discrete values, and such datasets are processed using matrix methods. As we will show below, linear algebra can also help show how a background subtraction affects the Fourier representation of a given dataset. 

Quantum oscillations are periodic in inverse field and are embedded within a magnetoresistive background that needs to be subtracted away. Consider a measured signal interpolated to a uniform grid in $x=1/B$:
\begin{equation}
    \bm y=(y_0,y_1,\ldots,y_{N-1})^T.
\end{equation}
A linear background estimate can be written as $\widehat{\bm b}=S\bm y$, where $S$ is a background fitting matrix. The background-subtracted signal is therefore
\begin{equation}
    \bm r=(I-S)\bm y.
    \label{eq:residual}
\end{equation}

The discrete Fourier transform (DFT) is also a matrix operation.  If $F$ is the unitary DFT matrix, then
\begin{equation}
    \widehat{\bm y}=F\bm y,
    \qquad
    \bm y=F^*\widehat{\bm y},
    \qquad
    F^*F=I,
\end{equation}
where the asterisk denotes a complex conjugate transpose (Hermitian).

Here, $\widehat{\bm y}$ represents the vector of complex FFT coefficients where
each entry corresponds to one discrete frequency bin.  Applying the FFT
to Eq.~\eqref{eq:residual} gives
\begin{equation}
    \widehat{\bm r}=K\widehat{\bm y},
    \qquad
    K=F(I-S)F^*.
    \label{eq:fourier_operator}
\end{equation}
Note that $K$ is the background-subtraction matrix after a change of basis into the Fourier domain. 

For the proceeding arguments, it is important to understand the difference between a diagonal and non-diagonal $K$. For example, consider a diagonal $K$ given by
\begin{equation}
    K=
    \begin{bmatrix}
    g_0&0&0\\
    0&g_1&0\\
    0&0&g_2
    \end{bmatrix};
\end{equation}
then, 
\begin{equation}
    \widehat r_j=g_j\widehat y_j,
\end{equation}
where each index $j$ represents a frequency bin. This means that each frequency bin is independently scaled by the diagonal terms in $K$, and that one frequency bin cannot influence the amplitude of another frequency bin. On the other hand, for a non-diagonal matrix:
\begin{equation}
    \widehat r_k=\sum_j K_{kj}\widehat y_j,
\end{equation}
such that an off-diagonal entry $K_{kj}$ with $k\ne j$ can transfer its spectral energy from one frequency  $j$ to another frequency $k$.  We refer to this transfer as \emph{frequency mixing}. Attenuation and frequency mixing are distinct because a diagonal filter can strongly suppress a peak without changing its frequency.

\subsection{Polynomial subtraction}
For a polynomial background of degree $p$, define the design matrix:
\begin{equation}
    X_p=
    \begin{bmatrix}
    1&x_0&x_0^2&\cdots&x_0^p\\
    1&x_1&x_1^2&\cdots&x_1^p\\
    \vdots&\vdots&\vdots&&\vdots\\
    1&x_{N-1}&x_{N-1}^2&\cdots&x_{N-1}^p
    \end{bmatrix}.
\end{equation}
Solving for least-squares, the background and residual terms are then given by
\begin{equation}
    \widehat{\bm b}_{\mathrm{poly}}=H_p\bm y,
    \qquad
    \bm r_{\mathrm{poly}}=(I-H_p)\bm y,
\end{equation}
where
\begin{equation}
    H_p=X_p(X_p^*X_p)^{-1}X_p^*.
    \label{eq:hat_matrix}
\end{equation}
Note that $H_p$ is a matrix that projects the measured data into the polynomial space represented by the functions $1,x,x^2,\ldots,x^p$. Therefore, the residual is orthogonal to every polynomial included in
the fit:
\begin{equation}
    X_p^*\bm r_{\mathrm{poly}}=0.
    \label{eq:orthogonal_residual}
\end{equation}

By observation, Equation~\eqref{eq:hat_matrix} shows that the projection matrix that fits the measured data is a function of the entire data set. In other words, every fitted value depends on the polynomial coefficients, which are themselves determined from all $N$ measurements. Therefore, a change to a single data point within the field interval can change the fitted background throughout the entire interval.

To help with simplification by factorization, consider a QR factorization $X_p=Q_pR$; note that the columns of $Q_p$ represent mutually orthogonal polynomial functions. Substituting for the QR factorization in $H_p$ gives $H_p=Q_pQ_p^*$. Defining $U_p=FQ_p$, the polynomial residual operator in the Fourier basis \(K_p\) is \begin{equation}
    K_p
    =F(I-H_p)F^*
    =I-U_pU_p^*
    =I-\sum_{\ell=0}^{p}u_\ell u_\ell^*,
    \label{eq:poly_fourier}
\end{equation}
where $u_\ell=Fq_\ell$ is the FFT of the $\ell$th sampled polynomial function.  Its entries are
\begin{equation}
    (K_p)_{kj}
    =\delta_{kj}
     -\sum_{\ell=0}^{p}u_{\ell,k}\overline{u_{\ell,j}},
    \label{eq:poly_entries}
\end{equation}
where the bar denotes a complex conjugate. 

The polynomial fit is calculated in a polynomial basis, but the resulting sampled curve can be re-expressed as a sum of the discrete sinusoidal components available to the FFT. In order to construct a non-trivial polynomial function therefore requires contributions from multiple FFT bins. In other words, each of the polynomial basis vectors in the Fourier domain $u_\ell$ generally contains many nonzero components, and the product $u_{\ell,k}\overline{u_{\ell,j}}$ in Eq.~\eqref{eq:poly_entries} is generally nonzero for many $k\ne j$. This means that polynomial subtraction is generally represented by a non-diagonal matrix in the Fourier basis and can consequently mix Fourier amplitudes across different frequency bins. Note that the existence of these off-diagonal terms establishes the possibility of frequency mixing, but does not make any statement regarding the magnitude of the mixing.

\subsection{Savitzky--Golay subtraction}
Unlike the polynomial fit, the SG algorithm performs a polynomial fit locally rather than over the entire data interval. Let the odd window length be $m=2h+1$. At an interior point $n$, the fitted background is a fixed weighted sum of the
$m$ neighboring measurements,
\begin{equation}
    (S_{\mathrm{SG}}\bm y)_n
    =\sum_{t=-h}^{h}c_ty_{n+t}.
    \label{eq:sg_local}
\end{equation}
The coefficients $c_t$ are obtained from the chosen local polynomial degree, but the same coefficient pattern is shifted from one interior point to the next. This repeated local operation is a convolution, and it can be intuitively understood as evaluating the polynomial fit based on the surrounding window of m points about the fitted point.

To see its action on frequency, apply Eq.~\eqref{eq:sg_local} to a
single complex sinusoid $y_n=e^{i\omega n}$:
\begin{align}
    (S_{\mathrm{SG}}\bm y)_n
    &=\sum_{t=-h}^{h}c_te^{i\omega(n+t)}\\
    &=\left(\sum_{t=-h}^{h}c_te^{i\omega t}\right)e^{i\omega n}\\
    &=\widehat c(\omega)e^{i\omega n}.
    \label{eq:sg_sinusoid}
\end{align}
This is the essence of the convolution theorem: the convolution of two functions is a product in Fourier space.
The output contains the same frequency $\omega$ multiplied by a scaling constant (gain) $\widehat c(\omega)$.  It does not contain a new frequency.  Equivalently, if $C$ denotes the periodic convolution matrix formed from the
coefficients $c_t$, then
\begin{equation}
    FCF^*=\diag(\widehat c_0,\ldots,\widehat c_{N-1}),
\end{equation}
where diag(...) represents a matrix with the arguments along the diagonal, and the ideal periodic SG residual operator is
\begin{equation}
    F(I-C)F^*
    =\diag(1-\widehat c_0,\ldots,1-\widehat c_{N-1}).
    \label{eq:ideal_sg}
\end{equation}
Ideal SG subtraction can therefore attenuate frequencies by different
amounts, but it does not mix Fourier bins.

Real measurements, however, are finite rather than periodic.  Near the first and last points, a centered SG window would extend outside the measured interval, so finite-data implementations use special endpoint fits. Consider a decomposition of the fitted SG background matrix
\begin{equation}
    S_{\mathrm{SG}}=C+E,
\end{equation}
where $C$ applies the repeated periodic rule and $E$ contains only the changes required at the endpoints.  In the Fourier basis,
\begin{equation}
    F(I-S_{\mathrm{SG}})F^*
    =\underbrace{\diag(1-\widehat c_k)}_{
       \text{independent attenuation of each bin}}
     -\underbrace{FEF^*}_{
       \text{mixing caused by endpoint treatment}}.
    \label{eq:finite_sg}
\end{equation}
Only the first and last $h$ rows of $E$ are nonzero, so the mixing contribution is supported on at most m of the N rows and its relative importance diminishes as the window occupies a smaller fraction of the record. A numerical evaluation of $A^{\mathrm{SG}}_j$ and $E^{\mathrm{SG}}_{\mathrm{mix},j}$ is shown in Fig.~\ref{FigS2}, where (a)--(b) give the amplitude response across a range of window sizes and (c)--(d) the corresponding off-bin mixing. 

The structural advantage of SG is not that it preserves every Fourier amplitude.  Its diagonal gain $1-\widehat c_k$ can strongly suppress a low-frequency oscillation if the window is chosen poorly.  Rather, the ideal interior SG operation treats each Fourier bin independently, whereas a polynomial subtraction generally couples different bins.  For finite data, SG also mixes frequencies through its endpoint rows, but this contribution becomes relatively small when the window contains only a small fraction of the full record.  The comparison does not establish that SG is less distortive for every signal and every parameter choice; the outcome also depends on polynomial degree, SG degree and window length, endpoint convention, fitting interval, and frequency of interest.
\subsection{Measuring attenuation and frequency mixing}
\label{sec:common_response}

To compare the polynomial and Savitzky--Golay (SG) background subtraction methods on equal footing, we introduce measures that apply to either method. Let \(S\) denote the linear background-fitting operator,
\begin{equation}
    S=
    \begin{cases}
        H_p, & \text{Polynomial fit}\\
        S_{\mathrm{SG}}, & \text{Savitzky--Golay fit}
    \end{cases}
\end{equation}
and define the corresponding background-subtraction operator as
\begin{equation}
    R=I-S.
    \label{eq:general_residual_operator}
\end{equation}
Thus, for sampled data \(\mathbf y\), the fitted background is \(S\mathbf y\)
and the background-subtracted residual is \(R\mathbf y\).

Throughout this discussion, Fourier amplitude refers to a complex coefficient in a particular frequency bin, whereas spectral energy refers to its squared magnitude. We take \(F\) to be the unitary discrete Fourier transform, so Parseval's theorem gives
\begin{equation}
    \|\mathbf y\|_2^2
    =
    \|F\mathbf y\|_2^2.
\end{equation}
Consequently, the total signal energy is the squared Fourier amplitude summed over all frequency bins. Background subtraction is a separate operation and does not generally preserve this energy.

Consider the normalized sampled Fourier mode
\begin{equation}
    \mathbf f_j=F^*e_j,
    \qquad
    \|\mathbf f_j\|_2=1,
    \label{eq:normalized_fourier_mode}
\end{equation}
where \(e_j\) is a unit vector representing a unit amplitude in Fourier bin \(j\) and zero amplitude in all other bins. After background subtraction, the output is
\begin{equation}
    \mathbf r_j=R\mathbf f_j.
\end{equation}
Equivalently, the subtraction operator in the Fourier basis is
\begin{equation}
    K=FRF^*,
\end{equation}
so that
\begin{equation}
    K e_j=F\mathbf r_j.
\end{equation}

The complex amplitude remaining in the original Fourier mode is obtained by projecting the output back onto \(\mathbf f_j\):
\begin{equation}
    g_j
    =
    \mathbf f_j^*R\mathbf f_j
    =
    e_j^*Ke_j
    =
    K_{jj}.
    \label{eq:same_bin_gain}
\end{equation}
We define the same-frequency amplitude response as
\begin{equation}
    A_j
    =
    |g_j|
    =
    \left|\mathbf f_j^*R\mathbf f_j\right|.
    \label{eq:same_bin_response}
\end{equation}
The value \(A_j=1\) indicates that the amplitude in the original mode is unchanged, whereas \(A_j=0\) indicates that no amplitude remains in that mode. These statements alone do not establish complete preservation or complete removal, because the operator may also transfer energy to other Fourier modes. Furthermore, a general linear operator does not require \(A_j\leq1\).

The total output energy after subtraction is
\begin{equation}
    E_{\mathrm{out},j}
    =
    \|R\mathbf f_j\|_2^2
    =
    \|Ke_j\|_2^2.
    \label{eq:total_output_energy}
\end{equation}
This quantity includes both the energy remaining in the original mode and the energy transferred to every other Fourier mode. The component orthogonal to the original mode is
\begin{equation}
    \left(I-\mathbf f_j\mathbf f_j^*\right)R\mathbf f_j.
\end{equation}
The corresponding off-bin mixing energy is therefore
\begin{equation}
    E_{\mathrm{mix},j}
    =
    \left\|
    \left(I-\mathbf f_j\mathbf f_j^*\right)R\mathbf f_j
    \right\|_2^2.
    \label{eq:mixing_definition}
\end{equation}
Because the same-mode and off-mode components are orthogonal,
\begin{equation}
    \boxed{
    E_{\mathrm{mix},j}
    =
    \|R\mathbf f_j\|_2^2
    -
    \left|\mathbf f_j^*R\mathbf f_j\right|^2
    }.
    \label{eq:general_mixing_energy}
\end{equation}
Equivalently, in the Fourier basis,
\begin{equation}
    E_{\mathrm{mix},j}
    =
    \sum_{k\neq j}|K_{kj}|^2.
\end{equation}
Thus, \(A_j\) measures the amplitude retained in the original frequency, whereas \(E_{\mathrm{mix},j}\) measures the output energy appearing in all other frequency bins.

\subsubsection*{Polynomial subtraction}

For a polynomial fit of degree \(p\), the fitting and subtraction
operators are
\begin{equation}
    S=H_p,
    \qquad
    R_p=I-H_p,
\end{equation}
where
\begin{equation}
    H_p
    =
    X_p(X_p^*X_p)^{-1}X_p^*
\end{equation}
is the orthogonal projection onto the sampled polynomial space. A projection matrix satisfies
\begin{equation}
    H_p^*=H_p,
    \qquad
    H_p^2=H_p.
    \label{eq:poly_projection_properties}
\end{equation}

For the normalized Fourier mode \(\mathbf f_j\), define
\begin{equation}
    d_j^{(p)}
    =
    \|H_p\mathbf f_j\|_2^2,
    \qquad
    0\leq d_j^{(p)}\leq1.
    \label{eq:mode_overlap}
\end{equation}
The quantity \(d_j^{(p)}\) is the fraction of the input mode's total energy absorbed by the polynomial projection. Using
Eq.~\eqref{eq:poly_projection_properties},
\begin{align}
    d_j^{(p)}
    &=
    \mathbf f_j^*H_p^*H_p\mathbf f_j\\
    &=
    \mathbf f_j^*H_p^2\mathbf f_j\\
    &=
    \mathbf f_j^*H_p\mathbf f_j.
    \label{eq:poly_overlap_identity}
\end{align}
Thus, although \(d_j^{(p)}\) is defined as the total absorbed energy, it is also numerically equal to the amplitude coefficient of the fitted component along the original Fourier mode. This equality is a special consequence of the orthogonal-projection property of \(H_p\).

If \(X_p=Q_pR\) is a QR factorization in which the columns of \(Q_p\) are orthonormal, then
\begin{equation}
    H_p=Q_pQ_p^*.
\end{equation}
The same quantity can therefore be represented in an orthonormal basis and  evaluated as
\begin{equation}
    d_j^{(p)}
    =
    \|Q_p^*\mathbf f_j\|_2^2
    =
    \sum_{\ell=0}^{p}
    \left|\mathbf q_\ell^*\mathbf f_j\right|^2.
    \label{eq:mode_overlap_qr}
\end{equation}

The same-frequency residual amplitude is
\begin{align}
    g_j^{(p)}
    &=
    \mathbf f_j^*(I-H_p)\mathbf f_j\\
    &=
    \mathbf f_j^*\mathbf f_j
    -
    \mathbf f_j^*H_p\mathbf f_j\\
    &=
    1-d_j^{(p)}.
    \label{eq:poly_same_bin_gain}
\end{align}
Because \(0\leq d_j^{(p)}\leq1\), the same-frequency amplitude response is
\begin{equation}
    A_j^{(p)}
    =
    1-d_j^{(p)}.
\end{equation}
The residual energy remaining specifically in the original Fourier mode is therefore
\begin{equation}
    \left|g_j^{(p)}\right|^2
    =
    \left[1-d_j^{(p)}\right]^2.
    \label{eq:poly_same_bin_energy}
\end{equation}

Because \(H_p\mathbf f_j\) and \((I-H_p)\mathbf f_j\) are orthogonal, their energies add:
\begin{equation}
    \|\mathbf f_j\|_2^2
    =
    \|H_p\mathbf f_j\|_2^2
    +
    \|(I-H_p)\mathbf f_j\|_2^2.
\end{equation}
Using \(\|\mathbf f_j\|_2^2=1\) gives the total residual energy
\begin{equation}
    E_{\mathrm{out},j}^{(p)}
    =
    \|(I-H_p)\mathbf f_j\|_2^2
    =
    1-d_j^{(p)}.
    \label{eq:poly_total_output_energy}
\end{equation}
This total residual energy consists of the energy remaining in the original mode and the energy redistributed to all other modes:
\begin{equation}
    \underbrace{1-d_j^{(p)}}_{\text{total residual energy}}
    =
    \underbrace{\left[1-d_j^{(p)}\right]^2}
    _{\text{energy in original mode}}
    +
    \underbrace{E_{\mathrm{mix},j}^{(p)}}
    _{\text{energy in all other modes}}.
    \label{eq:poly_energy_decomposition}
\end{equation}
Solving for the off-bin contribution gives
\begin{align}
    E_{\mathrm{mix},j}^{(p)}
    &=
    \left[1-d_j^{(p)}\right]
    -
    \left[1-d_j^{(p)}\right]^2\\
    &=
    d_j^{(p)}
    \left[1-d_j^{(p)}\right].
    \label{eq:poly_mixing_energy}
\end{align}

Low-frequency oscillations can have appreciable \(d_j^{(p)}\) when only a small number of periods occur within the fitting interval, because a short segment of a slowly varying sinusoid can resemble a low-degree polynomial. Increasing the polynomial degree enlarges the fitting space:
\begin{equation}
    \operatorname{span}\{1,x,\ldots,x^p\}
    \subseteq
    \operatorname{span}\{1,x,\ldots,x^p,x^{p+1}\}.
    \label{eq:nested_poly_spaces}
\end{equation}
For a fixed sampling grid, fitting interval, and least-squares inner product, an orthonormal basis for the enlarged space can be written as
\begin{equation}
    Q_{p+1}
    =
    \begin{bmatrix}
        Q_p & \mathbf q_{p+1}
    \end{bmatrix},
\end{equation}
where \(\mathbf q_{p+1}\) is orthogonal to all columns of \(Q_p\). It follows
that
\begin{align}
    d_j^{(p+1)}
    &=
    \|Q_{p+1}^*\mathbf f_j\|_2^2\\
    &=
    \|Q_p^*\mathbf f_j\|_2^2
    +
    \left|\mathbf q_{p+1}^*\mathbf f_j\right|^2\\
    &=
    d_j^{(p)}
    +
    \left|\mathbf q_{p+1}^*\mathbf f_j\right|^2\\
    &\geq
    d_j^{(p)}.
    \label{eq:nested}
\end{align}
Thus, increasing the polynomial degree cannot decrease the energy overlap of a given Fourier mode with the fitted polynomial space. Equality is possible when the mode is orthogonal to the newly added polynomial direction.

The off-bin mixing energy does not have to increase monotonically with polynomial degree. From Eq.~\eqref{eq:poly_mixing_energy}, it depends on the quadratic function
\begin{equation}
    E_{\mathrm{mix},j}^{(p)}
    =
    d_j^{(p)}\left[1-d_j^{(p)}\right].
\end{equation}
This value is equal to zero when the mode is completely preserved, \(d_j^{(p)}=0\), and when it is completely removed, \(d_j^{(p)}=1\). Its maximum value of \(1/4\) occurs at the intermediate overlap \(d_j^{(p)}=1/2\). Therefore, increasing the polynomial degree monotonically increases the absorbed energy, but it does not necessarily monotonically increase the absolute off-bin energy.

\begin{figure}[!htbp]
    \centering
    \includegraphics[width=6in]{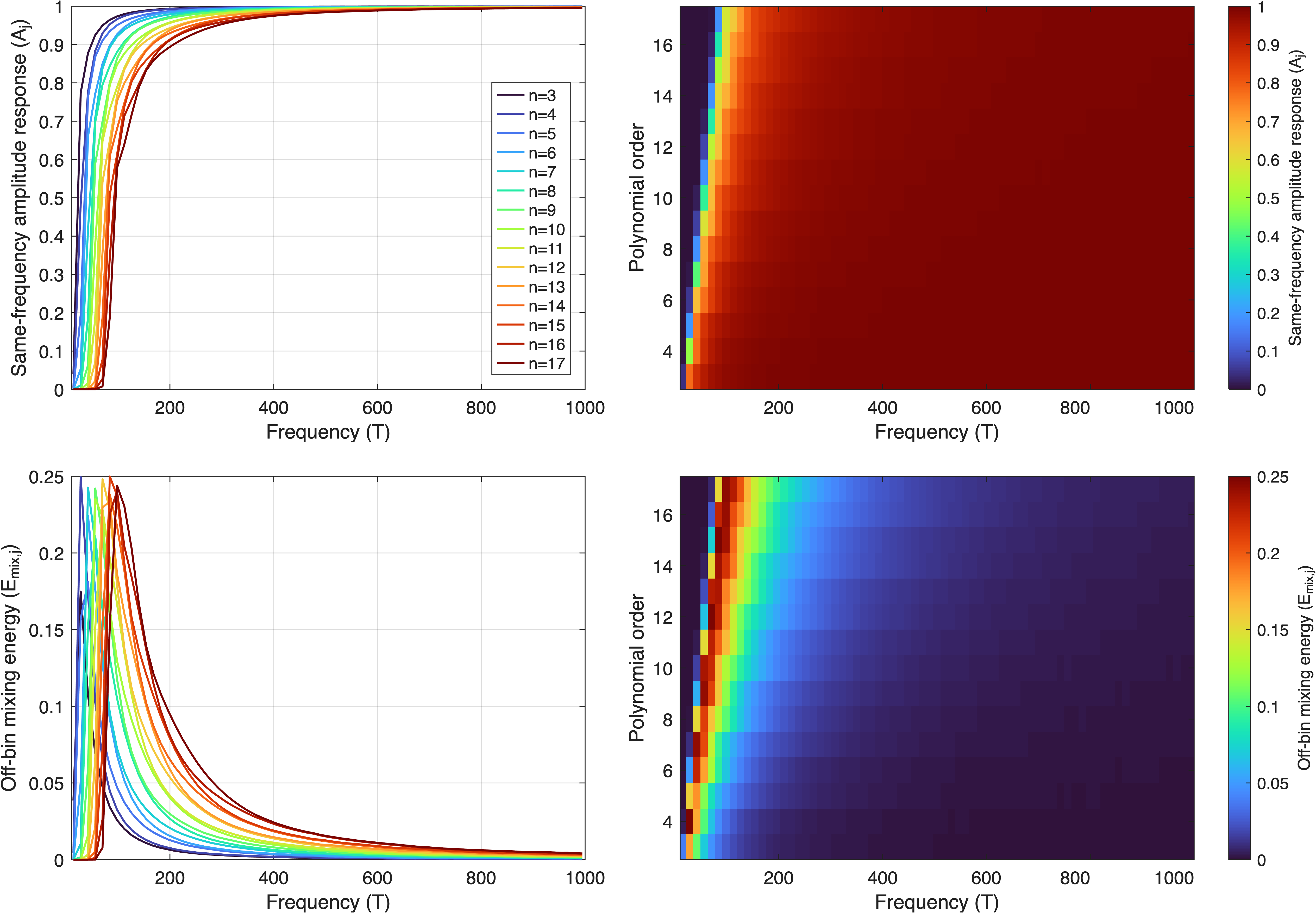}
    \caption{Spectral analysis of a polynomial background fitting under a fitted field of 7-14 T. \textbf{(a)} The same frequency amplitude response under different polynomial degrees. A lower value indicates a higher level of attenuation, and \textbf{(b)} its color map representation. \textbf{(c)} A measure of spectral energy leakage at a given frequency for different polynomial degrees and \textbf{(d)} its color map representation.}
    \label{FigS1}
\end{figure}

A numerical analysis of \(d_j^{(p)}\) and $E_{\mathrm{mix},j}^{(p)}$ for the polynomial background fitting is shown in Fig. \ref{FigS1}. As expected, Fig. \ref{FigS1}a-b shows that increasing the polynomial degree leads to increasing overfitting of the low frequency spectrum. Furthermore, Fig. \ref{FigS1}c-d shows the expected behavior where spectral leakage is non-monotonic and maximized at some intermediate value of frequency. As a practical numerical example, a fit to the sixth polynomial degree leads to a complete suppression of the $\sim$30T frequency, and the off-bin mixing is maximized at $\sim$60T frequency. 

\subsubsection*{Savitzky--Golay subtraction}

For SG subtraction,
\begin{equation}
    S=S_{\mathrm{SG}},
    \qquad
    R_{\mathrm{SG}}=I-S_{\mathrm{SG}}.
\end{equation}
The SG operator performs overlapping local polynomial fits and is generally
not an orthogonal projection:
\begin{equation}
    S_{\mathrm{SG}}^2\neq S_{\mathrm{SG}}.
\end{equation}
Consequently, there is no single SG analogue of \(d_j^{(p)}\) that obeys the same projection identities. In particular, the SG same-frequency response and off-bin mixing must be evaluated directly as
\begin{align}
    A_j^{\mathrm{SG}}
    &=
    \left|
    \mathbf f_j^*
    (I-S_{\mathrm{SG}})
    \mathbf f_j
    \right|,
    \label{eq:sg_same_bin_response}\\
    E_{\mathrm{out},j}^{\mathrm{SG}}
    &=
    \left\|
    (I-S_{\mathrm{SG}})\mathbf f_j
    \right\|_2^2,
    \label{eq:sg_output_energy}\\
    E_{\mathrm{mix},j}^{\mathrm{SG}}
    &=
    \left\|
    (I-S_{\mathrm{SG}})\mathbf f_j
    \right\|_2^2
    -
    \left|
    \mathbf f_j^*
    (I-S_{\mathrm{SG}})
    \mathbf f_j
    \right|^2.
    \label{eq:sg_mixing_energy}
\end{align}

In the limit where the data is periodic, SG smoothing acts as a convolution.
Each Fourier mode is then an eigenvector of the smoothing operator:
\begin{equation}
    S_{\mathrm{SG}}\mathbf f_j
    =
    \widehat c_j\mathbf f_j,
    \label{eq:sg_fourier_eigenvector}
\end{equation}
where \(\widehat c_j\) is the SG smoothing response at frequency \(j\).
The corresponding residual is
\begin{equation}
    (I-S_{\mathrm{SG}})\mathbf f_j
    =
    (1-\widehat c_j)\mathbf f_j.
    \label{eq:sg_ideal_residual}
\end{equation}
The ideal SG subtraction can therefore attenuate or amplify the
same-frequency amplitude according to
\begin{equation}
    A_j^{\mathrm{SG}}
    =
    |1-\widehat c_j|,
\end{equation}
while producing no off-bin mixing:
\begin{equation}
    E_{\mathrm{mix},j}^{\mathrm{SG}}=0.
\end{equation}

Near the ends of a finite data record, the SG fitting window cannot be centered on each point, so the fitting procedure must be modified. Consequently, the rows corresponding to the end points of the dataset can introduce frequency mixing. Nevertheless, this decomposition demonstrates that attenuation and frequency mixing are distinct effects. The convolution-like interior of an SG filter may strongly attenuate a low-frequency component without redistributing it among Fourier bins, whereas the finite-record endpoint treatment is responsible for deviations from this ideal diagonal response.

\begin{figure}[!htbp]
    \centering
    \includegraphics[width=6in]{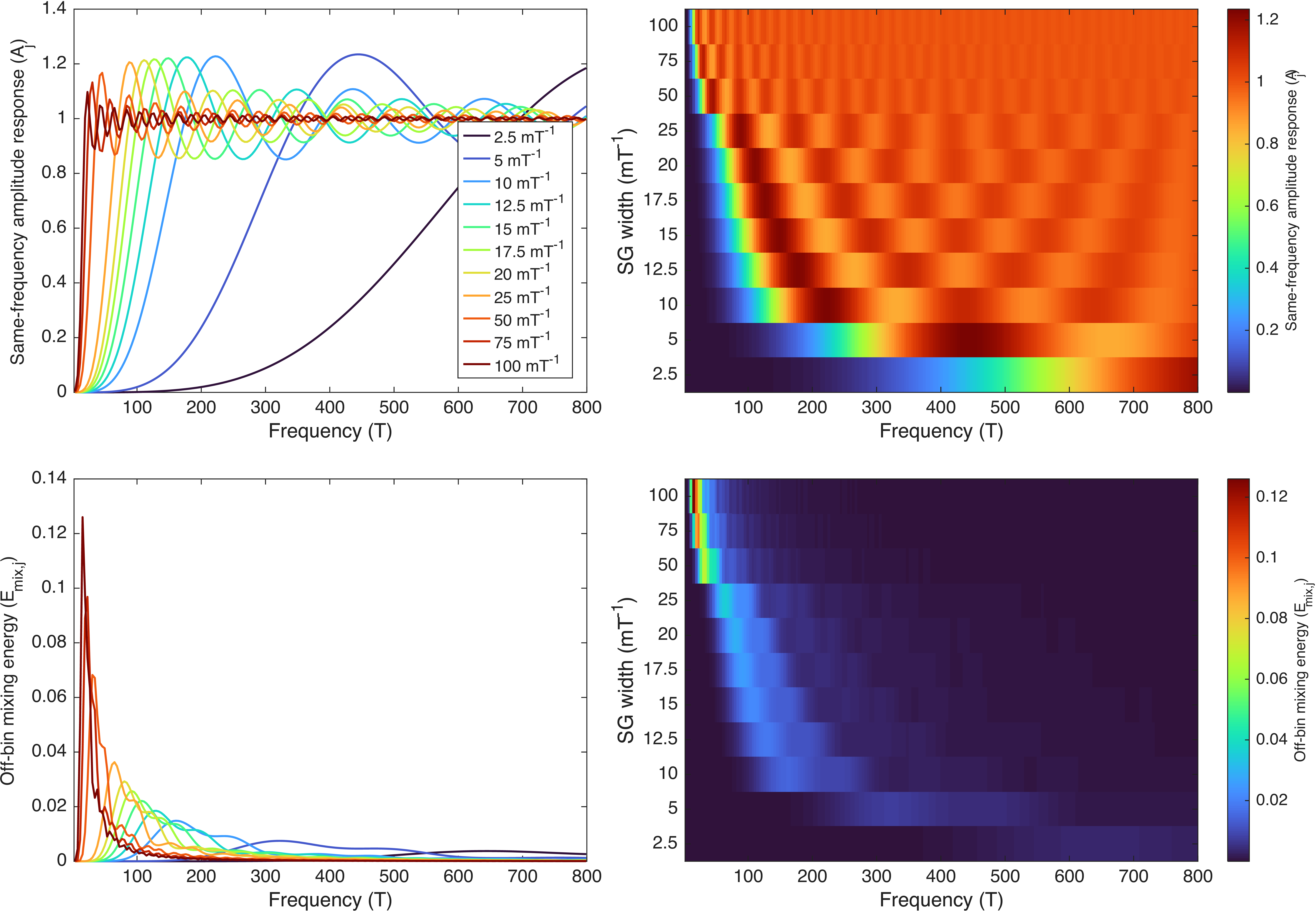}
    \caption{Spectral analysis of a SG background fitting under a field range of 7-14 T. \textbf{(a)} The same frequency amplitude response under SG fitting of various window sizes. A lower value indicates a higher level of attenuation, and \textbf{(b)} its color map representation. \textbf{(c)} A measure of spectral energy leakage at a given frequency for different window sizes and \textbf{(d)} its color map representation.}
    \label{FigS2}
\end{figure}



\clearpage
\newpage

\section{Coherent averaging}
\label{sec:si-coherent}

In this section, we illustrate the mechanism behind coherent averaging, largely following the formulism by Rompelman and Ros.\cite{rompelman1986coherent} For brevity, we consider only one frequency bin, so
\begin{equation}
    z_k=s+n_k,
\end{equation}
where the complex signal $s\in\mathbb{C}$ has the same phase in every measurement and $n_k$ is zero-mean noise. For $N$ repeated measurements, the coherent average is then:
\begin{equation}
    \bar z_N
    =\frac{1}{N}\sum_{k=1}^{N}z_k
    =s+\frac{1}{N}\sum_{k=1}^{N}n_k.
\end{equation}
Since $\E[n_k]=0$, $\E[\bar z_N]=s$. This means that the phase-aligned signal contributions add directly, and is the essence of coherent averaging. For complex noise, define:
\begin{equation}
    \Var(n_k)
    =\E\!\left[|n_k-\E[n_k]|^2\right].
\end{equation}
If the noise samples are independent and each has variance $\sigma^2$,
then
\begin{equation}
    \Var(\bar z_N)
    =\frac{1}{N^2}\sum_{k=1}^{N}\Var(n_k)
    =\frac{\sigma^2}{N}.
\end{equation}
Consequently, the signal-to-noise ratio of the signal amplitude improves as $\sqrt{N}$. This is because in the unnormalized sum, the coherent signal amplitude grows as $N$, whereas the root-mean-square amplitude of independent noise grows only as $\sqrt{N}$. Note that this serves as a general upper bound, and under real experimental conditions, effects such as inconsistent field ramp directions or imperfect contacts can limit the gain in signal amplitude.

This distinction can be made precise by decomposing the noise into a component common to every measurement and a component that varies independently between measurements,
\begin{equation}
    n_k = c + \epsilon_k, \qquad \Var(c)=\sigma_c^2,\quad \Var(\epsilon_k)=\sigma_\epsilon^2,
\end{equation}
where $c$ can be a fixed field-calibration offset, a background-subtraction residual, or hysteresis from always ramping in the same direction, and $\epsilon_k$ is random independent noise. The coherent average then has variance
\begin{equation}
    \Var(\bar z_N) = \sigma_c^2 + \frac{\sigma_\epsilon^2}{N}.
\end{equation}
Only the independent component averages down with $N$; the common component $\sigma_c^2$ remains regardless of how many measurements are combined, setting a noise floor below which additional averaging provides no further benefit. An example of coherent averaging on experimental data is shown in Fig. \ref{SI_TbTe3_2_CA}; several runs were averaged to increase the signal corresponding to the weak 500 T and 800 T features. Coherent averaging is therefore most effective when the dominant noise is random from run to run, and its practical benefit is set by how much of the total noise is instead common across runs.



\clearpage
\newpage

\section{Field-window bias in the extracted cyclotron mass}
\label{sec:si-fieldbias}
\begin{figure}[!htbp]
    \centering
    \includegraphics[width=6in]{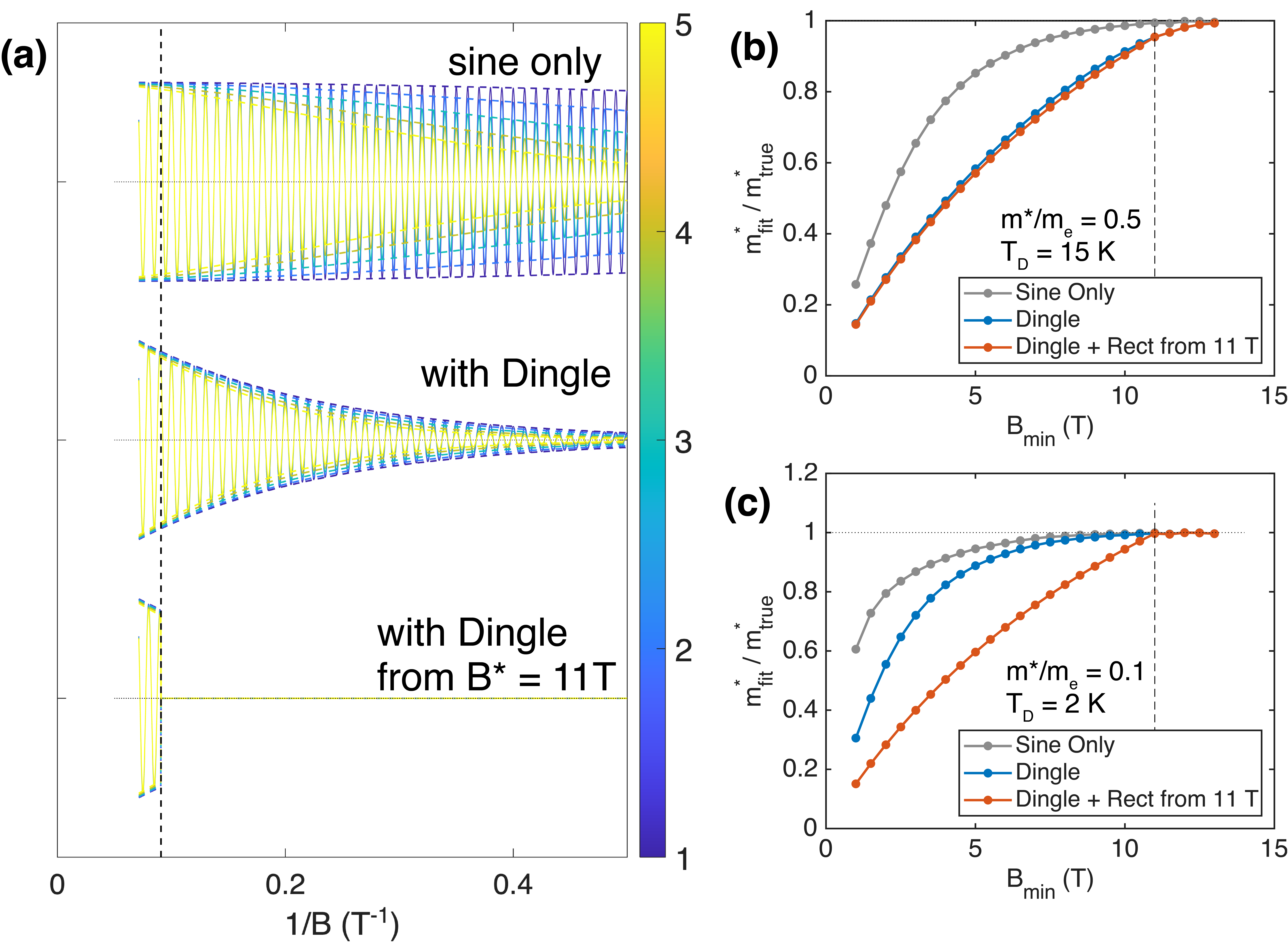}
    \caption{\textbf{(a)} The simulated waveforms representing the three different conditions at different temperatures. The top wave contains a sine wave subject to only thermal damping, the middle wave contains thermal and Dingle damping, and the bottom contains thermal and Dingle damping with oscillations that begin at B$^*$ = 11T. The simulation results are shown for \textbf{(b)} high m$^*$T$_D$ and \textbf{(c)} low m$^*$T$_D$, with the exact values shown in the figures.}
    \label{FigS3}
\end{figure}

In the main text, we discussed how the extracted cyclotron mass is sensitive field selection explored how the temperature and Dingle decay factors each can affect the final extracted quantity. Figure~\ref{FigS3} shows the waveform that reflects each of the three conditions in Fig.~\ref{Fig4}c of the main text, along with how changing the assumptions, such as the Dingle temperature, effective mass, and onset field $B^*$ in the simulation can affect the trends. 

Firstly, we note that Dingle damping is primarily affected by the product m$^*$T$_D$; therefore, we show two cases in subfigures b and c: with high and low m$^*$T$_D$, respectively. Interestingly, we find that a high m$^*$T$_D$ tends to result in a smaller m$^*_{\text{fit}}$. 

\clearpage
\newpage

\section{Tight-binding model of the RTe$_3$ Fermi surface}
\label{sec:si-tb-model}

\subsection*{Band structure of the Te square net}
\label{sec:si-bare-bands}

The low-energy electronic structure of $R\mathrm{Te}_3$ is dominated by a planar square net of Te atoms, in which the relevant orbitals are the in-plane $p_x$ and $p_z$ orbitals~\cite{Brouet2008,DiMasi1995}. Each orbital forms a one-dimensional chain along its own bond direction: $p_x$ chains run along $\hat{x}$, $p_z$ chains along $\hat{z}$, with the two sublattices coupled only weakly to one another. Retaining nearest-neighbor hopping along a chain ($t_\parallel$) and nearest-neighbor hopping perpendicular to a chain ($t_\perp$), the Bloch Hamiltonian for each orbital is obtained by Fourier transforming the real-space hopping Hamiltonian:
\begin{equation}
H_0^{(p_x)} = -t_\parallel \sum_{\langle i,j\rangle_x} c^\dagger_{i}c_j
              -t_\perp \sum_{\langle i,j\rangle_z} c^\dagger_i c_j
              + \mathrm{h.c.},
\end{equation}
and similarly for $p_z$ with $t_\parallel,t_\perp$ interchanged between the two crystallographic directions. With lattice constant $a$ (the Te--Te bond length) this gives the dispersion relations:
\begin{align}
\varepsilon_{p_x}(\mathbf{k}) &= -2t_\parallel \cos(k_x a) - 2t_\perp \cos(k_z a),
\label{eq:eps-px}\\
\varepsilon_{p_z}(\mathbf{k}) &= -2t_\perp \cos(k_x a) - 2t_\parallel \cos(k_z a).
\label{eq:eps-pz}
\end{align}
In the numerical implementation, momenta are expressed in the dimensionless combination $k_x a \to k_x$, $k_z a \to k_z \in [-\pi,\pi]$, so that Eqs.~\eqref{eq:eps-px}--\eqref{eq:eps-pz} are used directly with $a=1$.

The two orbitals hybridize at their band crossings, producing the diamond-shaped pocket observed by ARPES at the $X$ point~\cite{PhysRevB.77.235104,Moore2010}. This hybridization can be represented by the matrix element $t_{\mathrm{hyb}}$, giving the $2\times2$ bare Bloch Hamiltonian in the $\{p_x,p_z\}$ basis:
\begin{equation}
H_0(\mathbf{k}) =
\begin{pmatrix}
\varepsilon_{p_x}(\mathbf{k}) & t_{\mathrm{hyb}} \\
t_{\mathrm{hyb}} & \varepsilon_{p_z}(\mathbf{k})
\end{pmatrix},
\qquad
E_\pm(\mathbf{k}) = \frac{\varepsilon_{p_x}+\varepsilon_{p_z}}{2}
\pm \sqrt{\left(\frac{\varepsilon_{p_x}-\varepsilon_{p_z}}{2}\right)^2
+ t_{\mathrm{hyb}}^2}.
\label{eq:H0}
\end{equation}

\subsection*{Bilayer splitting}
\label{sec:si-bilayer}

$R\mathrm{Te}_3$'s Te net is structurally bilayer; the weak interlayer coupling splits each band into a bonding/antibonding pair, contributing to the multiple closely-spaced frequencies observed in quantum-oscillation experiments~\cite{Ru2008} beyond what the minimal single-layer model captures~\cite{PhysRevB.77.235104}. We treat this with the standard two-site bonding/antibonding construction: with intralayer Hamiltonian $H_0(\mathbf{k})$ identical in each layer and a $\mathbf{k}$-independent interlayer hopping $t_z$, the bilayer Hamiltonian block-diagonalizes into
\begin{equation}
H_\pm(\mathbf{k}) = H_0(\mathbf{k}) \pm t_z\, \mathbbm{1}.
\label{eq:bilayer-block}
\end{equation}

\subsection*{CDW-induced band folding}
\label{sec:si-cdw}

The effect of an incommensurate CDW on the tight-binding bands is treated within the standard mean-field, perturbative electron-phonon framework for density waves~\cite{Gruner1994}, as applied specifically to the $R\mathrm{Te}_3$ tight-binding model by Brouet et al.~\cite{PhysRevB.77.235104} and extended to the two-CDW case by Moore et al.~\cite{Moore2010}. Following this prescription, each CDW with wavevector $\mathbf{q}_i$ is represented by a fixed hybridization $\Delta_i$ connecting $\mathbf{k}$ to $\mathbf{k}\pm\mathbf{q}_i$, truncated at first order. For $R\mathrm{Te}_3$, $\mathbf{q}_1\approx \tfrac27\mathbf{b}^*$ couples the $p_z$ sector and $\mathbf{q}_2\approx \tfrac13\mathbf{a}^*$ couples the $p_x$ sector:
\begin{align}
H_{\mathrm{CDW1}} &= \Delta_1 \sum_{\mathbf{k}}
\left(c^\dagger_{p_z,\mathbf{k}+\mathbf{q}_1} c_{p_z,\mathbf{k}}
+ c^\dagger_{p_z,\mathbf{k}-\mathbf{q}_1} c_{p_z,\mathbf{k}}\right),
\label{eq:cdw1}\\
H_{\mathrm{CDW2}} &= \Delta_2 \sum_{\mathbf{k}}
\left(c^\dagger_{p_x,\mathbf{k}+\mathbf{q}_2} c_{p_x,\mathbf{k}}
+ c^\dagger_{p_x,\mathbf{k}-\mathbf{q}_2} c_{p_x,\mathbf{k}}\right).
\label{eq:cdw2}
\end{align}
$\Delta_1,\Delta_2$ set the primary and secondary CDW gap scales. \cite{Hu2014}

\subsection*{Extended-zone (shadow-band) Hamiltonian}
\label{sec:si-extended}

Truncating at first order in each CDW restricts the relevant momentum
replicas to $\mathcal{K}(\mathbf{k}) = \{\mathbf{k},\,
\mathbf{k}\pm\mathbf{q}_1,\, \mathbf{k}\pm\mathbf{q}_2\}$, each carrying
both orbitals, giving the $10\times10$ Bloch Hamiltonian:
\begin{equation}
H(\mathbf{k}) =
\begin{pmatrix}
H_0(\mathbf{k}) & V_1 & V_1 & V_2 & V_2 \\
V_1^\dagger & H_0(\mathbf{k}+\mathbf{q}_1) & 0 & 0 & 0 \\
V_1^\dagger & 0 & H_0(\mathbf{k}-\mathbf{q}_1) & 0 & 0 \\
V_2^\dagger & 0 & 0 & H_0(\mathbf{k}+\mathbf{q}_2) & 0 \\
V_2^\dagger & 0 & 0 & 0 & H_0(\mathbf{k}-\mathbf{q}_2)
\end{pmatrix},
\label{eq:H-extended}
\end{equation}
with coupling blocks $V_1 = \mathrm{diag}(0,\Delta_1)$, $V_2 = \mathrm{diag}(\Delta_2,0)$ (acting only on the $p_z$ sector for $\mathbf{q}_1$ and the $p_x$ sector for $\mathbf{q}_2$, per Eqs.~\eqref{eq:cdw1}--\eqref{eq:cdw2}). The $6\times6$ block using only the $\mathbf{q}_1$ replicas gives the single-CDW state; the full $10\times10$ form gives the double-CDW state; dropping all off-diagonal coupling recovers $H_0(\mathbf{k})$ for the undistorted state.

\clearpage
\newpage

\section{Experimental Methodology}
\label{sec:si-methods}
\subsection{Synthesis}
Single crystals of \textit{R}Te3 (R = Pr, Nd, Gd, Tb) were grown using a self-flux technique. Te (metal basis >99.999$\%$, Sigma-Aldrich) was mixed with a rare earth (>99.9$\%$, Sigma-Aldrich) in a starting molar ratio of 97:3. The mixture was sealed in an evacuated quartz ampule and heated to 900$^{\circ}$C over 12 hours and then slowly cooled down to 550$^{\circ}$C at a rate of 2$^{\circ}$C h$^{\text{-}1}$. The crystals were separated from the flux by centrifugation at 550$^{\circ}$C, and excess Te flux was removed using chemical vapor transport at 420$^{\circ}$C over 72 hours, with the crystal at the heat source and excess Te flux accumulating at the heat sink.

\subsection{Transport Measurements}
Bulk single crystals of \textit{R}Te3 were exfoliated to ensure minimal oxidation on the surface prior to attaching gold wires with silver paint. The devices were secured onto a sapphire substrate using Elmer's rubber cement to minimize self-exfoliation. A Quantum Design 14 T Physical Property Measurement System was used to perform resistance measurements in conjunction with the Stanford Research System SR830 or SR860 lock-in amplifier paired with a SR554 preamplifier, or a Lake Shore M81 synchronous source measure system. An excitation current of 20 mA was applied at a frequency of 10–30 Hz, and IV measurements were done to ensure that the device was still in the Ohmic regime. The raw data is available in Zenodo: 10.5281/zenodo.23702565.

\subsection{Data Processing}
The processing of Shubnikov-de Haas oscillations was implemented in MATLAB, with the general steps outlined below. The data were initially converted to inverse field and interpolated (using the spline method) onto a uniformly spaced inverse-field grid over $2^{14}$ points. The slowly varying background was then estimated either using a polynomial function or a Savitzky-Golay filter with a window size of 10 mT$^{-1}$ (unless otherwise specified) and to the second polynomial degree. This background was subtracted from the interpolated signal, and the mean of the resulting oscillatory component was removed. A Hann window may be applied before calculating the fast Fourier transform (FFT). Zero padding (up to 4x the number of interpolated points) was used to provide smoother sampling of the frequency spectrum.

At each temperature, the amplitude and position of each oscillation peak were determined from the maximum FFT amplitude within a predefined frequency interval. The temperature dependence of each peak amplitude was subsequently fitted to the thermal damping term of the Lifshitz--Kosevich expression to obtain the cyclotron effective mass. A 95\% confidence interval was estimated from the local nonlinear least-squares covariance matrix as
$m^* \pm t_{0.975,N-2}\sigma_{m^*}$, where $N$ is the number of fitted temperature points and $\sigma_{m^*}$ is the standard error of the fitted mass.

\clearpage
\newpage

\section{Experimental Data}
\label{sec:si-data}
In the section below, we present the experimental data for various \textit{R}Te$_3$ samples, including the RT curves, the energy dispersive X-ray spectroscopy (EDS) spectrum, along with the magnetoresistance curves and extracted oscillatory signals. Furthermore, we show the FFT spectrum of each sample in the field ranges of 7--14 T and 4--7 T; these ranges were chosen to highlight the peaks that exist in the high-field and low-field limits, respectively. Furthermore, the 4--7 T spectrum is effective at showing the spectral features that likely do not originate from magnetic breakdown. Some figures include insets to better show the high-frequency and low-frequency limits of the FFT spectrum. 

In addition, LK fits are provided for select samples. The fits are reported only for peaks satisfying the criteria developed in Sec.~\ref{sec:guidelines} of the main text: the peak frequencies are stable across more than one field window, and their extracted masses are reproducible across the field range and window functions we tested. Furthermore, we highlight how for most $R$Te$_3$ samples, the low-frequency spectrum tends to vary significantly across different samples of the same composition. All extracted masses shown below were extracted using a background subtraction and applied window function as noted in the figure captions. Therefore, the extracted masses may differ from existing reports. 
 
\clearpage
\newpage
\subsection{NdTe$_3$}
\subsubsection*{Sample 1}
\begin{figure}[!htbp]
    \centering
    \includegraphics[width=4in]{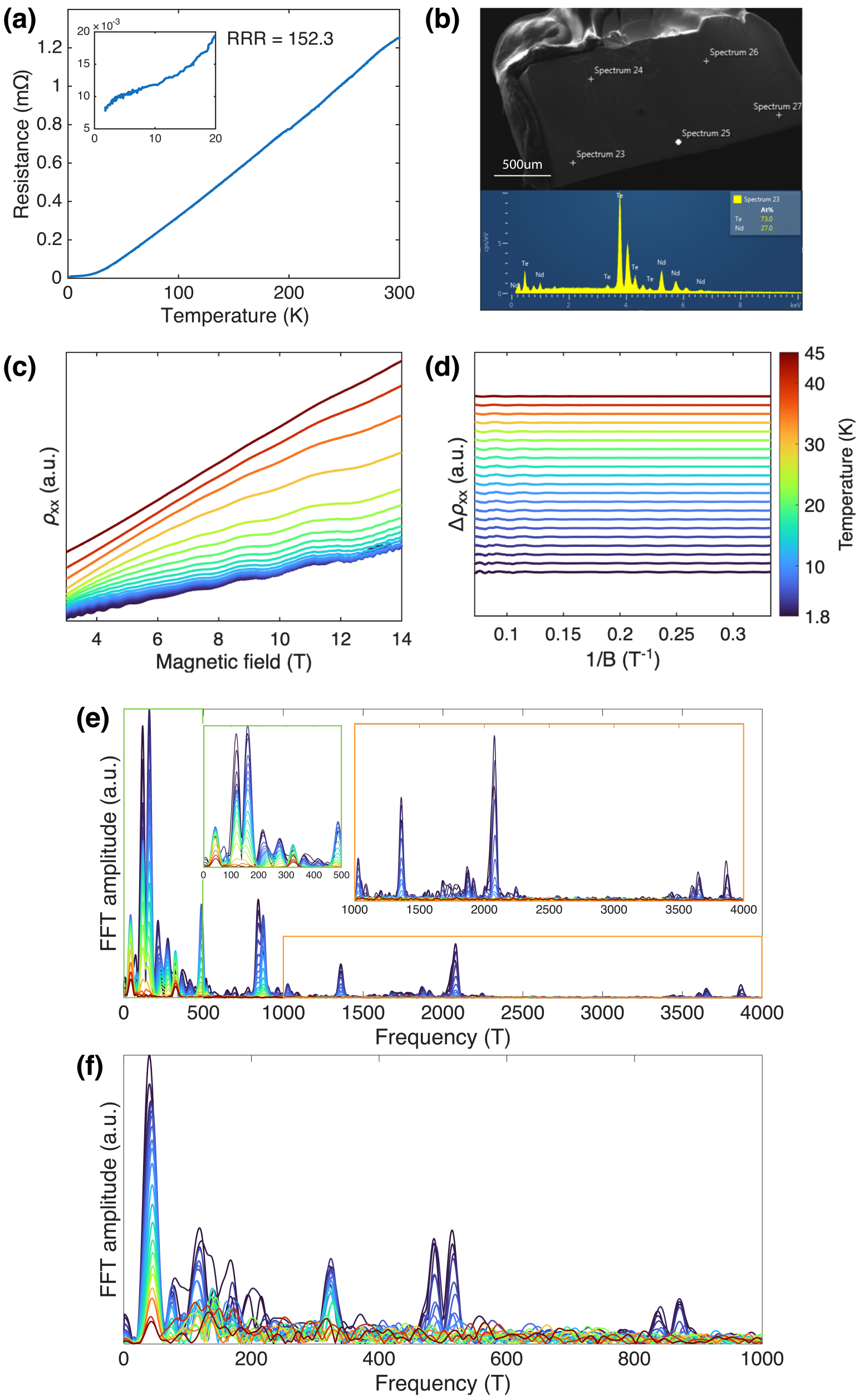}
    \caption{\textbf{(a)} RRR, \textbf{(b)} EDS, \textbf{(c)} magnetoresistance, and \textbf{(d)} oscillatory plots for this NdTe$_3$ sample. The FFT spectrum for NdTe$_3$ across various temperatures over a field range of \textbf{(e)} 7--14 T (with insets magnifying both the low and high frequency regions) and \textbf{(f)} 4--7 T. Both datasets were processed using a SG background subtraction of polynomial degree 2 and a 10 mT$^{-1}$ window with an applied Hann window function.}
    \label{Fig_SI_NdTe3_1}
\end{figure}

\setlength{\tabcolsep}{12pt}
\begin{table}[htbp]
\centering
\caption{Fitted quantum oscillation parameters for the observed peaks of sample 1 of NdTe$_3$. The parentheses denote the 95\% confidence interval.}
\label{tab:qo_pockets_NdTe3_1}
\renewcommand{\arraystretch}{1.2}
\begin{tabular}{cccccc}
\hline\hline
Field (T) & Pocket & $F$ (T) & $m^{*}/m_{e}$ & $T_{D}$ (K) & $\mu_{q}$ (cm$^{2}$V$^{-1}$s$^{-1}$) \\
\hline
\rowcolor{gray!12} 3--9 & $\alpha_1$ & 33 & 0.086(4) & n/a & n/a \\
3--9 & $\alpha_2$  & 39 & 0.115(8) & n/a & n/a \\
\rowcolor{gray!12} 3--9 & $\alpha_3$  & 45 & n/a & n/a & n/a \\
4--9 & $\beta_1$ & 487 & 0.146(11) & 6.2(8) & 2350(300) \\
\rowcolor{gray!12}4--9 & $\beta_2$  & 512 & 0.166(18) & 3.8(5) & 3400(500) \\
6--12 & $\gamma_1$ & 839 & 0.25(2) & 9.2(4) & 930(70) \\
\rowcolor{gray!12}6--12 & $\gamma_2$  & 873 & 0.174(9) & 14(2) & 850(100) \\
8--12 & $\eta_1$ & 2034 & 0.51(5) & 1.7(2) & 2500(300) \\
\rowcolor{gray!12}8--12 & $\eta_2$  & 2084 & 0.43(2) & 2.7(3) & 1900(200) \\
7--14 & $\delta_1$ & 3600 & 0.43(7) & 3.2(2) & 1500(200) \\
\rowcolor{gray!12}7--14 & $\delta_2$ & 3650 & 0.42(2) & 3.6(3) & 1400(100) \\
7--14 & $\delta_3$  & 3875 & 0.60(4) & 3.6(1) & 1000(100) \\
\hline\hline
\end{tabular}
\end{table}

\clearpage
\newpage
\begin{figure}[!htbp]
    \centering
    \includegraphics[width=5.5in]{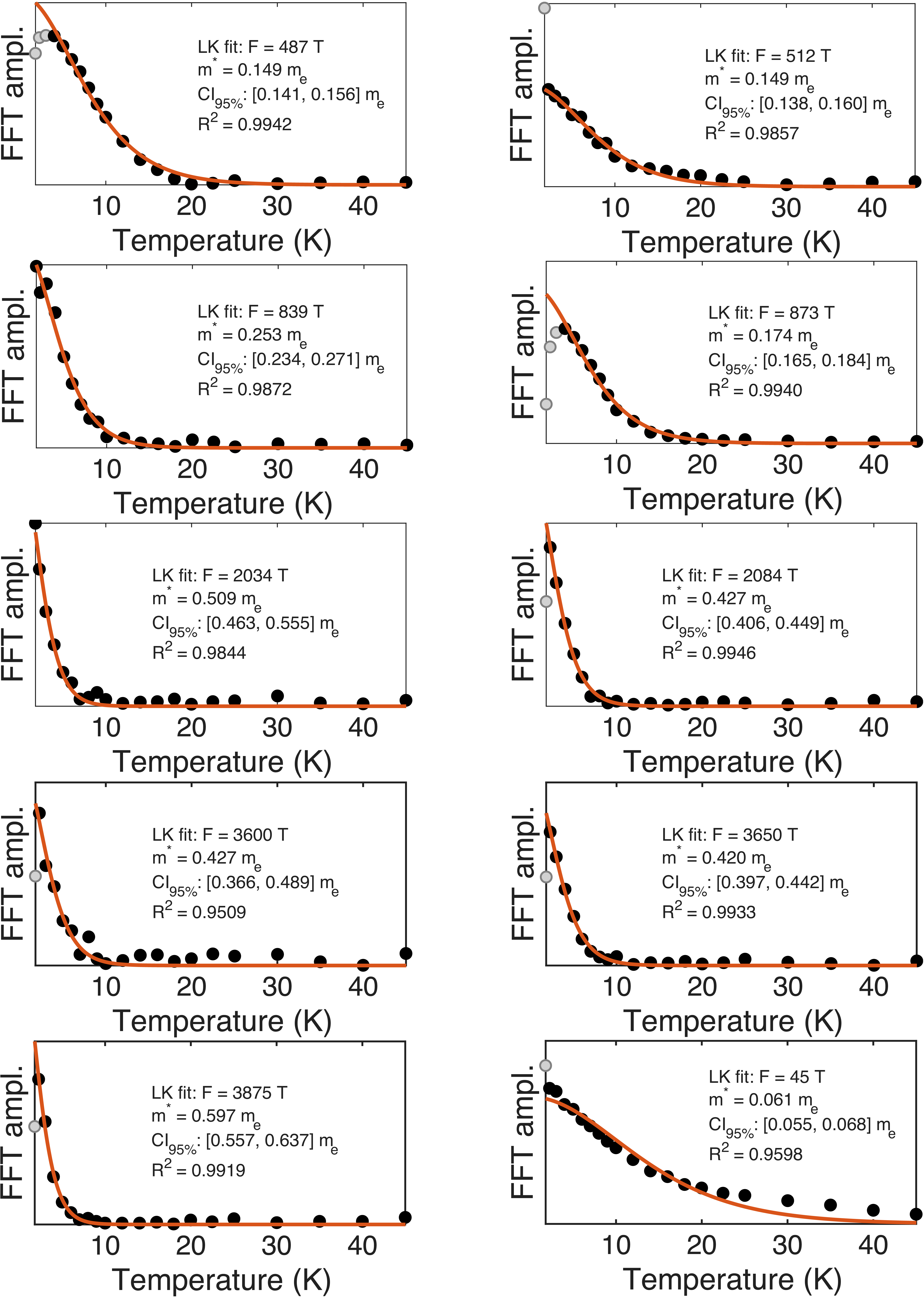}
    \caption{LK Fits of select NdTe$_3$ pockets in Table~\ref{tab:qo_pockets_NdTe3_1}. Several low-temperature points were excluded as the anomalies may be due to the magnetic transition temperatures that onset at 3.2 K. An attempted fit of the 45 T peak is shown, but is not included in Table~\ref{tab:qo_pockets_NdTe3_1} as the decay shape suggests the inclusion of unresolved peaks.}
    \label{Fig_SI_NdTe3_1_LK1}
\end{figure}

\clearpage
\newpage
\begin{figure}[!htbp]
    \centering
    \includegraphics[width=5.5in]{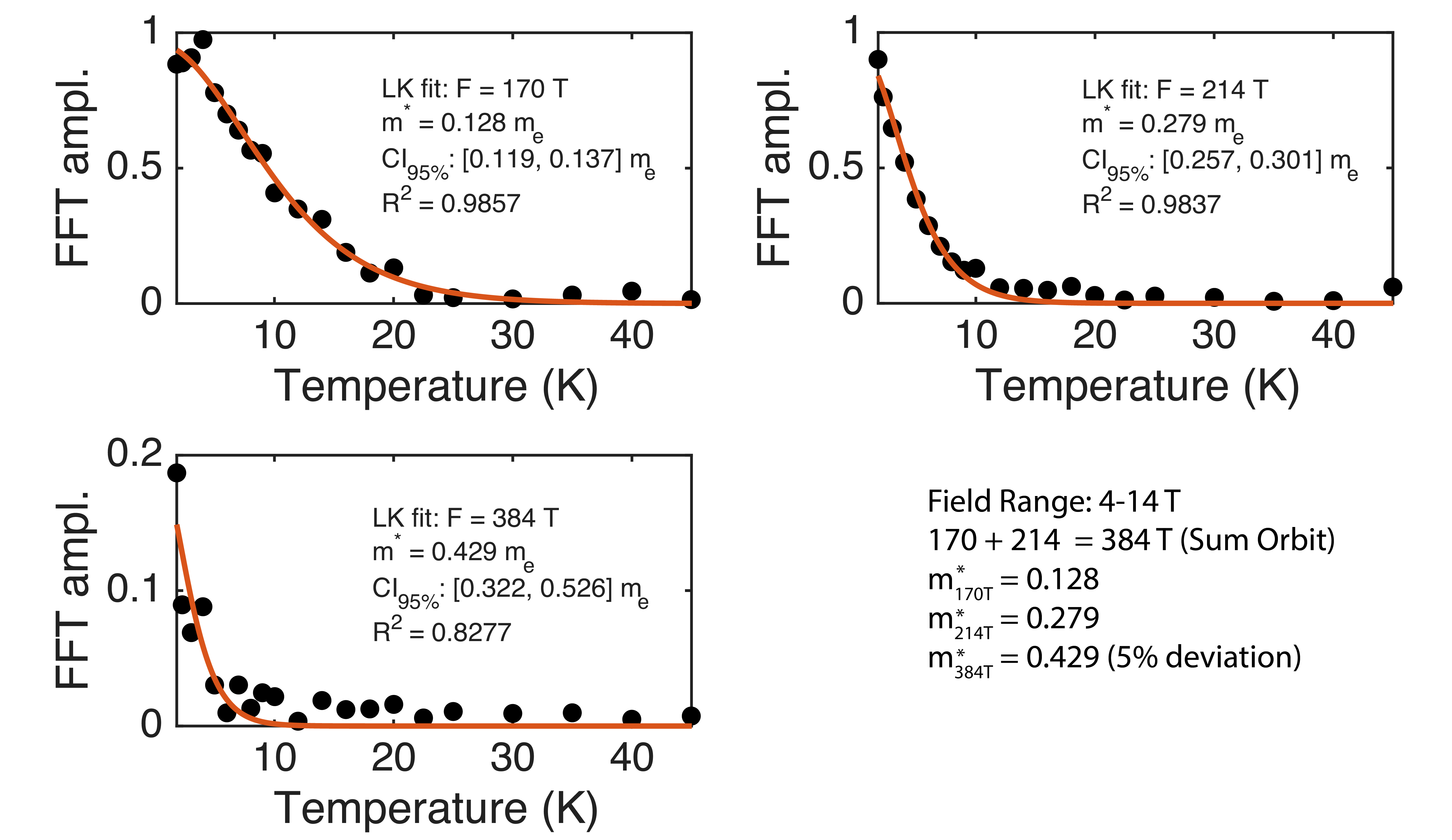}
    \caption{Lifshitz--Kosevich fits to three peaks from NdTe$_3$ extracted over a 4--14~T window ($\delta F = 5.6$~T). (a) $F = 170$~T, $m^* = 0.128\,m_e$; (b) $F = 214$~T, $m^* = 0.279\,m_e$; (c) $F = 384$~T, $m^* = 0.429\,m_e$. The frequencies satisfy $F_c \approx F_a + F_b$ to well within $\delta F$, and the fitted masses agree with $m^*_a + m^*_b = 0.407\, m_e$ to $5\%$. That prediction lies near the center of the $95\%$ confidence interval for (c), $[0.322, 0.526]\,m_e$, while $m^*_a$ alone, $m^*_b$ alone, $2m^*_a$, $2m^*_b$ and $m^*_b - m^*_a$ all fall outside it. The fit in (c) is the poorest of the three ($R^2 = 0.8277$) because its amplitude is both weaker and damped more quickly, so it reaches the noise floor within a few kelvin and the high-temperature points no longer track the thermal damping. The agreement identifies 170 and 214~T as a pair of components consistent with producing the 384~T feature; it does not establish that either is itself a fundamental extremal orbit.}
    \label{Fig_SI_NdTe3_BreakdownOrbit}
\end{figure}

\clearpage
\newpage

\subsubsection*{Sample 2}
\begin{figure}[!htbp]
    \centering
    \includegraphics[width=4in]{SI_Figures/NdTe3/Sample2/SI_NdTe3_sample2.png}
    \caption{\textbf{(a)} RRR, \textbf{(b)} EDS, \textbf{(c)} magnetoresistance, and \textbf{(d)} oscillatory plots for this NdTe$_3$ sample. The FFT spectrum for NdTe$_3$ across various temperatures over a field range of \textbf{(e)} 7--14 T (with insets magnifying both the low and high frequency regions) and \textbf{(f)} 4--7 T. Both datasets were processed using a SG background subtraction of polynomial degree 2 and a 10 mT$^{-1}$ window with an applied Hann window function.}
    \label{Fig_SI_NdTe3_2}
\end{figure}

\clearpage
\newpage

\subsection{GdTe$_3$}
\subsubsection*{Sample 1}

\begin{figure}[!htbp]
    \centering
    \includegraphics[width=4in]{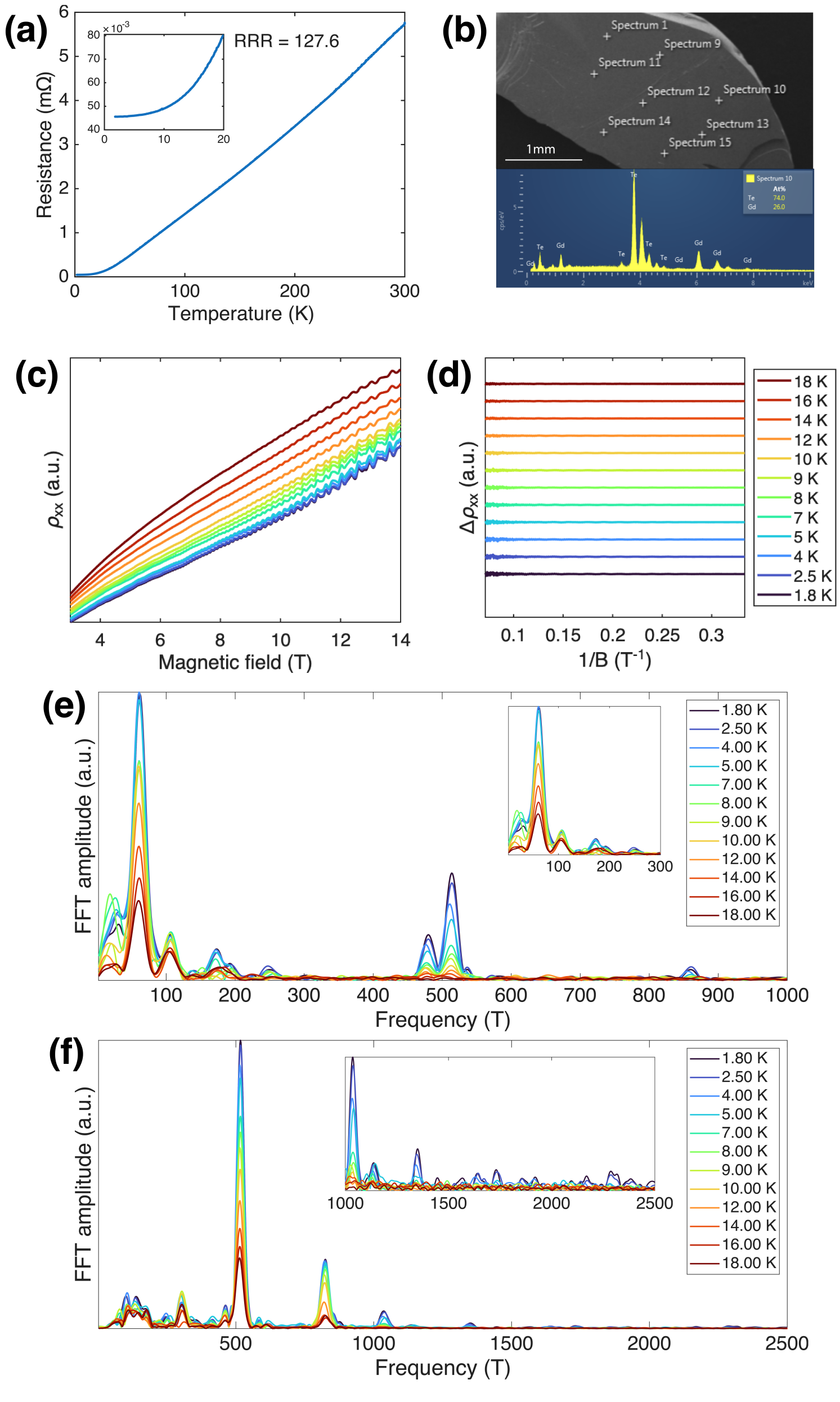}
    \caption{\textbf{(a)} RRR, \textbf{(b)} EDS, \textbf{(c)} magnetoresistance, and \textbf{(d)} oscillatory plots for this GdTe$_3$ sample. The FFT spectrum for GdTe$_3$ across various temperatures over a field range of \textbf{(e)} 7--14 T (with insets magnifying the 0--500 T and 800-2500 T regions) and \textbf{(f)} 4--7 T. Both datasets were processed using a SG background subtraction of polynomial degree 2 and a 10 mT$^{-1}$ window with an applied Hann window function.}
    \label{Fig_SI_GdTe3_1}
    
\end{figure}

\clearpage
\newpage

\subsubsection*{Sample 2}
\begin{figure}[!htbp]
    \centering
    \includegraphics[width=4in]{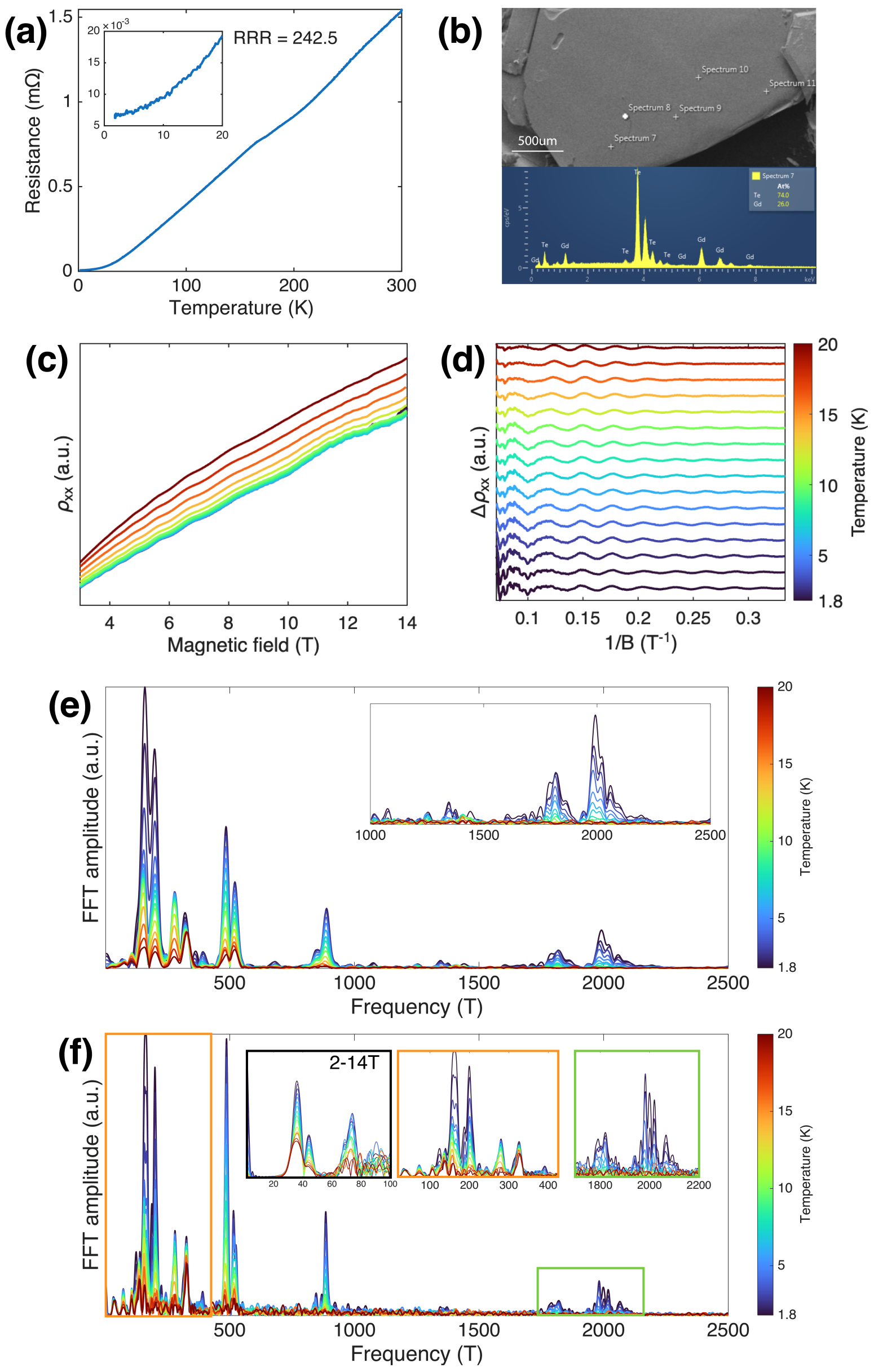}
    \caption{\textbf{(a)} RRR, \textbf{(b)} EDS, \textbf{(c)} magnetoresistance, and \textbf{(d)} oscillatory plots for this GdTe$_3$ sample. The FFT spectrum for GdTe$_3$ across various temperatures over a field range of \textbf{(e)} 7--14 T (with insets magnifying the high frequency regions) and \textbf{(f)} 4--7 T (with insets magnifying the low and high frequency regions). Another inset in (f) shows the FFT spectrum from 2--14 T, emphasizing the existence of a dual peak at 36T and 45T. Both datasets were processed using a SG background subtraction of polynomial degree 2 and a 10 mT$^{-1}$ window with an applied Hann window function.}
    \label{Fig_SI_GdTe3_2}
\end{figure}
\clearpage
\newpage

\setlength{\tabcolsep}{12pt}
\begin{table}[htbp]
\centering
\caption{Fitted quantum oscillation parameters for the observed peaks of sample 2 of GdTe$_3$. The parentheses denote the 95\% confidence interval.}
\label{tab:qo_pockets_GdTe3_2}
\renewcommand{\arraystretch}{1.2}
\begin{tabular}{cccccc}
\hline\hline
Field (T) & Pocket & $F$ (T) & $m^{*}/m_{e}$ & $T_{D}$ (K) & $\mu_{q}$ (cm$^{2}$V$^{-1}$s$^{-1}$) \\
\hline
\rowcolor{gray!12} 2--9 & $\alpha_1$ & 36 & 0.079(3) & n/a & n/a \\
2--9 & $\alpha_2$  & 45 & 0.083(4) & n/a & n/a \\
\rowcolor{gray!12} 6--14 & $\beta_1$  & 484 & 0.179(4) & n/a & n/a \\
6--14 & $\beta_2$ & 516 & 0.127(5) & n/a & n/a \\
\rowcolor{gray!12}6--14 & $\gamma_1$  & 845 & 0.19(2) & 11(1) & 1020(80) \\
6--14 & $\gamma_2$ & 868 & 0.16(2) & 14(1) & 970(50) \\
\rowcolor{gray!12}6--14 & $\gamma_3$  & 886 & 0.220(5) & 8.9(7) & 1070(70) \\
6--14 & - & 1790 & 0.34(6) & 6.8(4) & 930(50) \\
\rowcolor{gray!12}6--14 & - & 1820 & 0.29(2) & 7.2(4) & 1020(50)\\
6--14 & $\eta_1$ & 1982 & 0.39(2) & 5.2(3) & 1050(70) \\
\rowcolor{gray!12}6--14 & $\eta_2$  & 2020 & 0.43(3) & 5.0(3) & 990(60)\\
6--14 & $\eta_3$ & 2065 & 0.46(3) & 4.1(3) & 1120(70) \\

\hline\hline
\end{tabular}
\end{table}

\clearpage
\newpage

\begin{figure}[!htbp]
    \centering
    \includegraphics[width=4.5in]{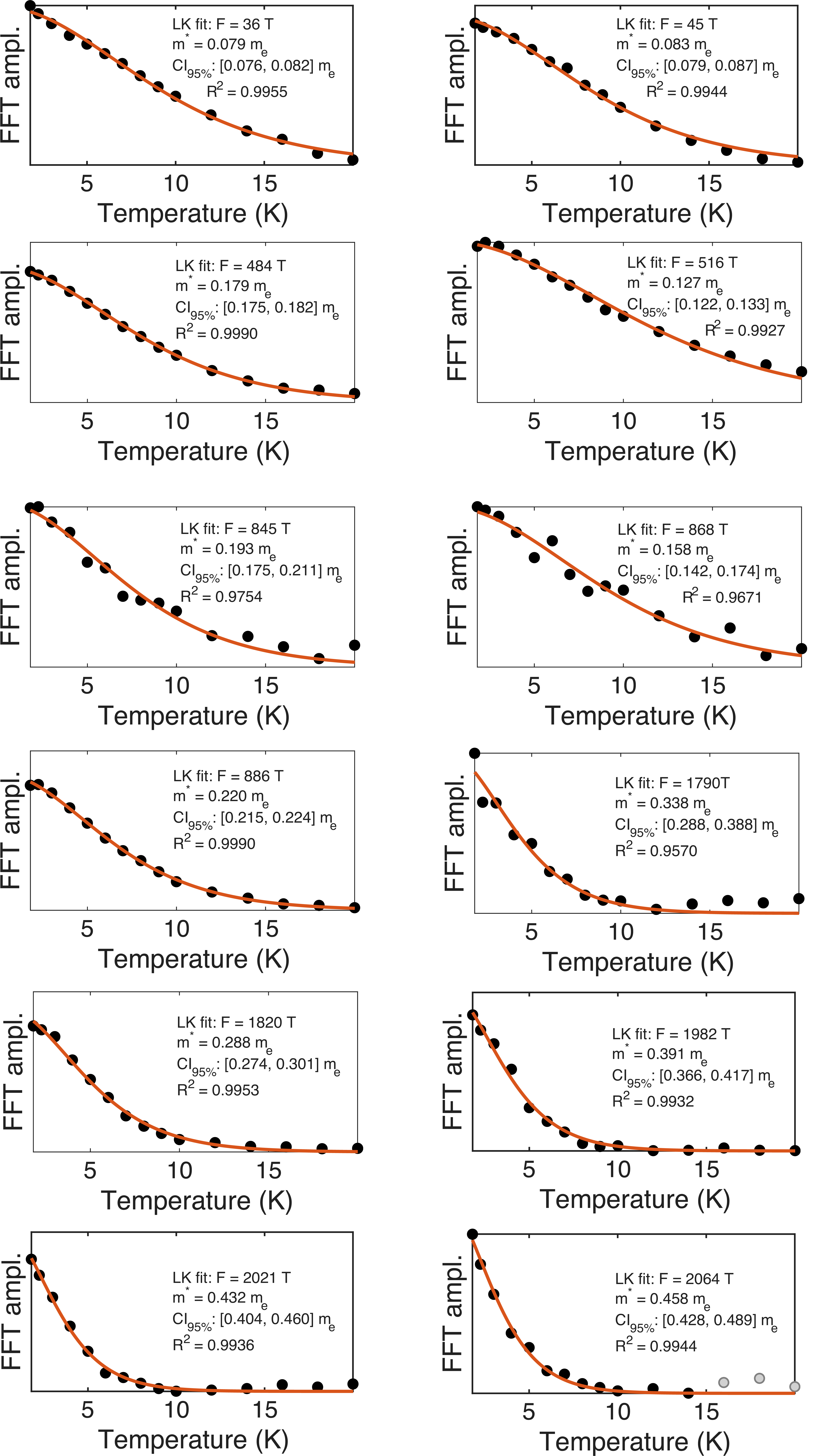}
    \caption{LK Fits of select GdTe$_3$ pockets in Table~\ref{tab:qo_pockets_GdTe3_2}. Some high-temperature amplitudes may be excluded as they may be under the noise floor.}
    \label{Fig_SI_GdTe3_1_LK1}
\end{figure}
\clearpage
\newpage

\subsection{PrTe$_3$}

\subsubsection*{Sample 1}
\begin{figure}[!htbp]
    \centering
    \includegraphics[width=4in]{SI_Figures/PrTe3/Sample1/SI_PrTe3_sample1.png}
    \caption{\textbf{(a)} RRR, \textbf{(b)} EDS, \textbf{(c)} magnetoresistance, and \textbf{(d)} oscillatory plots for this PrTe$_3$ sample. The FFT spectrum for PrTe$_3$ across various temperatures over a field range of \textbf{(e)} 7--14 T and \textbf{(f)} 4--7 T. Both datasets were processed using a SG background subtraction of polynomial degree 2 and a 10 mT$^{-1}$ window with an applied Hann window function.}
    \label{Fig_SI_PrTe3_1}
\end{figure}
\clearpage
\newpage

\subsubsection*{Sample 2}
\begin{figure}[!htbp]
    \centering
    \includegraphics[width=4in]{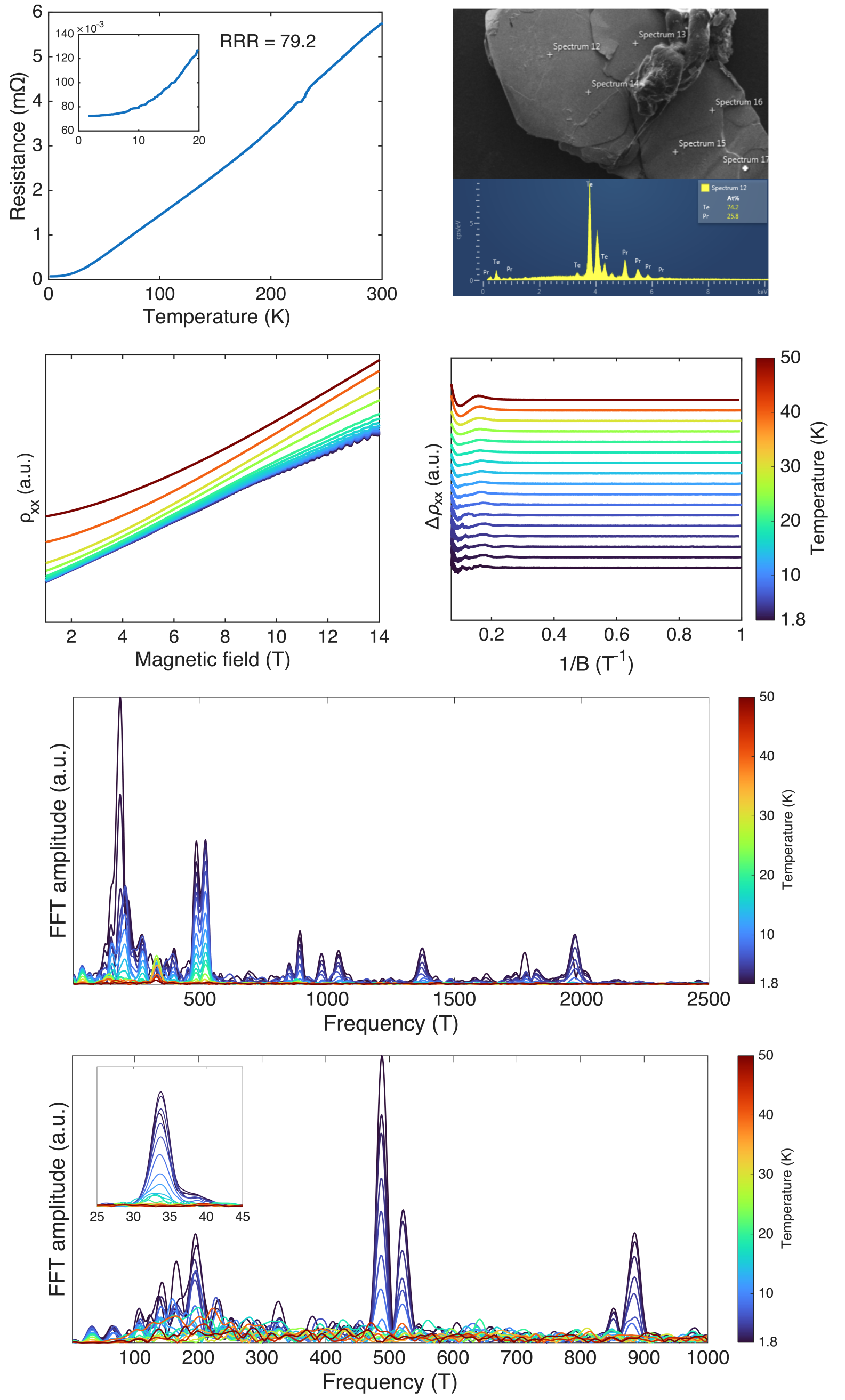}
    \caption{\textbf{(a)} RRR, \textbf{(b)} EDS, \textbf{(c)} magnetoresistance, and \textbf{(d)} oscillatory plots for this PrTe$_3$ sample. The FFT spectrum for PrTe$_3$ across various temperatures over a field range of \textbf{(e)} 7--14 T and \textbf{(f)} 4--7 T, with the inset emphasizing the low-frequency peaks fitted over 1--6 T. Both datasets were processed using a SG background subtraction of polynomial degree 2 and a 10 mT$^{-1}$ window with an applied Hann window function.}
    \label{Fig_SI_PrTe3_2}
\end{figure}
\clearpage
\newpage

\setlength{\tabcolsep}{12pt}
\begin{table}[htbp]
\centering
\caption{Fitted quantum oscillation parameters for the observed peaks of sample 2 of PrTe$_3$. The parentheses denote the 95\% confidence interval.}
\label{tab:qo_pockets_PrTe3_2}
\renewcommand{\arraystretch}{1.2}
\begin{tabular}{cccccc}
\hline\hline
Field (T) & Pocket & $F$ (T) & $m^{*}/m_{e}$ & $T_{D}$ (K) & $\mu_{q}$ (cm$^{2}$V$^{-1}$s$^{-1}$) \\
\hline
\rowcolor{gray!12} 1--6 & $\alpha_1$ & 33 & 0.057(5) & 5.6(2) & 6700(200) \\
1--6 & $\alpha_2$  & 39 & 0.059(11) & n/a & n/a \\
\rowcolor{gray!12} 6--14 & $\beta_1$  & 487 & 0.184(16) & n/a & n/a \\
6--14 & $\beta_2$ & 521 & 0.154(4) & n/a & n/a \\
\rowcolor{gray!12}6--14 & $\gamma_1$  & 850 & 0.239(18) & n/a & n/a \\
6--14 & $\gamma_2$ & 886 & 0.287(14) & n/a & n/a \\
\rowcolor{gray!12}6--14 & $2\beta_1$  & 974 & 0.36(3) & 2.9(1) & 2050(300) \\
6--14 & $2\beta_2$ & 1042 & 0.30(2) & 6.9(6) & 1030(70) \\
\rowcolor{gray!12}6--14 & $\beta_1+\gamma_1$ & 1371 & 0.37(2) & n/a & n/a\\
6--14 & $\eta_1$ & 1970 & 0.36(3) & 8.0(3) & 740(30) \\
\rowcolor{gray!12}6--14 & $\eta_2$  & 2020 & 0.50(9) & 5.3(2) & 800(30)\\

\hline\hline
\end{tabular}
\end{table}
\clearpage
\newpage

\begin{figure}[!htbp]
    \centering
    \includegraphics[width=5.5in]{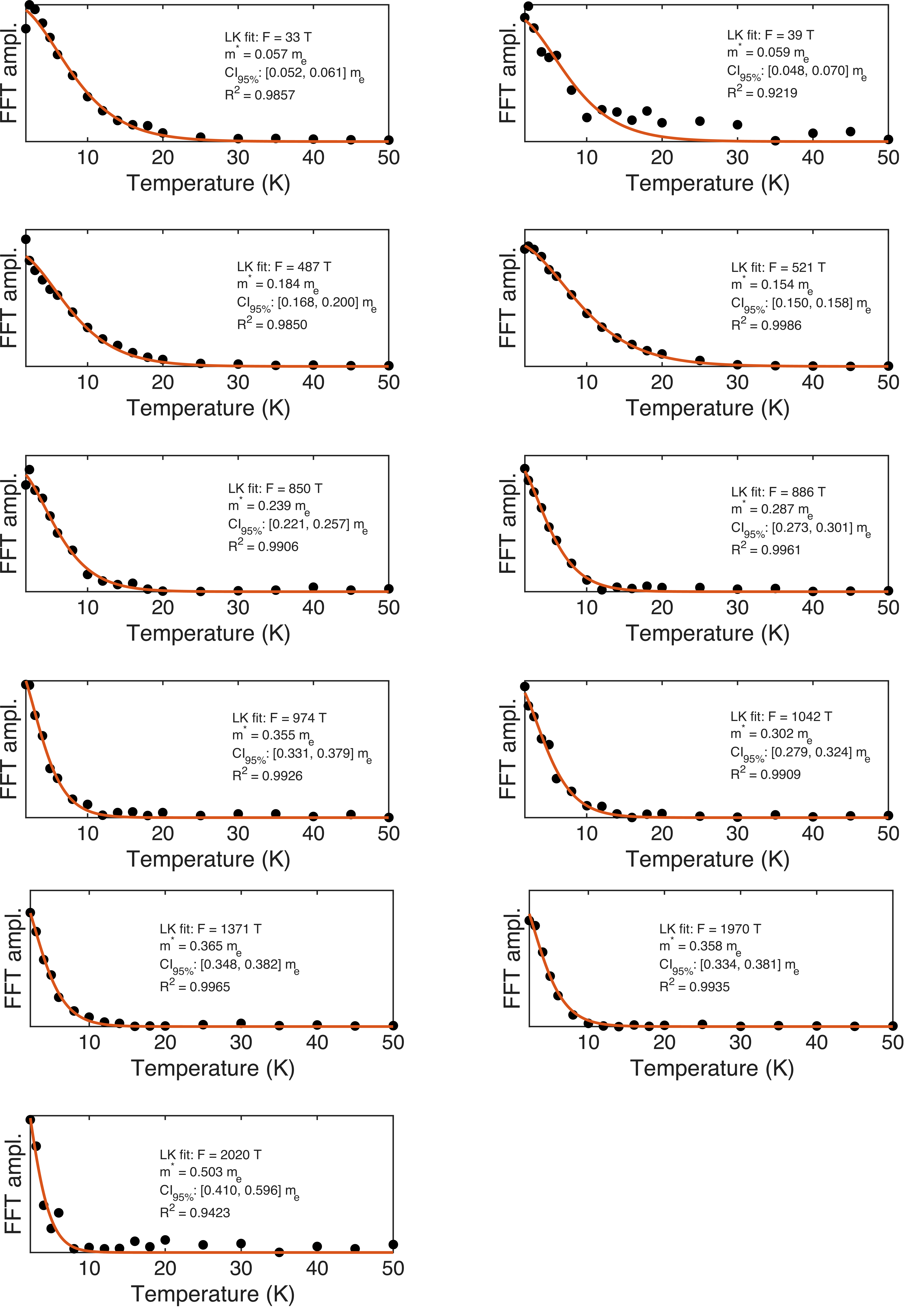}
    \caption{LK Fits of select PrTe$_3$ pockets in Table~\ref{tab:qo_pockets_PrTe3_2}. Some high-temperature amplitudes may be excluded as they may be under the noise floor.}
    \label{Fig_SI_PrTe3_2_LK}
\end{figure}
\clearpage
\newpage

\subsection{TbTe$_3$}
\subsubsection*{Sample 1}
\begin{figure}[!htbp]
    \centering
    \includegraphics[width=4in]{SI_Figures/TbTe3/Sample1/SI_TbTe3_sample1.png}
    \caption{\textbf{(a)} RRR, \textbf{(b)} EDS, \textbf{(c)} magnetoresistance, and \textbf{(d)} oscillatory plots for this TbTe$_3$ sample. The FFT spectrum for TbTe$_3$ across various temperatures over a field range of \textbf{(e)} 7--14 T and \textbf{(f)} 4--7 T. Both datasets were processed using a SG background subtraction of polynomial degree 2 and a 10 mT$^{-1}$ window with an applied Hann window function.}
    \label{Fig_SI_TbTe3_1}
\end{figure}

\clearpage
\newpage
\subsubsection*{Sample 2}
\begin{figure}[!htbp]
    \centering
    \includegraphics[width=4in]{SI_Figures/TbTe3/Sample2/SI_TbTe3_sample2.png}
    \caption{\textbf{(a)} RRR, \textbf{(b)} EDS, \textbf{(c)} magnetoresistance, and \textbf{(d)} oscillatory plots for this TbTe$_3$ sample. The FFT spectrum for TbTe$_3$ across various temperatures over a field range of \textbf{(e)} 7--14 T (with insets magnifying the high-frequency region) and \textbf{(f)} 4--7 T. Both datasets were processed using a SG background subtraction of polynomial degree 2 and a 10 mT$^{-1}$ window with an applied Hann window function.}
    \label{Fig_SI_TbTe3_2}
\end{figure}

\clearpage
\newpage
\begin{figure}[!htbp]
    \includegraphics[width=6in]{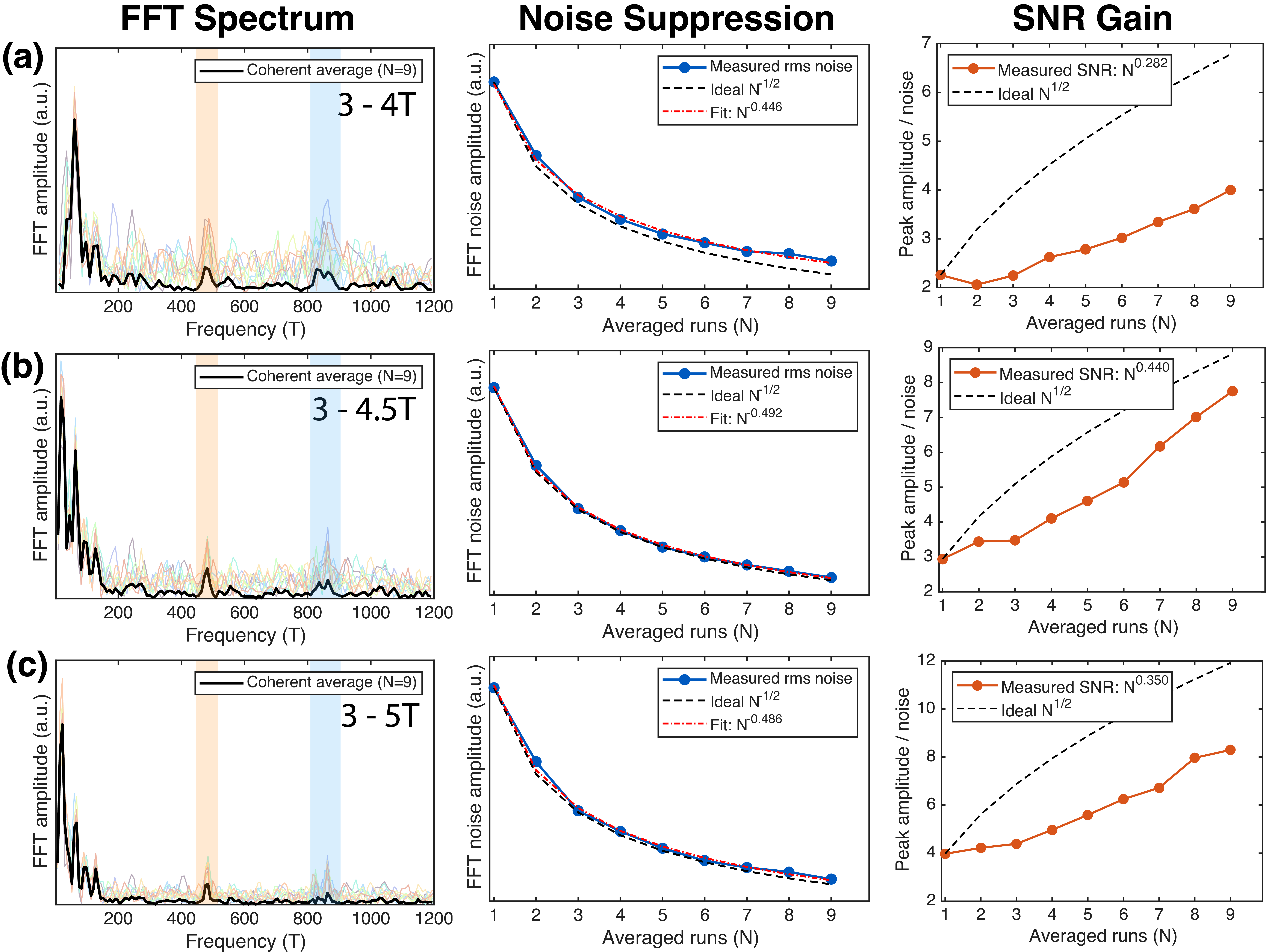}
    \caption{The FFT spectrum of TbTe$_3$ after coherent averaging over 9 runs over a field window of \textbf{(a)} 3--4 T, \textbf{(b)} 3--4.5 T, and \textbf{(c)} 3--5 T. Each row contains the FFT spectrum, the noise amplitude of the FFT calculated using the 650-800T window, and the peak amplitude of the 480T peak as highlighted in orange. The fitted noise exponents are close to the ideal $-1/2$ exponent, but the SNR exponents (0.28–0.44) are smaller than the ideal 1/2. The loss of the coherent peak amplitude may be due to inconsistent field sweep directions or inconsistent drift effects during experiment.}
    \label{SI_TbTe3_2_CA}
\end{figure}
\clearpage
\newpage

\begin{figure}[!htbp]
    \includegraphics[width=6.5in]{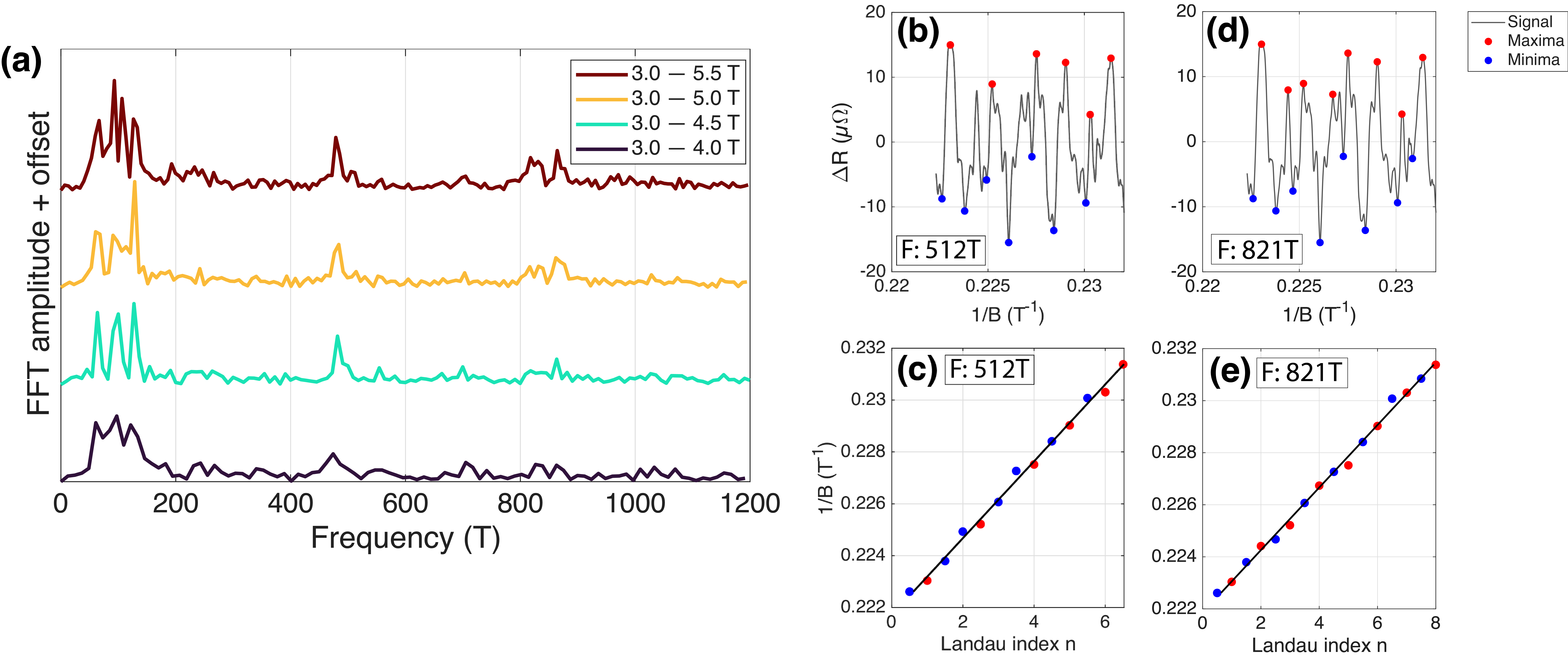}
    \caption{\textbf{(a)}The FFT spectrum of various low-field oscillations showing the early onset of the ~500T and ~800T peaks. The oscillations over a field of 4.3-4.5 T are shown, along with the Landau fits of the \textbf{(b,c)} 512 T peak and \textbf{(d, e)} 821 T peak. A small fit window was chosen to emphasize the small amplitude of oscillations.}
    \label{SI_TbTe3_2_FanPlot}
\end{figure}

\end{document}